\documentclass[aps,prx,twocolumn,colorlinks=true,linkcolor=black,urlcolor=blue,citecolor=black,pdfhttps://www.overleaf.com/project/624ef482c726789d7c1a8f5cencoding=auto,longbibliography,superscriptaddress]{revtex4-2}
\usepackage{graphicx}
\usepackage{color}
\usepackage{physics}
\usepackage{amsmath}
\usepackage{amsfonts}
\usepackage{bbold}
\usepackage{amssymb}
\usepackage{tikz}
\usepackage{bm}
\usepackage{braket}
\usepackage{pgfplots}
\usepackage{dsfont}
\usepackage{comment}
\usepackage{soul}
\usepackage[colorlinks=true,
            citecolor=blue,
            pdfencoding=auto]{hyperref}
\usepackage{blindtext}
\usepackage{mathrsfs}
\usepackage[scr=boondoxo,scrscaled=1.05]{mathalfa}
\usepackage{lineno}
\usepackage{dirtytalk}

\makeatletter
\newcommand{\overleftrightsmallarrow}{\mathpalette{\overarrowsmall@\leftrightarrowfill@}}
\newcommand{\overrightsmallarrow}{\mathpalette{\overarrowsmall@\rightarrowfill@}}
\newcommand{\overleftsmallarrow}{\mathpalette{\overarrowsmall@\leftarrowfill@}}
\newcommand{\overarrowsmall@}[3]{%
  \vbox{%
    \ialign{%
      ##\crcr
      #1{\smaller@style{#2}}\crcr
      \noalign{\nointerlineskip}%
      $\m@th\hfil#2#3\hfil$\crcr
    }%
  }%
}
\def\smaller@style#1{%
  \ifx#1\displaystyle\scriptstyle\else
    \ifx#1\textstyle\scriptstyle\else
      \scriptscriptstyle
    \fi
  \fi
}
\makeatother
\newcommand{\te}[1]{\overleftrightsmallarrow{#1}}

\begin{document}

\title{Anomalous and Topological Hall Effects with Phase-Space Berry Curvatures: Electric, Thermal, and Thermoelectric Transport in Magnets}

\author{Zachariah Addison}%
\affiliation{Department of Physics and Astronomy,
Wellesley College,
Wellesley, MA 02482, USA}
 \author{Lauren Keyes}%
\affiliation{Department of Physics,
The Ohio State University,
Columbus, OH 43210, USA}
\author{Mohit Randeria}
\affiliation{Department of Physics,
The Ohio State University,
Columbus, OH 43210, USA}

\date{\today}

\begin{abstract}
We develop a theory for the Hall, Nernst, and thermal Hall effects in magnetic materials that harbor topological spin textures
such as skyrmions. In addition to the ordinary response to an external magnetic field, there are two other contributions that have been described
by two distinct theories: an intrinsic anomalous response arising from momentum-space Berry curvature and 
a topological response from the real-space Berry curvature related to the topological charge density of the spin texture. 
We develop here a unified semiclassical theory that incorporates the effects of all phase-space Berry curvatures on an equal footing within a controlled calculation.
We analyze the electrical and thermal currents carried by electrons with arbitrary dispersion and spin-orbit coupling (SOC) 
of strength $\lambda$ interacting with arbitrary three-dimensional spin textures via an exchange coupling $J$.  
For small $\lambda/J$, we show that all linear response conductivities – electrical, thermoelectric, and thermal Hall – can be written as the 
sum of three contributions: ordinary, anomalous, and topological. We derive various general relations including the Weidemann-Franz, Kelvin, and Mott relations. 
We show that there is a topological phase transition as a function of $\lambda/J$ at which the semiclassical approach breaks down,
and the large $\lambda/J$ regime exhibits qualitatively different behavior with a vanishing topological Hall response.
\end{abstract}

\maketitle\

\section{Introduction}
\label{sec-intro}
Systems which break time-reversal symmetry host a variety of phases classified by their symmetry and topology. The Hall conductivity provides a useful probe of these systems. For instance, the Hall conductivity of a 2D insulator famously takes integer values in units of $e^2/h$; the integer, known as the Chern number, characterizes the momentum-space topology and is equal to the Brillouin zone integral of the momentum-space Berry curvature ~\cite{thouless1982quantized}. In contrast, the momentum-space topology in ferromagnetic metals results in an unquantized Hall signal. This anomalous Hall effect ~\cite{nagaosa2010anomalous} is proportional to the magnetization, and in many cases the dominant contribution is due the momentum-space Berry curvature ~\cite{nagaosa2010anomalous, yao2004first, xiao2010berry, haldane2004berry, onoda2006intrinsic} rather than scattering mechanisms. 

The Hall effect is even more interesting in systems that harbor topological spin textures. These systems exhibit a topological Hall effect~\cite{nagaosa2013topological,lee2009unusual, neubauer2009topological, kanazawa2011large, li2013robust, gallagher2017robust, ahmed2018chiral, ahmed2019spin, shao2019topological}, typically analyzed as a signal that appears in addition to the ordinary and anomalous Hall effect signals. The topological Hall effect arises from the real-space Berry phase~\cite{ye1999berry, bruno2004topological, nagaosa2012gauge, nagaosa2013topological, kim2013chirality, akosa2019tuning} acquired by conduction electrons moving in an ``emergent electromagnetic field" of topological spin textures.

Over the last two decades, a host of magnetic materials~\cite{roessler2006spontaneous, neubauer2009topological, nagaosa2013topological, fert2017magnetic, tokura2020magnetic} have been discovered to exhibit topologically nontrivial magnetic textures, including skyrmion crystals, disordered skyrmion arrays, and hedgehog crystals. These topological textures are most common in systems with broken inversion symmetry. In such systems, skyrmions are stabilized by the interplay between ferromagnetic exchange and bulk or interfacial Dzyaloshinskii-Moriya interaction. Some centrosymmetric crystals may also harbor skyrmions. In these crystals, other competing interactions are responsible for skyrmion stability ~\cite{tokura2020magnetic}. Although the topological Hall effect has by far been the best studied transport signal in these magnets, experiments suggest that thermoelectric and thermal transport also probe the topology of skyrmion phases in addition to a texture's scalar spin chirality~\cite{shiomi2013topological, hirschberger2020topological, kolincio2021large, scarioni2021thermoelectric, macy2021magnetic, zhang2021topological}.

We present here a comprehensive semiclassical theory of electrical, thermal, and thermoelectric transport in the presence of 
Berry curvatures in phase space, thus taking into account both real-space and momentum-space on an equal footing.
The main focus of our work is to understand the thermal currents that arise in magnets with topological textures when 
perturbed by an electrochemical or temperature gradient. This complements our previous work 
on the Hall effects~\cite{verma2022unified} and thermoelectric transport~\cite{addison2023theory}
in the presence of arbitrary Berry curvature.
Thermal transport currents are challenging to compute~\cite{luttinger1964theory, cooper1997thermoelectric}, and anomalous thermal transport has only been explored
theoretically in a few contexts~\cite{qin2011Energy,onoda2006theory,sugimoto2007gauge,shitade2014heat,shitade2014theory}. We build
upon the semiclassical theory~\cite{xiao2010berry} developed in ref.~\cite{xiao2020unified} to explore thermal transport in systems hosting 
topological spin textures.

Our most significant new results are as follows:
\\
(1) We determine the transport thermal current to linear order in the thermal and chemical potential gradients and the electric field. This requires us to
identify and subtract out the the ``bound" thermal currents that exist in equilibrium in the presence of a spatially varying magnetization.
\\
(2) We show, within a controlled semiclassical calculation, that the thermal Hall effect can be decomposed into a topological piece proportional to the skyrmion
(or topological charge) density, an anomalous contribution proportional to the momentum space Berry curvature, and an ordinary part
proportional to the magnetic field. 
\\
(3) We demonstrate the validity of the Kelvin and Wiedemann-Franz relations in the presence of arbitrary phase-space Berry curvatures.
This, together with the Mott relation deduced in \cite{addison2023theory}, completes the set of canonical relationships between the thermoelectric responses
in magnetic materials that harbor spin textures.
 \\
(4) Finally, we uncover a topological phase transition between weak and strong SOC relative to the exchange coupling, which has not been recognized earlier. 
 It separates two regimes with a qualitatively different topological Hall response. We show that
the real-space Berry curvature is the skyrmion density in the ``adiabatic" regime corresponding to the weak SOC phase, but its spatial average vanishes in the strong SOC phase.

We analyze in this paper a general 3D Hamiltonian with arbitrary dispersion and SOC, spin textures with arbitrary 3D spatial variation, and 
in the presence of an external magnetic field. In this sense, the results on the electrical and thermoelectric response presented here also go beyond
our previous works~\cite{verma2022unified,addison2023theory}, which were limited to quasi-2D systems without a magnetic field. 
Our new results are thus applicable to systems beyond the usual skyrmion materials. Examples of recent experimental interest
include the hedgehog lattice observed in 
$\text{MnSi}_{1-x}\text{Ge}_x$~\cite{fujishiro2019topological}, and systems exhibiting the anomalous or topological in-plane Hall effects \cite{you2019angular,wang2019magnetic,ge2020unconventional,tan2021unconventional,zhou2022heterodimensional,chen2023observation,cao2023plane}. 
None of these systems could have been described by our previous works.

\section{Methodology and Outline}

We describe here how our paper is organized, summarize our methodology, and discuss the underlying assumptions.

In Section \ref{sec-model}, we first introduce our model which describes the dynamics of conduction electrons interacting with local moments.
The electrons have an arbitrary band structure with general SOC, and the
local moments form a static spin texture with a specified but arbitrary (not necessarily periodic) spatial variation in 3D. 

It is useful to clearly discuss at the outset the small parameters that control our calculations.  
We focus on textures that vary slowly on a length scale $L_s$ that is much larger than the microscopic length scale of the
lattice spacing $a \sim 1/k_F$, the Fermi wavelength. This allows us to 
utilize semiclassical techniques~\cite{xiao2010berry, sundaram1999wave} to 
describe the motion of electron wave-packets centered about a point $(\bm{r}, \bm{k})$ in phase-space.
The semiclassical approach requires that the elastic mean free path
$\ell = v_F\tau$ satisfies $1/k_F\ll \ell$, where $v_F$ is the Fermi velocity, and that the magnetic field is such that the cyclotron energy 
$\hbar\omega_c \ll E_F$, the Fermi energy. In principle, $\ell/L_s$ and $\omega_c\tau$ can take on any values consistent with the 
semiclassical inequalities. In practice, however, the Boltzmann equation is much easier to solve when $\omega_c\tau \ll 1$ and $\ell \ll L_s$, and this is 
the regime that we will focus on in this paper. These assumptions are realistic for many materials~\cite{tokura2020magnetic}  that exhibit topological spin textures
where $10 \lesssim L_s \lesssim 500$ nm, while $1 \lesssim \ell \lesssim 100$ nm given that  $10 \lesssim k_F\ell \lesssim 100$.

The SOC strength $\lambda$ of the conduction electrons plays an important role in our analysis.
Since we are modeling systems where the magnetism arises from 3d transition metal ions,
we work in the regime where $\lambda \ll E_F\!\sim t$, the hopping that sets the bandwidth scale.
However, our results depend crucially on the ratio of $\lambda$ to the exchange coupling $J$ between conduction 
electron spins and the local moments forming the spin texture. 

We mainly focus on $\lambda \ll J \lesssim E_F$, sometimes called the ``adiabatic regime" in the literature. We prefer 
to call this the ``weak SOC regime'' since ``adiabatic" has a variety of meanings in different contexts.
For weak SOC, we show that the real-space Berry curvature coincides with the
topological charge density $\hat{\bf m}(\bm{r}) \cdot \partial_{r_i}\hat{\bf m}(\bm{r}) \times \partial_{r_j}\hat{\bf m}(\bm{r})$
for the arbitrary spin texture $\hat{\bf m}(\bm{r})$, and that the 
topological and anomalous responses are of comparable magnitude.
We show that the ``strong SOC" regime $J \ll \lambda \ll E_F$ is qualitatively different
and separated from the weak SOC regime by a topological transition. In fact, we will show that the semiclassical approximation
breaks down at this topological transition.

In Section \ref{sec-eom} we derive the equations of motion that govern the dynamics of $({\bm{r}}(t), {\bm{k}}(t))$
in the presence of arbitrary phase-space Berry curvatures and external electric and magnetic fields.  We also derive, using tools from differential geometry, the phase-space volume measure that satisfies Liouville's theorem.

Next we show in Section \ref{sec-boltzmann} how we determine the electronic distribution function
$f(\bm{r}, \bm{k})$ by solving the Boltzmann equation with a relaxation time $\tau$
to linear order in external perturbations, the electric field ${\bf E}$, and gradients of the chemical potential $\mu$ and temperature $T$.

\begin{figure*}
\includegraphics[width=.9 \textwidth]{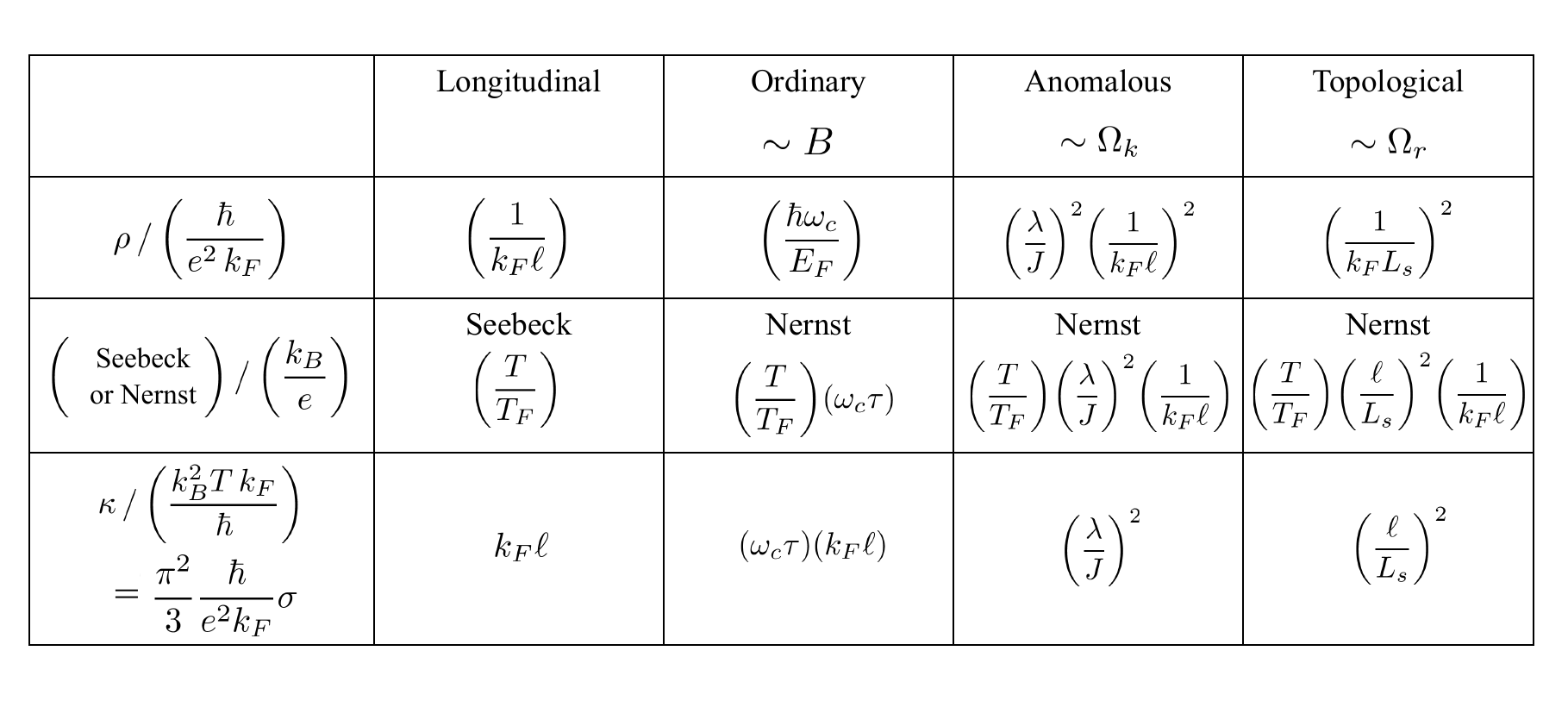}
\caption{{\bf Summary of results.}  Table summarizing the semiclassical results for the resistivity, Seebeck or Nernst signal, and thermal conductivity in 
3D systems in the regime where the microscopic length scale $k_F^{-1}\sim a\ll$ elastic mean free path $\ell \ll L_s$ the scale of variation of the spin texture, and SOC $\lambda \ll J$, the exchange coupling  between electrons and spin texture, and $\lambda \ll$ Fermi energy $E_F = k_B T_F$. 
Results for the longitudinal response is shown in the second column. For the ordinary response we work in the weak field limit where the cyclotron frequency satisfies $\omega_c\tau  \ll 1$ where $\tau = \ell/v_F$.
We show that, within the domain of validity of our theory, the transverse response can be decomposed as the
sum of three terms -- the ordinary, anomalous, and topological contributions -- whose scaling with various parameters is shown in separate columns.
The thermal conductivity scales like the electric conductivity as a result of the Weidemann-Franz relation that we derive. Numerical estimates of the anomalous and topological 
thermal Hall response and their comparison with experiments are presented in Section~\ref{sec-conclusions}.}
\label{scaling table}
\end{figure*}

In Section \ref{sec-currents}, we compute the transport currents that flow in response to an external perturbation.
The transport currents, which are measured or applied in experiments, are distinct from the total currents, which 
include contributions from ``bound magnetization currents" that exist in equilibrium.
The electrical transport current $\bm{j}_{\text{tr}}^e$ has been discussed extensively in the 
literature~\cite{cooper1997thermoelectric, xiao2010berry, verma2022unified, addison2023theory},
where classical electromagnetism is used to identify the bound currents arising from the curl of a magnetization.  
 
The thermal transport current $\bm{j}_{\text{tr}}^Q$ is theoretically more challenging to calculate~\cite{luttinger1964theory, cooper1997thermoelectric, qin2011Energy}.
It is not as simple as in the electrical case to identify the bound current, in part due to the lack of a canonical definition for an analogous ``energy magnetization".
Attempts have been made to identify its correct form via derivatives of the free energy with respect to a  gravito-magnetic field \cite{onoda2006theory,sugimoto2007gauge,shitade2014heat,shitade2014theory}. 
While in principle this does provide a method for calculating the energy magnetization, in practice these calculations can 
be cumbersome even in the simplest crystalline systems. 
We instead build on ref.~\cite{xiao2020unified} and identify the bound current from the equilibrium expectation value of the 
total current operator. This allows us to determine $\bm{j}_{\text{tr}}^Q$ without direct calculation of an energy magnetization. 

Using this approach we calculate the electric, thermal, and thermoelectric conductivity tensors, 
defined by
\begin{align}
\begin{pmatrix}
\bm{j}^{e}_{\text{tr}} \\
\bm{j}^{Q}_{\text{tr}}
\end{pmatrix} = 
\begin{pmatrix}
\te{\sigma} & \te{\alpha} \\
\te{\beta} & \te{\kappa}
\end{pmatrix}
\begin{pmatrix}
\bm{E}-\frac1e\bm{\nabla}_r\mu \\
\bm{-\nabla}_r T
\end{pmatrix}.
\label{transport matrix}
\end{align}
\noindent
The Hall effect in the electric $\te{\sigma}$, thermoelectric $\te{\alpha}$, electrothermal $\te{\beta}$, and thermal conductivities $\te{\kappa}$ is identified
as the antisymmetric part of each tensor.  This eliminates time reversal (TR) even contributions that may exist in the 
off-diagonal components, since
Onsager reciprocity $\sigma_{ij}({\bf B}) = \sigma_{ji}(-\,{\bf B})$ implies that its symmetric part is TR-even while the antisymmetric part is TR-odd.

We thus define the antisymmetric ``Hall pseudovector", $\zeta_i \equiv \epsilon_{ijk} \zeta_{jk}/2$ (repeated indices summed), for $\zeta\in\{\sigma, \alpha, \kappa\}$.  
For instance, the Hall response measured in the $xy$-plane is represented by the $z$ component of the Hall pseudovector 
$\zeta_z = (\zeta_{xy}-\zeta_{yx})/2$. 
In what follows, we do not discuss $\te{\beta}$ because we find that 
$\te{\beta}=T\, \te{\alpha}$
and thus it is not an independent transport coefficient.

In Section \ref{sec-conductivities} we show that
the electric, thermoelectric, and thermal Hall conductivities can be written as the sum of the ordinary, anomalous, and topological contributions, namely
\begin{equation}
    \bm{\zeta} = \bm{\zeta}^O + \bm{\zeta}^A + \bm{\zeta}^T,\ \quad \quad \ \zeta\in\{\sigma, \alpha, \kappa\}.
     \label{sumOAT}
\end{equation}
\noindent
The ordinary term $\bm{\zeta}^O$ arises from the usual Lorentz force,
the anomalous term $\bm{\zeta}^A$ from the momentum-space Berry curvature,
and the topological contribution $\bm{\zeta}^T$ from the real-space Berry curvature.
All other contributions, including those arising from mixed real-momentum space Berry curvature, are negligible in the regime
where our calculations are controlled. 

In Section \ref{sec-transport} we derive that the Wiedemann-Franz relation between the electrical and thermal conductivities and the Mott relation between the 
thermoelectric and electrical conductivities at low temperatures, and show that these are valid even in the presence of phase-space Berry curvatures.
We then express our results for the conductivities in terms of transport coefficients that are more directly related to experimental measurements, namely 
the resistivity, and the Seebeck and Nernst coefficients. The scaling of various transport coefficients in a 3D system with the parameters of our theory are 
summarized in the table in Fig.~1. We end this Section with a discussion of the in-plane Hall effect.

In Section \ref{sec-SOCRegime} we demonstrate how the strong SOC regime ($J \ll \lambda \ll t$)
is qualitatively different from the weak SOC regime ($\lambda \ll J \lesssim t$) that we focus on in 
the bulk of the paper. We analyze the example of a 2D skyrmion crystal (SkX) in detail and analytically show that 
there is a topological transition at an intermediate value of $\lambda/J \sim O(1)$ separating these 
two regimes. The semiclassical approach is shown to break down at this topological transition
due to the gap closing.
We show that the real-space Berry curvature integrated over the SkX unit cell, which is 
a topological invariant, changes discontinuously across this transition: it equals the non-zero skyrmion number 
for weak SOC, but vanishes for strong SOC. 

In Section \ref{sec-conclusions} we present order-of-magnitude numerical estimates of the topological and anomalous contributions of thermal Hall signal
and show that our results compare favorably with recent experiments.
Finally, we conclude with open questions and directions for future work.
In addition, there are eight Appendices A through H, which contain details of derivations or an elaboration on the results presented in the paper.

\section{Model}
\label{sec-model}
We consider a 3D system of itinerant electrons interacting with local moments that form a topological spin texture.
Our formalism can be used to model bulk metallic magnets and magnetic multilayers that harbor topological spin 
textures~\cite{nagaosa2013topological, fert2017magnetic, tokura2020magnetic}, 
as well as heavy metal/magnetic insulator bilayers~\cite{ahmed2019spin, shao2019topological}.

We consider a Hamiltonian of the form
\begin{equation}
    \widehat{H} = \widehat{H}_{\text{t}}+\widehat{H}_{\lambda}+\widehat{H}_{\text{J}}
    \label{H full}
\end{equation}
\noindent
where $\widehat{H}_{\text{t}}$ is a tight-binding description of conduction electrons, 
$\widehat{H}_{\lambda}$ describes the effect of spin-orbit coupling (SOC) on the itinerant electrons, and
$\widehat{H}_{\text{J}}$ takes into account the coupling of the electrons to the magnetic texture. 
We next elaborate on each of these terms.

The electron kinetic energy is given by 
\begin{equation}
    \widehat{H}_{\text{t}}=-\sum_{ij,\sigma}t_{ij}\hat{c}^{\dagger}_{i\sigma}\hat{c}_{j\sigma},
        \label{H ke}
\end{equation}
\noindent
where $\sigma$ labels up/down spins, and the hopping amplitudes $t_{ij}$ between sites $i$ and $j$ describes a single band with arbitrary dispersion. The electron density is determined by the Fermi energy $E_F$.

We will see below that SOC splits the single band of eq.~\eqref{H ke} into two and gives rise to ${\bf k}$-space Berry curvature
in the presence of a net magnetization. It is well known that in a multi-band system one can get ${\bf k}$-space Berry curvature even in the absence of SOC, but
here we restrict attention, for simplicity, to the single band case. This is also reasonable since SOC is an essential ingredient in many 
materials of interest where it stabilizes skyrmions via the Dzyaloshinskii-Moriya interaction.

We consider a general SOC of the form
\begin{equation}
    \widehat{H}_{\lambda}=\dfrac{\lambda\hbar}{at} \sum_{ ij, \mu\nu,\sigma\sigma'}\hat{c}^{\dagger}_{i\sigma}(\chi^{\phantom{\dagger}}_{\mu\nu}\sigma_\mu^{\sigma\sigma'}v_\nu^{ij})\hat{c}^{\phantom{\dagger}}_{j\sigma'}.
    \label{H SOC}
\end{equation}
\noindent
It describes an arbitrary linear coupling between the velocity $\bm{v}^{ij}=i (\bm{r}_i-\bm{r}_j)t_{ij} /\hbar$ on a bond $(i,j)$
and the conduction electron spin $\bm{\sigma}$, defined by a vector of Pauli matrices with matrix elements labeled by 
$\sigma,\sigma'\in \{\uparrow,\downarrow\}$.
$\lambda$ is the strength of the SOC with dimensions of energy, $a$ is the lattice spacing, and $t$ the scale of the hopping.

We can take into account various different types of SOC in our analysis with an appropriate choice of the dimensionless matrix 
$\chi_{\nu\mu}$ with $\mu,\nu\in \{x,y,z\}$. These include the following two familiar examples. If we choose
$\chi_{xy}= - \chi_{yx} = 1$, with all other components zero, then eq.~\eqref{H SOC} reduces to 
$\sum_{ij,\sigma \sigma'}(\sigma_x^{\sigma \sigma'}v_y^{ij}-\sigma_y^{\sigma \sigma'}v_x^{ij})\hat{c}^{\dagger}_{i\sigma}\hat{c}^{\phantom{\dagger}}_{j\sigma'}$
and we obtain a Rashba SOC. Similarly, linear Dresselhaus SOC is obtained by choosing $\chi_{xx}= -\chi_{yy}= 1$ with all other components zero.

The third term in the Hamiltonian of eq.~\eqref{H full}  
\begin{equation}
    \widehat{H}_{\text{J}}=-J \sum_{i,\sigma\sigma'}\hat{\bm{m}}(\hat{\bm{r}}_i)\cdot (\hat{c}^{\dagger}_{i\sigma}\boldsymbol{\sigma}^{\sigma\sigma'}\hat{c}^{\phantom{\dagger}}_{i\sigma'})
\label{H J} 
\end{equation}
\noindent
describes the coupling of strength $J$ between the itinerant electron spin $\bm{\sigma}$ and a magnetic texture $\hat{\bm{m}}(\bm{r})$, with $\hat{m}_x^2({\bf r}) + \hat{m}_y^2({\bf r}) + \hat{m}_z^2({\bf r}) = 1$ at each ${\bf r}$. We assume that the local moments are static, but can have an arbitrary spatial variation in three dimensions.

We will apply external electric and magnetic fields to our system in the next section when we derive the semiclassical equations of motion
for the conduction electrons described by the Hamiltonian of eq.~\eqref{H full}.
At the end, we will also take into account elastic scattering of the conduction electrons from impurities,
characterized by a scattering time $\tau$ or equivalently a mean free path $\ell = v_F \tau$.
The last step in our semiclassical transport calculation ($a \ll \ell$) will be the solution of the Boltzmann equation within a relaxation time approximation,
and it is here that the time scale $\tau$ will govern the approach to local equilibrium.

\section{Semiclassical Equations of Motion} 
\label{sec-eom}
As noted in Section \ref{sec-intro}, our goal is to take into account both the momentum space Berry curvature and the real space Berry curvature on an equal footing, and the semiclassical approach is ideally suited for this purpose. We begin by considering
electronic wave-packets centered around the point $\bm{\xi} = (\bm{r}_c, \bm{q}_c)$ in six-dimensional phase space, and 
start by obtaining the equations of motion for $\bm{\xi}$. The semiclassical analysis is justified when there is a clear separation of length scales, so that 
the lattice spacing and Fermi wavelength $a \sim k_F^{-1} \ll L_s$, the characteristic scale of variation the spin texture $\hat{\bm{m}}(\bm{r})$ of the local moments,
as well as the length scales associated with the external electric and magnetic fields (which are spatially uniform in our problem).
The time scale of the magnetic dynamics is also assumed to be much longer than the electronic time scales of the system, which justifies the static magnetic texture treated here. We note in passing that the ``skyrmion Hall effect"~\cite{nagaosa2013topological}, which involves the motion of spin textures
is only seen within non-linear response once the skyrmions are depinned. We focus here only on linear response.

We will find the equations of motion through a standard semiclassical procedure~\cite{xiao2010berry, sundaram1999wave}; see Appendix \ref{app eom} 
for details. We start by performing an expansion of $\widehat{H}$ around an arbitrary real-space point $\bm{r}_c$, to obtain
\begin{equation}
    \widehat{H}=\widehat{H}_c+\Delta\widehat{H}
\end{equation}
\noindent
where $\widehat{H}_c=\widehat{H}(\hat{\bm{r}}\rightarrow \bm{r}_c)$ and $\Delta\widehat{H}$ is the first order correction, smaller than 
$\widehat{H}_c$ by a factor of $a/L_s$. 
We will find that the leading order contributions to the transverse electric and thermal responses of the system arise at first order in this expansion,
and we neglect terms that are higher order in $a/L_s$. 

\begin{figure}
    \centering    \includegraphics[width=0.45\textwidth]{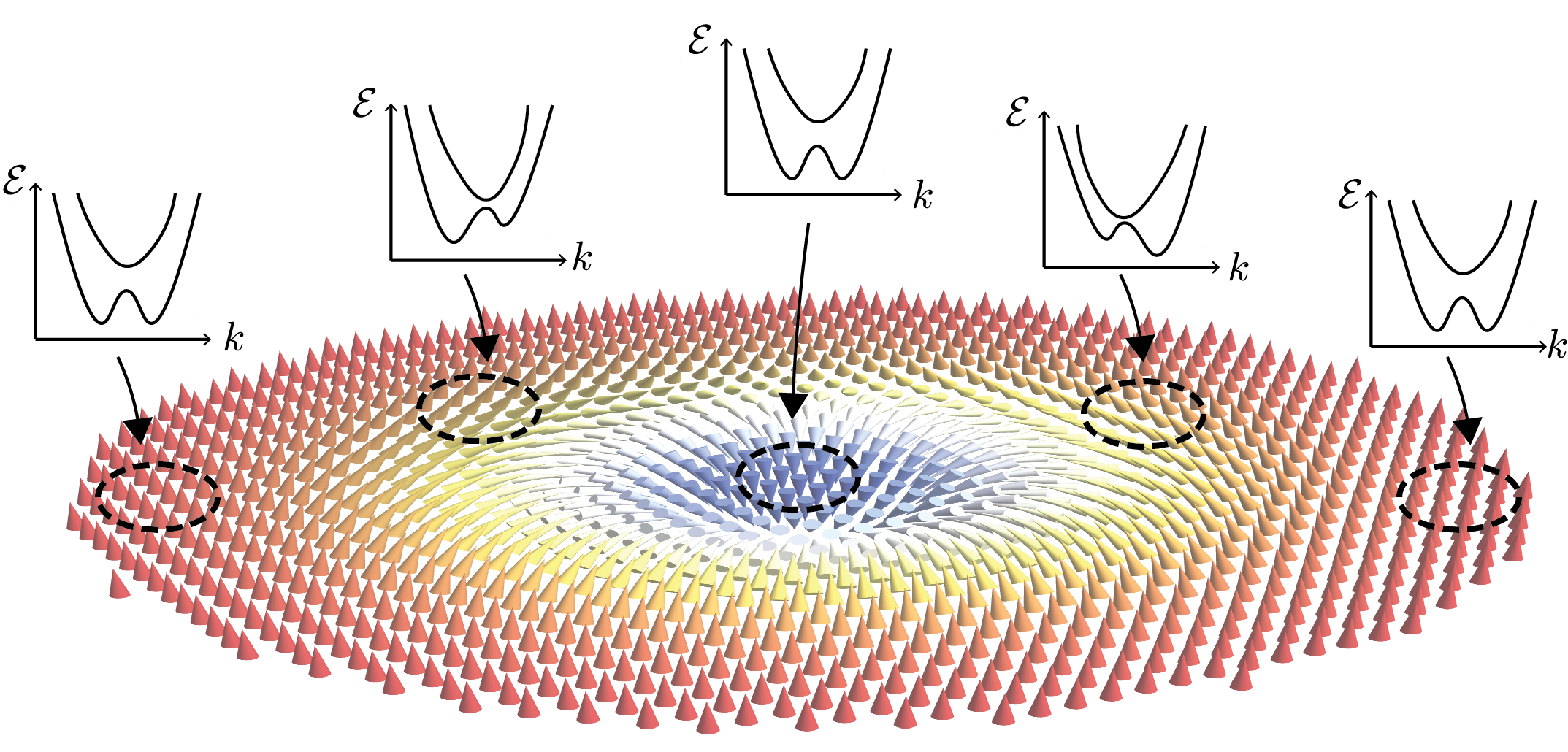}
    \caption{{\bf Local Band Structure Schematic.}  Illustration of the spatial variation of the local semiclassical band structure at 
    various positions within a topological skyrmion magnetic texture.}
    \label{fig:SkBand}
\end{figure}

$\widehat{H}_c$ has discrete translation symmetry, and may be written in the Bloch basis with eigenstates $\ket{\psi_{\pm}(\bm{r}_c, \bm{q})}$ indexed by a crystal momentum $\bm{q}$ and a band index $\pm$. It takes the form 
\begin{align}
\widehat{H}_c &= \varepsilon(\bm{q})\mathbb{1} + \bm{d}(\bm{r}_c,\bm{q})\cdot \bm{\sigma},    \nonumber \\
    d_\nu(\bm{r}_c,\bm{q}) &= \dfrac{\lambda}{a\,t}\,\chi^{\phantom{\dagger}}_{\nu\mu}\partial_{q_\mu}\varepsilon(\bm{q}) - J \hat{m}_\nu(\bm{r}_c),
    \label{bloch d}
\end{align}
\noindent
where $\mu,\nu\in \{x,y,z\}$.
The eigenenergies are $\mathcal{E}_{\pm}(\bm{r}_c, \bm{q}) = \varepsilon(\bm{q})\pm |\bm{d}(\bm{r}_c, \bm{q})|$. 
Denoting the eigenvalues of $\Delta\widehat{H}$ by $\Delta \mathcal{E}$, 
the eigenvalues of the full Hamiltonian $\widehat{H}$ can be written as $\widetilde{\mathcal{E}}_{\pm}=\mathcal{E}_{\pm}+\Delta\mathcal{E}$
to first order in $a/L_s$. In Figure~\ref{fig:SkBand} we illustrate the spatial variation of the semiclassical eigenvalues at 
various positions within a topological skyrmion magnetic texture.

We next include the effect of the electromagnetic scalar and vector potentials $\phi$ and $\bm{A}$
on the conduction electrons. This leads to a shift $\widetilde{\mathcal{E}}_{\pm}(\bm{\xi})-e\phi(\bm{r}_c, t)$ in the energies
and $\bm{k}_c\equiv \bm{q}_c+(e/\hbar)\bm{A}(\bm{r}_c,t)$ in the momenta, where $e > 0$.

Using the Bloch states, we construct wave-packets $\ket{W_{\pm}(\bm{r}_c, \bm{k}_c)}$ centered about the phase-space point $(\bm{r}_c, \bm{k}_c)$.
From now on, we drop the ``$c$" subscript to simplify notation and denote points in phase-space by
\begin{equation} 
\bm{\xi}\equiv(r_x,r_y,r_z,k_x,k_y,k_z).
\end{equation}

Using the semiclassical Lagrangian
\begin{equation}
L_\pm(\bm{\xi})=\bra{W_\pm(\bm{\xi})}\left( i\hbar\dfrac{d}{d t}-\widehat{H}\right)\ket{W_\pm(\bm{\xi})}
\end{equation}
\noindent
we obtain the equations of motion~\cite{xiao2010berry, sundaram1999wave}
\begin{equation}
    \dot{\xi}_{\alpha} = (1/\hbar)(\Gamma^{\pm})^{-1}_{\alpha\beta} \ \partial_{\xi_{\beta}}[\widetilde{\mathcal{E}}_{\pm}(\bm{\xi})-e\phi(\bm{r}, t)],
    \label{EOM}
\end{equation}
\noindent
where the indices $\alpha,\beta$ label the six components of $\bm{\xi}$.
The equations of motion involve the inverse of $\te{\Gamma}^{\pm}$, a $6\times 6$ antisymmetric matrix that encodes information 
about how the phase-space Berry curvatures and external magnetic field impact the dynamics:

\begin{align}
    &\te{\Gamma}^{\pm}(\bm{\xi})=\begin{pmatrix}
\te{\Omega}^{\pm}_{rr} - {e\over\hbar}\te{F}& \te{\Omega}^{\pm}_{rk} - {\mathbb{1}}\\
\te{\Omega}^{\pm}_{kr}  + {\mathbb{1}}& \te{\Omega}^{\pm}_{kk} \\
\end{pmatrix}
\label{Gamma matrix}
\end{align}

\noindent
Here $F_{ij}$ are the spatial components of the electromagnetic field tensor 
\begin{equation}
F_{ij} = \partial_i A_j - \partial_j A_i,
    \label{B field}
\end{equation}
\noindent  
with $i,j \in \{x,y,z\}$ so that, e.g., the magnetic field $B_z=F_{xy}$. 
We note that, in addition to the familiar ${\bf k}$-space and ${\bf r}$-space Berry curvatures $\te{\Omega}_{kk}$ and $\te{\Omega}_{rr}$, 
the $\te{\Gamma}^{\pm}$ matrix also includes ``mixed-space" Berry curvatures $\te{\Omega}_{rk}$ and $\te{\Omega}_{kr}$. The components of the phase-space Berry curvatures, in general, 
are given by 
\begin{align}
    \Omega_{\xi_{\alpha}\xi_{\beta}}^{\pm}(\bm{\xi}) & = \ s\,\Omega_{\xi_{\alpha}\xi_{\beta}}, \ \ \quad s = \pm 1    \label{curv} \\
    \Omega_{\xi_{\alpha}\xi_{\beta}} &= \frac12\hat{\bm{d}}\cdot(\partial_{\xi_{\alpha}}\hat{\bm{d}}\times\partial_{\xi_{\beta}}\hat{\bm{d}}).
    \nonumber
\end{align}
\noindent
Using $\hat{\bm{d}} = {\bm{d}}/|{\bm{d}}|$, it is straightforward to rewrite
the above result as
$\Omega_{\xi_{\alpha}\xi_{\beta}} = {\bm{d}}\cdot(\partial_{\xi_{\alpha}}{\bm{d}}\times\partial_{\xi_{\beta}}{\bm{d}})/2|{\bm{d}}|^3$,
which greatly simplifies the Berry curvature calculation for the Hamiltonian of eq.~\eqref{bloch d}.

We have to solve the equations of motion for each band separately but, to simplify notation, 
we will drop the band index $(\pm)$ in the intermediate steps unless it is essential to show it explicitly.
We write $\te{\Gamma}^{-1}$ in the compact block form 
\begin{align}
    &\te{\Gamma}^{-1}(\bm{\xi})=\dfrac{1}{\mathcal{D}(\bm{\xi})}\begin{pmatrix}
\te{\mathbf{K}}(\bm{\xi}) & \te{\mathbf{S}}(\bm{\xi})\\
-\te{\mathbf{S}}^T(\bm{\xi}) & \te{\mathbf{R}}(\bm{\xi})\\
\end{pmatrix}
\label{gamma inverse}
\end{align}
\noindent
where 
\begin{equation}
\mathcal{D}(\bm{\xi}) \equiv \text{pf}\ \te{\Gamma}(\bm{\xi})
\end{equation}
\noindent
is the pfaffian. We discuss in Appendix \ref{app eom} why this particular form, with the pfaffian factored out, is natural for
the inverse of an even-dimensional antisymmetric matrix. Specifically for our analysis, we will see that writing $\te{\Gamma}^{-1}$ in this form 
makes transparent the cancellation with the same pfaffian that appears in the phase space volume (see below).

The full expressions for the block matrices appearing in eq.~\eqref{gamma inverse} are shown in detail in Appendix \ref{app eom} 
 (see eq.~\eqref{full block matrices}), and involve terms up to quadratic order in Berry curvatures and magnetic field. 
 Some intuition for these matrices can be obtained by examining each to lowest order in Berry curvatures: 
 $\mathbf{K}_{ij} \approx s \Omega_{k_i k_j}$, $\mathbf{R}_{ij} \approx s\Omega_{r_i r_j} - ({e}/{\hbar})F_{ij}$, and $\mathbf{S}_{ij} \approx \delta_{ij}$.

We can now rewrite the equations of motion \eqref{EOM} as
\begin{align}
    \dot{\bm{r}} &= \dfrac{1}{ \hbar\mathcal{D}}\bigg[  \overleftrightarrow{\mathbf{K}} \bigg(\bm{\nabla}_r \widetilde{\mathcal{E}} + e\bm{E}\bigg)+\overleftrightarrow{\mathbf{S}} \bm{\nabla}_k\widetilde{\mathcal{E}} \bigg], \nonumber \\  \dot{\bm{k}} &= \dfrac{1}{\hbar\mathcal{D}}\bigg[ \overleftrightarrow{\mathbf{R}} \bm{\nabla}_k \widetilde{\mathcal{E}} - \overleftrightarrow{\mathbf{S}}^T\bigg(\bm{\nabla}_r \widetilde{\mathcal{E}} + e\bm{E} \bigg)\bigg]
    \label{xidot}
\end{align}
\noindent
where $\bm{E}=-\partial_t \bm{A}-\bm{\nabla}_r\phi$ is the electric field, and 
we have suppressed the explicit $\bm{\xi}$ dependence of various factors for simplicity.
These equations dictate how the phase-space Berry curvatures and external magnetic field 
impact the semiclassical dynamics of conduction electrons.
Here is how some familiar effects appear in these equations.
The usual Lorentz force contribution to $\dot{\bm{k}}$ appears 
through $\te{\mathbf{R}}\bm{\nabla}_k\widetilde{\mathcal{E}}$
with the magnetic field coming from $\te{\mathbf{R}}$.
Since at leading order in curvatures $\mathbf{R}_{ij} \approx s \Omega_{r_i r_j} - ({e}/{\hbar})F_{ij}$, 
we might expect that the real space Berry curvature will have an effect analogous to a magnetic field,
except for the sign change $s\,=\pm$ between the two bands.
The anomalous velocity term, arising from ${\bm{k}}$-space Berry curvature  is contained in 
$\dot{\bm{r}}$ through $\te{\mathbf{K}}\bm{\nabla}_r\widetilde{\mathcal{E}}$.

In the remainder of the paper we use equation \eqref{xidot} together with the 
Boltzmann equation in a controlled calculation to determine the electric and thermal currents. 
This calculation will require integration over semiclassical phase space weighted by the phase space measure, so we must 
determine the correct measure before we can proceed.

\medskip
\noindent
\textbf{Liouville's Theorem:}
According to Liouville's theorem, the phase space volume element must remain invariant under time evolution. In the absence of Berry curvatures, this volume element is 
simply $\mathrm{d}r_x\wedge \mathrm{d}r_y \wedge \mathrm{d}r_z \wedge  \mathrm{d}k_x \wedge  \mathrm{d}k_y \wedge  \mathrm{d}k_z$. However, in the presence of Berry curvatures, $\bm{r}$ and $\bm{k}$ no longer satisfy the 
canonical Poisson bracket relations, i.e., $\{r_i,k_j\}\neq\delta_{ij}$, and the invariant volume element must be modified~\cite{xiao2005berry}.
We show here that in general the modified volume element is given by
\begin{equation}
\mathcal{V}=\mathcal{D}(\bm{\xi})\,\mathrm{d}r_x\wedge \mathrm{d}r_y \wedge \mathrm{d}r_z \wedge  \mathrm{d}k_x \wedge  \mathrm{d}k_y \wedge  \mathrm{d}k_z, 
\label{phase space volume}
\end{equation}
where $\mathcal{D}(\bm{\xi}) \equiv \text{pf}\, \te{\Gamma}(\bm{\xi})$ is the pfaffian of the matrix introduced in eq.~\eqref{Gamma matrix}.

For the special case of a system with {\it only} ${\bm{k}}$-space Berry curvature, the result 
$\mathcal{D}(\bm{\xi})=1+(e/\hbar)\bm{B}\cdot\bm{\Omega}_k$, with $\Omega_{k_l}=(1/2)\epsilon_{lmn}\Omega_{k_mk_n}$,
is well known~\cite{xiao2005berry} and widely used.
The more general result \eqref{phase space volume} for a system with arbitrary phase-space Berry curvatures 
is stated without proof in ref.~\cite{xiao2010berry}.  We present here a simple derivation using standard tools of differential geometry. 
These may be unfamiliar to some readers and the remainder of this Section may be skipped without impeding understanding of the rest of the paper.

To satisfy Liouville's theorem, the Lie derivative of eq.~\eqref{phase space volume} along flows generated by the 
velocity field $\dot{\bm{\xi}}\equiv \dv{\bm{\xi}}{t}$ must vanish: $\mathcal{L}_{\dot{\bm{\xi}}} \mathcal{V}=0$. Since
$\dot{\bm{\xi}} = \sum_\alpha \dot{\xi}_\alpha \partial_{\xi_\alpha} = \dv{}{t}$, 
we simplify the notation by writing $\mathcal{L}_t\equiv \mathcal{L}_{\dot{\bm{\xi}}}$
to indicate that this Lie derivative measures the change of the volume form in time.

We define the two-form $\omega=\sum_{ij}\Gamma_{ij}\mathrm{d}\xi_i\wedge \mathrm{d}\xi_j/2$ and relate it to $\mathcal{V}$ by noting that
$ \mathcal{V}= (\omega\wedge\omega\wedge\omega)/{3!}$ as shown in Appendix \ref{app measure}.  The Lie derivative is $\mathcal{L}_t \mathcal{V} = \omega \wedge \omega \wedge \mathcal{L}_t \omega$.
We next use the Cartan formula
\begin{equation}
    \mathcal{L}_t \omega= i_t (\text{d} \omega) + \text{d}  (i_t \omega),
    \label{cartan}
\end{equation}
where $\text{d}$ is the exterior derivative and $i_t$ is the interior product~\cite{nakahara2003}.  
In Appendix \ref{app measure} we show that each term in eq.~\eqref{cartan} vanishes; the idea behind this is explained in brief here.
We show that $\omega$ is an exact form by finding an $\eta$ with the property $\omega = \text{d}\eta$, and thus 
the first term obviously vanishes since $\text{d}^2=0$.
For the second term, we compute $\text{d}(i_t\omega)$ and use the equation of motion \eqref{EOM} to
write it as a product of a symmetric double derivative with an antisymmetric two-form, which also vanishes.
This fixes the measure up to an overall multiplicative constant that is determined by noting that $\mathcal{D}(\bm{\xi}) = 1$
in the absence of any curvatures.

\section{Boltzmann equation} 
\label{sec-boltzmann}
Now that we have obtained the equations of motion, 
the next step in calculating observables within the semiclassical framework is to solve the 
Boltzmann equation. We will thus obtain the electronic distribution functions $f^\pm(\bm{\xi})$ that
describes the phase space occupation of each band $(\pm)$ in the presence of external driving fields.
In the semiclassical approach, the band index is a good quantum number and there are no inter-band transitions. 
We will thus solve for the distribution functions of each band separately, but to simplify notation, 
we will drop the band index $(\pm)$ in the intermediate steps and reinstate it at the end.

In thermodynamic equilibrium, the distribution function is just the Fermi-Dirac distribution: 
$f_{\text{eq}}(\bm{\xi}) = (1 + \exp(\widetilde{\mathcal{E}}(\bm{\xi})-\mu_0)/k_B T_0)^{-1}$, 
where the chemical potential $\mu_0$ and temperature $T_0$ determine the number 
and energy density of the homogeneous system in the absence of any external perturbations.
In a system subjected to thermal or chemical potential gradients, currents are driven by deviations 
from \textit{local} equilibrium. The local equilibrium distribution $f_{\text{leq}}(\bm{\xi})$ also takes the form of the Femi-Dirac distribution $f_{\text{eq}}(\bm{\xi})$, 
but with a spatially varying chemical potential $\mu_0\rightarrow\mu(\bm{r})$ and temperature $T_0\rightarrow T(\bm{r})$. 

The full non-equilibrium distribution $f(\bm{\xi})$ can be written as $f(\bm{\xi})=f_{\text{leq}}(\bm{\xi}) + \delta f(\bm{\xi})$
 where $\delta f(\bm{\xi})$ describes deviations from local equilibrium. 
To calculate currents, we must determine $\delta f(\bm{\xi})$, and to this end we solve the Boltzmann equation 
\begin{equation}
\dot{\bm{r}}\cdot\bm{\nabla}_r f(\bm{\xi})+\dot{\bm{k}}\cdot\bm{\nabla}_k f(\bm{\xi}) =  -\ \dfrac{\left[f(\bm{\xi})-f_{\text{leq}}(\bm{\xi})\right]}{\tau}.
    \label{boltz}
\end{equation}
\noindent
within a relaxation time approximation. As discussed in the paragraph below eq.~\eqref{H J}, 
the mean free path $\ell = v_F \tau$ must satisfy $a \sim k_F^{-1} \ll \ell$, or equivalently 
the scattering rate $\hbar/\tau \ll E_F$ or the bandwidth $\sim t$, to 
ensure the validity of the semiclassical approach.

We are interested here in the linear response to an electric field, temperature gradients, and chemical potential gradients.
Solving the Boltzmann equation to first order in external perturbations, we find
\begin{align}
    \delta f = &-\tau(1+\mathbb{P})^{-1} \bigg[\big(\dot{\bm{k}}^{(1)}\cdot \bm{\nabla}_k \widetilde{\mathcal{E}} + \dot{\bm{r}}^{(1)}\cdot \bm{\nabla}_r \widetilde{\mathcal{E}} \nonumber\\
    &- \dot{\bm{r}}^{(0)}\cdot\bm{\nabla}_r \mu\big)\partial_{\widetilde{\mathcal{E}}}+ \dot{\bm{r}}^{(0)}\cdot\bm{\nabla}_r T\partial_T \bigg] f_{\text{eq}}
    \label{boltz soln}
\end{align}
\noindent 
where 
\begin{equation}
\mathbb{P}=\tau(\dot{\bm{r}}^{(0)}\cdot\bm{\nabla}_r+ \dot{\bm{k}}^{(0)}\cdot\bm{\nabla}_k), 
\label{operator-P}
\end{equation}
and the superscripts on $\dot{\bm{r}}$ and $\dot{\bm{k}}$ 
refer to the (zeroth or first) order in $\bm{E}$ terms in the equations of motion. 
A detailed derivation of eq.~\eqref{boltz soln} is given in Appendix \ref{app boltzmann}. 
We note that, once the ${\nabla}_r T$ and ${\nabla}_r \mu$ terms are explicitly extracted from $f_{\text{leq}}$, 
within linear response we can set  $T(\bm{r})\rightarrow T_0$ and $\mu(\bm{r})\rightarrow\mu_0$ in the rest of the expression. This is why 
$f_{\text{eq}}$ appears on the right hand side of eq.~\eqref{boltz soln}.

Formally, the semiclassical result obtained above is valid for arbitrary $\omega_c\tau$, provided both the 
cyclotron energy $\hbar\omega_c$ and $\hbar/\tau$ are  $\ll E_F$, and for arbitrary $\ell/L_s$, provided that 
the lattice spacing $a \sim k_F^{-1} \ll$ than both the the mean free path $\ell$ and the spin texture length scale $L_s$. The solution of eq.~\eqref{boltz soln} greatly simplifies in the limit of weak fields $\omega_c \tau \ll1$ and ``slowly varying" spin textures 
with $\ell \ll L_s$. 

We proceed as follows.
Using $a \ll L_s$ we can ignore the $\Delta{\mathcal{E}}$ correction and write
$\widetilde{\mathcal{E}} \simeq {\mathcal{E}} = \varepsilon({\bf k}) + |{\bf d}({\bf r}, {\bf k})|$.
We then use the scaling of the derivatives $\bm{\nabla}_r\sim(1/L_s)$ and $\bm{\nabla}_k\sim a$
together with eq.~\eqref{xidot} with ${\bf E} = 0$ to estimate how $\dot{\bm{r}}^{(0)}$ and $\dot{\bm{k}}^{(0)}$
scale in terms of the various small parameters.
Substituting this in eq.~\eqref{operator-P}, we see that the contributions to the operator $\mathbb{P}$ scale at most as
$\omega_c \tau$ or $\ell/L_s$.
Thus, in the weak-field limit with slowly-varying spin textures, we can make the expansion
$(1 + \mathbb{P})^{-1} = (1 - \mathbb{P} + \dots)$, a generalization of the classic Zener-Jones analysis, 
as explained in detail in Appendix \ref{app boltzmann}. 

We end this Section by presenting results in the regime of weak SOC $\lambda \ll J \lesssim E_F\sim t$ or bandwidth,
which is the main focus of this paper.
(See Section~\ref{sec-SOCRegime} where we discuss the opposite regime of strong SOC).
In the weak SOC regime we can set $\lambda\!=\!0$, at
leading order in $a/L_s$ and $\omega_c\tau$,  in eq.~\eqref{boltz soln} to obtain

\begin{align}
    \delta f\approx&\bigg[\tau v_i-\hbar\tau^2 \big(s\, \Omega_{r_mr_n}-\frac{e}{\hbar}F_{mn}\big) v_n M_{mi}^{-1} \bigg]\times \nonumber\\
    &\bigg[\dfrac{\varepsilon - \mu_0}{T_0}\partial_{r_i} T + (eE_i - \partial_{r_i}\mu)\bigg]\partial_{\varepsilon}f_{\text{eq}}.
    \label{boltz soln approx}
\end{align}
\noindent
We note that here neither the momentum space Berry curvature nor the mixed space Berry curvature appear 
in the equation above. They are both proportional to $\lambda/J \ll 1$ and therefore do not contribute at leading order to the distribution function $\delta f$. 
In fact, one can see from dimensional analysis that $\Omega_{kr} \sim 1/(k_F L_S) \ll 1$ and serves as
an additional small parameter that allows us to neglect $\Omega_{kr}$ relative to the $\mathbb{1}$ in the
semiclassical equations of motion; see eq.~\eqref{Gamma matrix}.

The terms proportional to $\tau^2$ are the leading order contributions to the transverse conductivites, while terms linear in $\tau$ contribute to the longitudinal conductivities (see Appendix \ref{app boltzmann}). 
Consistent with this limit, we have replaced the energies $\widetilde{\mathcal{E}}_{\pm}(\bm{\xi})$ with 
\begin{equation}
\varepsilon_{\pm}(\bm{k}) = \varepsilon(\bm{k})\pm J. 
\end{equation}
In particular, $f_{\text{eq}}\big(\widetilde{\mathcal{E}}_{\pm}\big)\rightarrow f_{\text{eq}}\big(\varepsilon_{\pm}\big)$,
and the velocity and inverse mass tensor are given by
\begin{equation}
    v_i \equiv \partial_{k_i}\varepsilon(\bm{k})/\hbar;\quad M^{-1}_{ij}\equiv \partial_{k_i}\partial_{k_j}\varepsilon(\bm{k})/\hbar^2.
    \label{v and M}
\end{equation}

\section{Currents}
\label{sec-currents}

In the absence of external perturbations (temperature gradients,  chemical potential gradients, and electric fields), there can still 
be circulating ``bound magnetization currents" {\it in equilibrium}, that do not contribute to the currents that are either measured or applied in most transport experiments. 
Following refs.~\cite{cooper1997thermoelectric, xiao2020unified}, we therefore define the electric (thermal) \textit{transport} current 
\begin{align}
    \bm{j}^{e(Q)}_{\text{tr}}(\bm{r}) &= \bm{j}^{e(Q)}(\bm{r}) - \bm{j}^{e(Q)}_{\text{mag}}(\bm{r}).
    \label{j tr}
\end{align}
\noindent
as the difference between the total current, given by the expectation values $\bm{j}^{e(Q)}(\bm{r})=\langle\widehat{\bm{j}}^{e(Q)}(\bm{r})\rangle$ 
of the electrical $(e)$ or thermal $(Q)$ current operators, and the corresponding ``magnetization currents''. 
Note that these are all current densities, but for simplicity we call them currents. In global equilibrium, we expect the transport currents to vanish, 
and thus $\langle\widehat{\bm{j}}^{e(Q)}(\bm{r})\rangle_{\rm eq} =\bm{j}^{e(Q)}_{\text{mag}}(\bm{r})$. The name ``magnetization current" derives from
the fact that $\bm{j}_{\text{mag}}^e=\bm{\nabla}_r\times\bm{M}$, where $\bm{M}$ is the magnetization given by the usual thermodynamic 
relation $M_i= - \partial F/\partial B_i$. 
In the following, we will identify an analogous ``heat magnetization current".

We will calculate $\bm{j}_{\text{tr}}^{e(Q)}$ as follows.
(1) Identify the current operators.
(2) Evaluate their expectation values in global equilibrium to find $\bm{j}_{\text{mag}}^{e(Q)}$. 
(3) Calculate their expectation values under non-equilibrium conditions and subtract $\bm{j}_{\text{mag}}^{e(Q)}$. What remains is the transport current.

We now proceed with finding the expressions for $\bm{j}^{e(Q)}(\bm{r})$. The electrical (e) and number (N) current operators are defined by 
\begin{align}
   \widehat{\bm{j}}^e(\bm{r})&=- {\delta{\widehat{H}}}/{\delta{\bm{A}(\bm{r})}} = -\,e\, \widehat{\bm{j}}^N(\bm{r}),
    \label{current operators}
\end{align}
\noindent
and the thermal current operator is given by
\begin{equation}
\widehat{\bm{j}}^Q(\bm{r})=\widehat{\bm{j}}^E(\bm{r})-\mu(\bm{r})\widehat{\bm{j}}^N(\bm{r})
\end{equation}
with
\begin{equation}
    \widehat{\bm{j}}^E(\bm{r})=\dfrac{1}{2}\{\widehat{H}-e\phi(\bm{r}), \widehat{\bm{j}}^N(\bm{r})\}.
\end{equation}

The expectation values of operators are calculated with the electron wave-packet as described in Appendix \ref{app eom}. We use the shorthand notation 
\begin{equation}
\int_{\bm{k}} \equiv \int \frac{d^3k}{(2\pi)^3}, \ \ \int_{\bm{r}} \equiv \frac{1}{V}\int d^3r, \ {\rm and} \ \int_{\bm{\xi}}\equiv \int_{\bm{r}}\int_{\bm{k}}
\end{equation} 
where $V$ is the volume of the system.
We also implicitly sum over repeated indices and suppress the $\bm{\xi}$-dependence of quantities. The expectation values are given by
\begin{align}
\bm{j}^e(\bm{r})&=-e\sum_{\pm}\int_{\bm{k}}\bigg[\mathcal{D}\, f \, \dot{\bm{r}}+\bm{\nabla}_r \times \big( \mathcal{D}\,f\,\bm{\mathfrak{m}}\big)\bigg] \nonumber \\
\bm{j}^Q(\bm{r})&=\sum_{\pm}\int_{\bm{k}}\bigg[\mathcal{D}\, f\, \dot{\bm{r}}\, \tilde{\mathcal{E}}(\bm{\xi})
+\bm{\nabla}_{\bm{r}}\times\big(\mathcal{D}\, f\, \bm{\mathfrak{m}}\, \tilde{\mathcal{E}}(\bm{\xi})\big)\bigg]\nonumber\\
&-\big(e\phi(\bm{r})+\mu(\bm{r})\big)\bm{j}^N(\bm{r})
\label{currents}
\end{align}
\noindent
where
\begin{equation}
\mathfrak{m}_n=\dfrac{i}{2\hbar}\epsilon_{nij}\bra{\partial_{k_i}u}(\widehat{\mathcal{H}}_c-\mathcal{E})\ket{\partial_{k_j}u}
\end{equation}
\noindent
times the electric charge is the orbital magnetic moment of a wave-packet and $\bm{j}^N(\bm{r})=-\bm{j}^e(\bm{r})/e$. The terms containing $\bm{\mathfrak{m}}(\bm{\xi})$ in eq.~\eqref{currents} originate from the finite width of the electron wave-packet that allow contributions to the current deriving from its rotation; see Appendix \ref{app eom} and ref.~\cite{xiao2010berry}. 

\medskip
\noindent
{\bf Bound Magnetization Currents:} In global equilibrium, when $f(\bm{\xi})=f_{\text{eq}}(\bm{\xi})$, $\mu(r)=\mu_0$, $T(r)=T_0$, and $\phi(r)=\phi_0$, 
the transport current must vanish, and the equilibrium expectation value $\bm{j}_{\text{eq}}^{e(Q)}$ is equal to the bound or magnetization current.  
We thus solve for $\bm{j}^{e(Q)}_{\text{mag}}(\bm{r})$ by evaluating eq.~\eqref{currents} in global equilibrium, without reference to any 
thermodynamic definitions of the local magnetizations.  

Following ref.~\cite{xiao2020unified}, the global equilibrium currents are
\begin{align}
    \bm{j}_{\text{eq}}^e(\bm{r})&=-e\bm{\nabla}_r\times \bm{M}^N_{\text{eq}}(\bm{r})\nonumber\\
    \bm{j}_{\text{eq}}^Q(\bm{r})&=\bm{\nabla}_r\times\bm{M}^E_{\text{eq}}(\bm{r})-(e\phi_0+\mu_0)\bm{\nabla}_r\times \bm{M}^N_{\text{eq}}(\bm{r}).
    \label{eq currents}
\end{align}
\noindent
with
\begin{align}
 \bm{M}^N_{\text{eq}}(\bm{r}) &= \sum_{s=\pm}\int_{\bm{k}}\bigg(\mathcal{D}f_{\text{eq}}(\bm{\xi})\bm{\mathfrak{m}}(\bm{\xi})+g_{\text{eq}}(\bm{\xi})\,s\,\bm{\Omega}_k\bigg)\nonumber \\
\bm{M}^E_{\text{eq}}(\bm{r})&=\sum_{s=\pm}\int_{\bm{k}}\bigg(\mathcal{D}f_{\text{eq}}(\bm{\xi})\tilde{\mathcal{E}}(\bm{\xi})\bm{\mathfrak{m}}(\bm{\xi})+h_{\text{eq}}(\bm{\xi})\,s\,\bm{\Omega}_k\bigg)
    \label{magnetizations}
\end{align}
\noindent
where $\bm{\Omega_k}$ is a pseudovector constructed from the ${\bf k}$-space Berry curvature, $\Omega_{k_i}\equiv \epsilon_{ijk}\Omega_{k_j k_k}/2$ and $s = \pm 1$ for the two bands. Note that for simplicity we have suppressed the band label on all other quantities.
See Appendix \ref{app jeq} for a detailed derivation.

We have also introduced the auxiliary functions $g_{\text{eq}} = -k_B T_0 \ln(1 + e^{-(\tilde{\mathcal{E}}-\mu_0)/k_BT_0})$ and $h_{\text{eq}} = -\int_{\tilde{\mathcal{E}}}^{\infty}d\eta\ \eta f_{\text{eq}}$ which have the following relationships to $f_{\text{eq}}$:
\begin{align}
    \frac{\partial g_{\text{eq}}}{\partial \tilde{\mathcal{E}}} =-\frac{\partial g_{\text{eq}}}{\partial \mu_0}= f_{\text{eq}} \,\,,\,\,
    \frac{\partial h_{\text{eq}}}{\partial \tilde{\mathcal{E}}} = -\frac{\partial h_{\text{eq}}}{\partial \mu_0}+g_{\text{eq}}=\tilde{\mathcal{E}}f_{\text{eq}}.
    \label{auxID}
\end{align}
\noindent

Eq.~\eqref{magnetizations} follows from eq.~\eqref{currents} by noting that to first order in the Berry curvatures $-\partial_{k_j}\mathbf{S}_{ij} = \partial_{r_j}\mathbf{K}_{ij}$ where we recall that the $\mathbf{K}$ and $\mathbf{S}$ matrices were $3\times 3$ blocks of 
$\te{\Gamma}^{-1}$ in eq.~\eqref{gamma inverse}. 
We would like to emphasize again the crucial role, already noted in ref.~\cite{xiao2020unified}, played by the mixed space Berry curvatures in ${\mathbf{S}}$
in determining the form of the equilibrium magnetization and associated equilibrium currents.

By evaluating the currents in global equilibrium, we have identified the magnetization currents. Equilibrium (bound) electric currents take the familiar form as the curl of the magnetization.  The heat magnetization current is less familiar~\cite{gromov2015thermal, qin2011Energy, zhang2020Thermodynamics}, but in a similar fashion is defined in terms of the curl of the energy and number magnetizations. 

\medskip
\noindent
{\bf Transport Currents:} 
We next evaluate eq.~\eqref{currents} under non-equilibrium conditions, i.e., in the presence of a spatially dependent temperature, chemical potential, and electric potential. 
As before, the non-equilibrium distribution function deviates from local equilibrium: $f(\bm{\xi})=f_{\text{leq}}(\bm{\xi}) + \delta f(\bm{\xi})$. 
We find that the non-equilibrium electric current includes a term  $\bm{j}_{\text{mag}}^e=-e\bm{\nabla}_r\times \bm{M}^N_{\text{leq}}(\bm{r})$ and the heat current contains a term $\bm{j}_{\text{mag}}^Q=\bm{\nabla}_r\times\bm{M}^E_{\text{leq}}(\bm{r})-(e\phi(\bm{r})+\mu(\bm{r}))\bm{\nabla}_r\times \bm{M}^N_{\text{leq}}(\bm{r})$. 
We identify these as the magnetization currents existing under non-equilibrium conditions. 
Note that they have the same form as the equilibrium currents in eq.~\eqref{eq currents}, but with the crucial difference that we must use
spatially dependent fields $\mu(\bm{r})$ and $\phi(\bm{r})$, and \textit{local equilibrium} magnetizations, $\bm{M}_{\text{leq}}^{N}$ and $\bm{M}_{\text{leq}}^{E}$. 
The local equilibrium magnetizations are similar to those in eq.~\eqref{magnetizations}, except that $f_{\text{eq}}\rightarrow f_{\text{leq}}$, and the auxiliary functions are similarly altered, $g_{\text{leq}}=g_{\text{eq}}\big(\mu_0\rightarrow \mu(\bm{r}), T_0\rightarrow T(\bm{r})\big)$ and 
$h_{\text{leq}}=h_{\text{eq}}\big(\mu_0\rightarrow \mu(\bm{r}), T_0\rightarrow T(\bm{r})\big)$. 

Subtracting the magnetization currents from the total nonequilibrium currents allows us to calculate the transport currents to 
linear order in $(e\bm{E}-\bm{\nabla}_r \mu)$ and $\bm{\nabla}_r T$. 
Explicitly writing the band labels $\pm$ that we had suppressed thus far, we find

\begin{widetext}
\begin{align}
    \bm{j}_{\text{tr}}^e(\bm{r}) &= -\frac{e}{\hbar}\sum_{\pm}\int_{\bm{k}} \bigg[\delta f^{\pm}\, \bm{\nabla}_k\widetilde{\mathcal{E}}_{\pm}
    + \bigg(s\, \bm{\Omega}_k\times\bm{\nabla}_r T\bigg) \partial_{T_0} g^{\pm}_{\text{eq}}+s\,  \bm{\Omega}_k\times(e\bm{E}-\bm{\nabla}_r\mu)\partial_{\mu_0}
    g^{\pm}_{\text{eq}}\bigg]
     \label{jtr e}
    \\
    \bm{j}_{\text{tr}}^Q(\bm{r})&= \frac{1}{\hbar}\sum_{\pm}\int_{\bm{k}} \bigg[(\widetilde{\mathcal{E}}_{\pm} -\mu_0)\,
    \delta f^{\pm}\,\bm{\nabla}_k\widetilde{\mathcal{E}}_{\pm}
    +\bigg(s\, \bm{\Omega}_k\times \bm{\nabla}_rT\bigg)\partial_{T_0} (h^{\pm}_{\text{eq}}-\mu_0 g^{\pm}_{\text{eq}})+ s\,\bm{\Omega}_k\times 
    (e\bm{E}-\bm{\nabla}_r\mu)T_0\partial_{T_0} g^{\pm}_{\text{eq}} \bigg].
    \label{jtr q}
\end{align}
\end{widetext}

Up to this point we have not made any assumption about $\lambda/J$ in this Section. We conclude 
by focusing on the weak SOC regime with $\lambda \ll J \lesssim E_F$ or bandwidth.
(The strong SOC regime is discussed in Section~\ref{sec-SOCRegime}).
In this limit, we can make the replacement $\widetilde{\mathcal{E}}_{\pm}(\bm{\xi})\rightarrow \varepsilon_{\pm}(\bm{k})=\varepsilon(\bm{k})\pm J$ 
just as we did in eq.~\eqref{boltz soln approx}.
The equilibrium functions $f_{\text{eq}}, g_{\text{eq}}$, and $h_{\text{eq}}$ are all functions of $\varepsilon_{\pm}(\bm{k})$, 
but from eq.~\eqref{v and M} we see that the velocity $v_i$ and inverse mass tensor $M_{ij}^{-1}$ are independent of the band label.

From the Boltzmann solution (eq. \eqref{boltz soln approx}) we thus see that the effect of the ``emergent magnetic field" $\pm(\hbar/e)\Omega_{r_m r_n}$ from the real space Berry curvature is
analogous to the (real) magnetic field $F_{mn}$ of eq.~\eqref{B field}, except for the sign change between the two bands.
These $\delta f^{\pm}$'s are then substituted into the currents of eqs.~\eqref{jtr e}, \eqref{jtr q} to find the conductivities in the next Section.

We note that only the real- and momentum-space Berry curvatures enter the currents defined by eqs.~\eqref{jtr e}, \eqref{jtr q}, and \eqref{boltz soln approx}. 
The mixed space Berry curvature enters the equations of motion through the off-diagonal block in eq.~\eqref{Gamma matrix}, 
where $\delta_{ij}$ dominates over $\Omega_{r_ik_j} \sim (a/L_s)(\lambda/J) \ll 1$.

The real-space Berry curvature contribution enters $\delta f^\pm$ at zeroth order in SOC. The next correction, 
arising from the real space derivative of the eigenenergies $\mathcal{E}_{\pm}$, scales like $(\ell/L_s)(\lambda/t)$.
This leads to a term in $\delta f^\pm$ that is proportional to $\nabla\cdot{\bf {\hat m}}$ as shown in \eqref{div m}. 
We neglect this contribution in our analysis of the conductivities below, since its volume integral 
can be rewritten as a boundary term that vanishes for an arbitrary spin texture in the thermodynamic limit. 
For a periodic structure, such as a skyrmion crystal, this contribution vanishes in every unit cell; see
ref.~\cite{verma2022unified}.

\section{Conductivities}
\label{sec-conductivities}

We now use spatial average of the local transport currents \eqref{jtr e} and \eqref{jtr q}, together with the solution \eqref{boltz soln approx} of the Boltzmann equation
to determine the long wavelength electrical ($\sigma$), thermoelectric ($\alpha$), and thermal ($\kappa$) conductivities defined in \eqref{transport matrix}. Note that in the next two Sections we will focus on the weak SOC regime. To present our results compactly, we will use the notation 
\begin{equation}
\zeta \in \{\sigma, \alpha, \kappa\}
\end{equation}
\noindent
to represent any one of the three conductivity tensors.  
We do not include here the $\te{\beta}$ tensor of eq.~\eqref{transport matrix}, because it does not contain any
independent information. It is easy to see that our results obey the Kelvin relation between the 
electrothermal and thermoelectric transport coefficients, namely $\te{\beta}=T_0\,\te{\alpha}$. This consistency with the Onsager reciprocal relations also serves as a nontrivial check on 
our analysis, especially on our expression for the thermal transport current.

Let us first briefly look at the longitudinal response which to leading order is independent of the 
magnetic field and of all Berry curvature effects in the regime where our results are valid. 
We find that 
$\zeta_{xx} = \tau\sum_{\pm}\int_{\bm{\xi}}v_x^2(\partial_{\varepsilon}X^{\zeta}_{\pm})$
where the function

\begin{align}
    X^{\zeta}_{\pm}=
    \begin{cases}
        -e^2f_{\text{eq}}^{\pm} & \text{if } \zeta = \sigma \\
        -e\,\partial_{T_0}g_{\text{eq}}^{\pm} & \text{if } \zeta = \alpha \\
        \partial_{T_0}(h_{\text{eq}}^{\pm}-\mu_0 g_{\text{eq}}^{\pm}) & \text{if } \zeta = \kappa
    \end{cases}
    \label{dist fns}
\end{align}

\noindent
Using the the relations between the functions $g_{\text{eq}}$ and $h_{\text{eq}}$ to the 
Fermi function $f_{\text{eq}}$, we see that 
\begin{equation}
 \partial_{\varepsilon} X^{\zeta}_{\pm} = \big(- {\partial f_{\text{eq}}^{\pm}}/{\partial \varepsilon}\big)\,\, Y^{\zeta}_{\pm}
 \end{equation}
 where
 \begin{equation}
 Y^{\zeta}_{\pm}=
    \begin{cases}
        e^2 & \text{if } \zeta = \sigma \\
        -e\, \big(\varepsilon_{\pm} - \mu_0\big)/T_0 & \text{if } \zeta = \alpha \\
        \big(\varepsilon_{\pm} - \mu_0\big)^2/T_0 & \text{if } \zeta = \kappa \\
    \end{cases}
    \label{df-de}
\end{equation}

\noindent
These are the natural prefactors for the velocities in the charge-charge, charge-heat, and heat-heat current correlations.
We thus recover the standard results
\begin{equation}
   \zeta_{xx} = \tau\sum_{\pm}\int_{\bm{k}} \bigg(\!- \frac{\partial f_{\text{eq}}^{\pm}}{\partial \varepsilon}\bigg )\, Y^{\zeta}_{\pm}\, v_x^2
   \label{longitudinal cond}
\end{equation}
where $\int_{\bm{\xi}}$ reduces to $\int_{\bm{k}}$ since there is no spatial dependence to the integrand.
Here and below, $\big(\!- {\partial f_{\text{eq}}^{\pm}}/{\partial \varepsilon}\big) \simeq \delta \big( \varepsilon_{\pm} - \mu_0\big)$ restricts 
$\int_{\bm{k}}$ to the Fermi surface in the low temperature regime $k_BT_0 \ll \mu_0$.

We next turn to the transverse response, the main focus of our work.
We see that the transport currents have two distinct types of contributions: 
(i) an ``extrinsic" contribution that involves the scattering time $\tau$, which enters through $\delta f^{\pm}$ in the first term of
eqs.~\eqref{jtr e} and \eqref{jtr q}, and (ii) an ``intrinsic" contribution independent of $\tau$, coming from the 
second and third terms in each of the expressions. 
The external magnetic field and real space Berry curvature enter $\bm{j}_{\text{tr}}^{e(Q)}$ only through $\delta f^{\pm}$
and the corresponding ``extrinsic" terms give rise to the ordinary and topological responses respectively.  The ${\bf k}$-space Berry curvature, on the other hand,  enters through the ``intrinsic" terms and leads to the anomalous response.  The form of eqs.~\eqref{jtr e}, \eqref{jtr q}, and \eqref{boltz soln approx} clearly shows that there are three contributions 
-- ordinary, topological , and anomalous -- to the currents and hence to the conductivities as well.
We now show this in detail.

As discussed in Section \ref{sec-intro}, the Hall conductivities can be compactly represented by the pseudovector $\bm{\zeta}$. From  eqs.~\eqref{jtr e} and \eqref{jtr q} we see that

\phantom{...}

\begin{widetext}
\begin{equation}
    \zeta_i = \sum_{\pm}\int_{\bm{\xi}}\bigg[\frac{\hbar\tau^2}{2}\epsilon_{ijk}\epsilon_{lmn}(\partial_{\varepsilon}X^{\zeta}_{\pm}) v_j M_{km}^{-1}v_n\bigg(s\,\Omega_{r_l} - \frac{e}{\hbar} B_l\bigg) + \frac{X^{\zeta}_{\pm}}{\hbar}s\,\Omega_{k_i}\bigg]
    \label{full Hall}
\end{equation}
\end{widetext}
\noindent
where $i, j, k, l, m, n \in \{x, y, z\}$,  we have written both real and momentum-space Berry curvatures as pseudovectors, and where 
$X^{\zeta}_{\pm}$ is defined in eq.~\eqref{dist fns}. 
We thus see that the Hall conductivities in eq.~\eqref{full Hall} can be written as the sum of three pieces
\begin{equation}
    \bm{\zeta} = \bm{\zeta}^O + \bm{\zeta}^A + \bm{\zeta}^T.
    \label{sumOAT1}
\end{equation}
\noindent
where the ordinary $\bm{\zeta}^O$ term depends upon the external magnetic field $\bm{B}$, the
anomalous $\bm{\zeta}^A$ term on the momentum space Berry curvature, $\bm{\Omega_{k}}$, and the 
the topological $\bm{\zeta}^T$ term on the real space Berry curvature, $\bm{\Omega_{r}}$.
This justifies the additive decomposition of the experimental signals into these three parts
in the regime of validity of our analysis: $k_F \sim a \ll \ell\!=\!v_F\tau \ll L_s$ with $\hbar\omega_c \ll\hbar/\tau \ll E_F$ and $\lambda \ll J \sim E_F$.
The reason why mixed Berry curvatures make a negligible contribution to the final result for transport in this regime was discussed at the end of 
the previous section; see eq.~\eqref{boltz soln approx}.

We next show that these three terms can be written as
\begin{align}
    \bm{\zeta}^O &= \tau^2\, \te{G}^{\zeta}\,\overline{\bm{B}}, \nonumber \\
    \bm{\zeta}^A & = \te{F}^{\zeta}\,\bm{\Lambda},    
    \label{transport coeffs} \\    
    \bm{\zeta}^T &= {\hbar}\tau^2\, \te{F}^{\zeta}\, \bm{n}^{\text{sk}}. \nonumber 
\end{align}
\noindent
Here each expression is written as a product of a tensor $\te{G}^{\zeta}$ or $\te{F}^{\zeta}$ that depends only on the band structure, with an appropriate pseudovector 
$\overline{\bm{B}}$, $\bm{\Lambda}$, or $\bm{n}^{\text{sk}}$, all of which which are defined below. 
Let us discuss each contribution in turn.

\medskip
\noindent
{\bf Ordinary response}: $\bm{\zeta}^O$ is the term in $\bm{\zeta}$ that is proportional to the external magnetic field, e.g., related to the ``ordinary"
$\sigma_{xy}$, $\alpha_{xy}$, or $\kappa_{xy}$. 
In eq.~\eqref{transport coeffs}, the spatial average $\overline{\bm{B}}= \int_{\bm{r}} B(\bm{r})$ is simply $\bm{B}$ for a uniform magnetic field.
Here we have split the $\int_{\bm{\xi}}$ appearing in eq.~\eqref{full Hall} into an $\int_{\bm{r}}$ that enters the definition of the pseudovector, and 
 $\int_{\bm{k}}$  that enters $\te{G}^{\zeta}$ defined below, since the leading order contributions to the latter are only 
 functions of $\bm{k}$ through the energies $\varepsilon_{\pm}(\bm{k})$. (This separation of the real and momentum space integrals 
works in exactly the same way in the topological and anomalous terms as well). We thus see that
\begin{equation}
    G^{\zeta}_{il}=\frac{e}{2} \epsilon_{ijk}\epsilon_{lmn}\sum_{\pm}\int_{\bm{k}} \bigg(\!- \frac{\partial f_{\text{eq}}^{\pm}}{\partial \varepsilon}\bigg)\, Y^{\zeta}_{\pm}\,   v_{j}M_{km}^{-1}v_n
\label{g integrals}
\end{equation}
\noindent
which depends only on the band structure through the band velocity $v_i$ and band curvature $M_{km}^{-1}$, defined in eq.~\eqref{v and M},
and $Y^{\zeta}_{\pm}$ defined in eq.~\eqref{df-de}. We note that the first equation in eq.~\eqref{transport coeffs} together with eq.~\eqref{g integrals}
reproduces the standard weak field results $(\omega_c\tau \ll 1)$ in semiclassical transport theory.

\medskip
\noindent
{\bf Topological response}: 
The topological contribution $\bm{\zeta}^T$ is similar to the ordinary one, but with some crucial differences. 
The ``emergent electromagnetic field"  is given by ($2\pi$ times) the flux quantum $h/e$ multiplied by 
the pseudovector $\bm{n}^{\text{sk}}$ whose components are given by the topological charge -- or skyrmion -- density in the $yz$, $zx$, and $xy$ planes:
\begin{equation}
    n_i^{\text{sk}} = \epsilon_{ijk}\epsilon_{lmn}{1 \over V}\int {d^3{r}}\ \hat{m}_l(\bm{r}) \partial_{r_j}\hat{m}_m(\bm{r})\partial_{r_k}\hat{m}_n(\bm{r}).
    \label{skyrmion number}
\end{equation}
\noindent
The tensor $F^{\zeta}_{il}$ appearing in eq.~\eqref{transport coeffs} is given by 
\begin{equation}
    F^{\zeta}_{il}=\frac12 \epsilon_{ijk}\epsilon_{lmn}\sum_{\pm}\, s\, \int_{\bm{k}} 
    \bigg(\!- \frac{\partial f_{\text{eq}}^{\pm}}{\partial \varepsilon}\bigg)\, Y^{\zeta}_{\pm}\,   v_{j}M_{km}^{-1}v_n.
\label{f integrals}
\end{equation}
\noindent
with the ${\bm{k}}$-space integrals dominated by states near the Fermi surface. In fact the
integrals in $F^{\zeta}_{il}$ are identical to those in $G^{\zeta}_{il}$,  however, unlike eq.~\eqref{g integrals} there is 
a relative sign $s = \pm 1$ between the contributions of the two bands originating from the sign change of the Berry curvature in eq.~\eqref{curv}.
This sign change between bands in $F^{\zeta}_{il}$ has important observable consequences.
It leads to, for instance, the avoidance of a Sondheimer cancellation in the topological (and also anomalous) Nernst effects, as noted in Ref. \cite{addison2023theory}.

\medskip
\noindent
{\bf Anomalous response}: 
There are two important questions about the form of $\bm{\zeta}^A$ in eq.~\eqref{transport coeffs}.
What is the pseudovector $\bm{\Lambda}$? And why does $\bm{\zeta}^A$ involve the same $F^{\zeta}_{il}$ tensor as the topological term? 
The latter is far from from obvious by looking at the anomalous and topological contributions in eq.~\eqref{full Hall}.

We briefly sketch here how we obtain the anomalous response in eq.~\eqref{transport coeffs},
with details of the lengthy algebra relegated to Appendix \ref{app anom}. For concreteness, we focus on the anomalous Hall effect $\sigma_i^A$;
the strategy for $\alpha_i^A$ and $\kappa_i^A$ is similar.
We see from the last term in eq.~\eqref{full Hall}  that 
\begin{equation}
\sigma_i^A = - (e^2/\hbar) \sum_{\pm} s \int_{\bm{\xi}}f^{\pm}_{\text{eq}}\Omega_{k_i}.
\label{an hall}
\end{equation}
We substitute into this result the Berry curvature  
$\Omega_{k_i} = \frac14 \epsilon_{ijk}\epsilon_{lmn} \hat{d}_l (\partial_{k_j}\hat{d}_m)(\partial_{k_k}\hat{d}_n)$
calculated using the $\hat{\bm{d}}$-vector of eq.~\eqref{bloch d} for $\lambda \ll J$. This leads to
\begin{align}
    \sigma_{i}^A = &\frac{e^2\hbar^3}{4}  \bigg(\frac{\lambda}{Jat}\bigg)^2 \epsilon_{ijk}\, \epsilon_{lmn}\, \chi_{m \mu}\, \chi_{n \nu} \nonumber\\
   & \times\, \int_{\bm{r}} \hat{m}_l(\bm{r}) \sum_{\pm} s \int_{\bm{k}}  f^{\pm}_{\text{eq}} M_{j \mu}^{-1}M_{k\nu}^{-1}
\end{align}
where the effective masses arise from the ${\bm{k}}$-space derivatives of band velocities.
We then manipulate the $\int_{\bm{k}}$ term using integration by parts and exploiting the antisymmetric $\epsilon$-tensors
as shown in Appendix \ref{app anom} we obtain
\begin{align}
    \sigma_{i}^A =& \frac{e^2\hbar^3}{8}\bigg(\frac{\lambda}{Jat}\bigg)^2 \epsilon_{ijk}\epsilon_{lmn}
    \epsilon_{\gamma\alpha\beta}\epsilon_{\gamma\mu\nu}\chi_{m \alpha}\chi_{n \beta}\int_{\bm{r}}\hat{m}_l(\bm{r})  \nonumber\\
    & \times \sum_{\pm} s \int_{\bm{k}} \bigg(\!- \frac{\partial f_{\text{eq}}^{\pm}}{\partial \varepsilon}\bigg)\, v_j M_{k \nu}^{-1} v_{\mu}
    = F^\sigma_{ij}\Lambda_j,
\end{align}
\noindent
with the $F^\sigma_{ij}$  of eq.~\eqref{f integrals} and $\Lambda_j$ defined below. A general prescription for writing the
anomalous Hall result \eqref{an hall} in a metallic ferromagnet as a Fermi surface integral was given in ref.~\cite{haldane2004berry}. The derivation
above achieves the same goal for a system with an arbitrary spatially varying magnetization.

Using a similar procedure, all three anomalous transverse conductivities can be written in the form 
$\bm{\zeta}^A = \te{F}^{\zeta} \bm{\Lambda}$ where the pseudovector 
\begin{equation}
    \Lambda_i = \frac{\hbar^3}{4} \bigg(\frac{\lambda}{Jat}\bigg)^2\epsilon_{ijk}\epsilon_{lmn}\chi_{mj}\chi_{nk}\int_{\bm{r}} \hat{m}_l(\bm{r}).
    \label{lambda}
\end{equation}
\noindent
To get a better feel for $\bm{\Lambda}$, let us look at some specific examples.
For Rashba SOC, described below eq.~\ref{H SOC}, we find 
$\bm{\Lambda}^{\text{R}} = \hbar^3 (\lambda/Jat)^2 \bar{m}_z \hat{z}$, where $\bar{m}_z$ indicates a spatial average 
of the z component of the magnetization, a result reported earlier~\cite{verma2022unified}. 
For the linear Dresselhaus SOC, with $\chi_{xx}=1$, $\chi_{yy}=-1$, and zeroes elsewhere,
$\bm{\Lambda}^{\text{D}}=-\hbar^3 (\lambda/Jat)^2 \bar{m}_z \hat{z}$. 
In order for the $x$ and $y$ components of the magnetization to come into play, there must be nonzero SOC along the z-direction. 
For instance, Ising SOC combined with Rashba SOC would require the nonzero component $\chi_{zz}=1$ in addition to the Rashba components. 
One then finds $\bm{\Lambda}^{\text{R}+\text{I}}=\hbar^3 (\lambda/Jat)^2( \bar{m}_y \hat{x}-\bar{m}_x \hat{y}+\bar{m}_z \hat{z})$. 
In essence, $\bm{\Lambda}$ relates the anomalous conductivities to components of the magnetization, in accordance with the SOC present in the system.

\section{Transport coefficients}
\label{sec-transport}

We now discuss various aspects of our results. First, we discuss the general relations between various transport coefficients that are well known in
the theory of metals. We show that these continue to remain valid in the presence of momentum and real space Berry curvatures.
Second, we relate our results for the electrical and thermoelectric conductivities, which are the natural quantities to compute theoretically, to 
the resistivity, and the Seebeck and Nernst coefficients, which are the natural quantities that experimentalists measure.  Using these relations, the complete set of thermoelectric susceptibilities can be determined from the electrical conductivity itself.  In Appendix \ref{hallC} we calculate the chemical potential (or density) dependence of the electrical Hall conductivity for nearest neighbor interactions on the square lattice and highlight these relationships. 

\medskip
\noindent
\textbf{Kelvin, Mott, and Wiedemann-Franz relations:}  We discuss here some general relations between the electrical, thermal, and thermoelectric 
transport coefficients that can be seen from our results. First, Onsager reciprocity mandates that the Kelvin relation $\te{\beta}=T_0\te{\alpha}$ be obeyed between the 
electrothermal and thermoelectric tensors defined in eq.~\eqref{transport matrix}.
It is straightforward to check that this relation is indeed obeyed in our calculations.

Second, ordinary metals exhibit the Mott relation between the thermoelectrical and electrical conductivity tensors at low temperatures. We see from our results 
that, even in the presence of arbitrary Berry curvatures, the anomalous and topological terms individually obey the Mott relation
\begin{equation}
    \te{\alpha}^{T(A)} =-\frac{\pi^2}{3}\frac{k_B^2 T_0}{e}\frac{\partial}{\partial \mu}\te{\sigma}^{T(A)}
    \label{mott}
\end{equation}
at low temperatures as noted in ref.\cite{addison2023theory}; see Appendix \ref{app Wiedemann} for details.

Next, we show the validity of the 
Wiedemann-Franz (WF) relation between the thermal and electrical transport coefficients.
The anomalous and topological terms individually satisfy
\begin{equation}
    \te{\kappa}^{T(A)} = \frac{\pi^2}{3}\frac{k_B^2 T_0}{e^2}\te{\sigma}^{T(A)}
    \label{wf}
\end{equation}
\noindent
at low temperatures. We sketch here a brief outline of this proof with details described in Appendix \ref{app Wiedemann}.

We find it convenient here to write the conductivity as a sum of ``intrinsic" and ``extrinsic" terms,
in parallel with contributions to the transport currents described in the paragraph below eq.~\eqref{longitudinal cond}.
As noted there, the ``intrinsic" term depends on the ${\bf k}$-space Berry curvature and leads to the anomalous response,
while the ``extrinsic" term depends on the external magnetic field and real space Berry curvature and leads to the
sum of the ordinary and topological responses.

For both types of contributions, it is best to change variables to write the response functions in terms of integral functions of an energy scale $\eta$ rather than just an integral over the phase space variables $\bm{\xi}$.  The response functions each take the form of a phase space integral of the product of some $\bm{\xi}$ dependent function, $S(\bm{\xi})$, and some function that depends only on the energy, $P(\varepsilon(\bm{\xi}))$. We may rewrite this product as

\begin{equation} P(\varepsilon(\bm{\xi}))S(\bm{\xi})=\int d\eta\, \partial_\eta \Theta(\eta-\varepsilon(\bm{\xi})) P(\eta)S(\bm{\xi}).
\end{equation}

\noindent
For the ``extrinsic" contributions, $P(\eta)$ is peaked around $\mu_0$ for $k_BT_0\ll E_F$. For the ``intrinsic" contributions we integrate by parts and note that $\partial_\eta P(\eta)$ is also peaked around the Fermi energy at low temperatures.  Expansions of these terms can be performed around $\mu_0$ to show that the WF relation is satisfied for both the intrinsic and extrinsic pieces (see Appendix \ref{app Wiedemann} for details).

\medskip
\noindent
{\bf Resistivity and Nernst Coefficient:} 
It is straightforward to express our results in terms of the experimentally measured resistivity, Seebeck coefficient, and Nernst coefficient.
The dependence of the resistivity, Seebeck and Nernst effects, and the thermal conductivity on the parameters of our theory is
derived in Appendix \ref{app scaling} and summarized in Section \ref{sec-intro}; see Fig.~\ref{scaling table}. 
The resistivity tensor is given by $\te{\rho}=\te{\sigma}^{-1}$, and thus in 2D, for instance, we get $\rho_{xx} \simeq 1/\sigma_{xx}$
and $\rho_{xy} \simeq \sigma_{xy}/\sigma_{xx}^2$ since $\sigma_{xx} \gg \sigma_{xy}$.

The Seebeck and Nernst coefficients are given by components of the matrix $\te{S}=\te{\sigma}^{-1}\te{\alpha}$. Using the argument of the previous paragraph and the Mott relation, we find that $N_{xy}\approx -(\pi^2/3)(k_B^2 T_0/e)(\partial/\partial\mu_0)(\sigma_{xy}/\sigma_{xx})$ for each of the ordinary, anomalous, and topological Nernst coefficients. As pointed out in \cite{addison2023theory}, the ordinary Nernst effect is highly suppressed in the relaxation time approximation. This Sondheimer cancellation \cite{sondheimer1948theory} is avoided in the anomalous and topological Nernst effects due to the opposite signs of Berry curvatures in the two bands of this spin-split model.

\medskip
\noindent
{\bf In-plane Hall effects:} We now turn to a discussion of the ordinary, anomalous, and topological {\it in-plane Hall effects} (IPHE). These Hall effects arise from a magnetic field ${\bf B}$
or effective magnetic field that lies parallel to the plane in which the Hall effect is measured, namely, the plane spanned by a unit vector that points in the direction $\bm{\nabla}T$, $\bm{\nabla}\mu$, or $\bm{E}$ and a unit vector that points in the direction of the applied currents. This is in contrast to the usual ``out of plane'' Hall effect (OPHE), in which $\bm{B}$ is applied perpendicular to this measurement plane. We would like to emphasize that the IPHE is a genuine Hall effect deriving from the antisymmetric part of the conductivity tensor that is odd under time reversal (${\bf B}\to - \,{\bf B}$). It is distinct from what is often termed a ``planar Hall effect", which contributes to the symmetric part, $\zeta_{ij}+ \zeta_{ji}$, and is even under time reversal.

The IPHE has garnered great experimental interest in recent years. The ordinary IPHE has been reported in $\text{RuO}_2$ \cite{cui2024antisymmetric}, but most experimental studies have focused on the anomalous and topological IPHEs \cite{you2019angular,wang2019magnetic,ge2020unconventional,tan2021unconventional,zhou2022heterodimensional,chen2023observation,cao2023plane}. We shall therefore briefly discuss the conditions which are required for these two effects, as captured by Eq. \eqref{transport coeffs}.

Both the anomalous and topological IPHEs require that there be an effective magnetic field ($\bm{\Lambda}$ or $\bm{n}^{\text{sk}}$, respectively) lying in the plane. In the case of the topological IPHE, this amounts to requiring that there be a nonzero skyrmion density lying in the planes perpendicular to the plane of measurement. This depends solely on the magnetic texture, and could be satisfied, for example, by skyrmion tubes lying in the measurement plane or by 3D magnetic hedgehogs \cite{yokouchi2015formation,kanazawa2017noncentrosymmetric,wolf2022unveiling}. In the anomalous IPHE, the orientation of $\bm{\Lambda}$ depends on both the SOC and the magnetic texture, as revealed by the expression for $\bm{\Lambda}$ given in Eq. \eqref{lambda}. For example, Rashba SOC, $(\bm{k}\times\hat{z})\cdot\bm{\sigma}$, combined with a ferromagnetic texture $\hat{\bm{m}}=\hat{z}$, results in $\bm{\Lambda}\parallel \hat{z}$. However, Rashba SOC which breaks both x- and z-mirror symmetries, such as $(\bm{k}\times(\hat{z}+\hat{x}))\cdot\bm{\sigma}$, coupled to $\hat{\bm{m}}=\hat{z}$, results in nonzero in-plane components of $\bm{\Lambda}$.

In addition to an in-plane effective magnetic field, the anomalous and topological IPHEs require nonzero off-diagonal components of $\te{F}^{\zeta}$. By inspecting Eq. \eqref{f integrals}, we see that this IPHE arises from an effective Lorentz force caused by in-plane components of the effective magnetic field. The IPHE described by this theory can only appear in 3D materials, because this effective Lorentz force is forbidden in 2D. In addition, the off-diagonal components of $\te{F}^{\zeta}$ are only permitted in low symmetry crystals \cite{kurumaji2023symmetry}.

Going beyond symmetry constraints on the IPHE, one might wonder about the relative strength of the in-plane effects and the more traditional OPHE.  For the electrical Hall conductivity, the IPHE vanishes for a spherical Fermi surface, and a small
perturbation to the dispersion that breaks the isotropy of the Fermi surface leads to a ratio of the IPHE to OPHE that scales with the size of this perturbation.  Therefore, low symmetry crystals with elongated elliptical Fermi surfaces possess a greater opportunity to observing large IPHEs \cite{keyesThesis, keyes2025IPHE}. We leave for future work the full exploration of the parameter space where IPHEs can be significant.

\section{Weak vs.~strong SOC and Topological transition}
\label{sec-SOCRegime}
Up to this point, we have focused on the weak SOC (``adiabatic") regime $\lambda \ll J \lesssim t \sim E_F$.
We will now show how our results change as a function of $\lambda/J$, revealing a topological phase transition 
in the electronic structure at which the Berry curvatures diverge and the semiclassical approach breaks down. 
Beyond this transition, we find that the topological Hall response vanishes in the strong SOC regime $J \ll \lambda \ll t \sim E_F$.

To be explicit, we focus on the simplest example that allows us to analyze the problem in detail, though we believe that the 
qualitative conclusions are more general. We look at 
2D electrons with parabolic dispersion with effective mass $m^*$, bandwidth cutoff of order $t \sim \hbar^2/m^*a^2$, 
and Rashba SOC of strength $\lambda$, interacting with a skyrmion crystal spin texture $\hat{\bf m}({\bf r})$.
Its semiclassical Hamiltonian \eqref{bloch d} is defined in terms of
\begin{equation}
    {\bf d} = \left(\lambda k_y a \!-\!J \hat{m}_x({\bf r}), -\!\lambda k_x a\!- \!J \hat{m}_y({\bf r}), -\!J \hat{m}_z({\bf r})\right).
    \label{Rashba d}
\end{equation}
\noindent
The eigenvalues are ${\mathcal{E}}_{\pm} = \hbar^2 k^2/2m^* \pm d$ where 
$k^2 = k_x^2 + k_y^2$ and the gap separating the two bands is
$d = |{\bf d}({\bf r}, {\bf k})|$.

Our analysis proceeds in three steps. (1) First, we
examine the $(\lambda/J)$-dependence of the \textbf{r}-space Berry curvature
$\Omega_{r_xr_y}\!=\!{\bm{d}}\cdot(\partial_{r_x}{\bm{d}}\times\partial_{r_y}{\bm{d}})/{2d^3}$.
Its integral ${\cal Q} = \int_{\rm{u.c.}}\!d^2r\, \Omega_{r_xr_y}/{2\pi}$
over a region with periodic boundary conditions, say the  the skyrmion lattice unit cell (u.c.), is an integer: the Chern number.  In contrast to the small SOC regime, we will find that ${\cal Q}$ is in general distinct from one of the other topological invariants of the system, 
$N_{sk} = \int_{\rm{u.c.}}\!d^2r\ \hat{\bf m}(\bm{r}) \cdot (\partial_{r_i}\hat{\bf m}(\bm{r}) \times \partial_{r_j}\hat{\bf m}(\bm{r}))/4\pi$,
the skyrmion number. Obviously, $N_{sk}\in \mathbb{Z}$ independent of $(\lambda/J)$, for a skyrmion crystal.
We will show that ${\cal Q}$ vanishes at strong SOC, even though it
coincides with $N_{sk}$ in the weak SOC regime, as we saw in our analysis of the topological Hall effect above.
(2) Clearly, this discontinuous change in ${\cal Q}$ implies a topological phase transition in the band structure.
We identify this transition at a critical value $(\lambda/J)_c$ of order unity where the gap closes,
the Berry curvature diverges and the semiclassical approach breaks down.
(3) Finally, we briefly discuss the large $(\lambda/J)_c$, where we argue that one can again use the 
semiclassical method, but now the topological Hall effect vanishes, and anomalous Hall response arising
from the \textbf{k}-space Berry curvature dominates.

We write $\Omega_{r_xr_y}$ as the sum of two terms,
\begin{align}
    \Omega_{r_xr_y} =  -& {J^3 \over 2 d^3}\ \hat{\bf m}(\bm{r}) \cdot (\partial_{r_x}\hat{\bf m}(\bm{r}) \times \partial_{r_y}\hat{\bf m}(\bm{r})) 
        \label{Omega-xy} \\
   +&  {\lambda J^2 \over 2 d^3} \left(k_y a, - k_x a, 0\right) \cdot (\partial_{r_x}\hat{\bf m}(\bm{r}) \times \partial_{r_y}\hat{\bf m}(\bm{r})). \nonumber 
\end{align}
In the weak SOC limit $d \approx J$, the first term dominates and $\Omega_{r_xr_y}$ is just the skyrmion density,
${\cal Q} = N_{sk}$ as noted above. However, in the opposite limit of strong SOC, $d \approx \lambda k a$ independent of $\bm{r}$, the second term dominates, 
and $\Omega_{r_xr_y}$ can no longer be identified with the skyrmion density. 
We also see that now $\Omega_{r_xr_y}$ scales like $(J/\lambda)^2 L_s^{-2} \ll L_s^{-2}$, the scale of the 
\textbf{r}-space Berry curvature for weak SOC.

We next show that ${\cal Q}=0$ in the strong SOC limit. 
From the second term of eq.~\eqref{Omega-xy}, we see that
$\Omega_{r_xr_y}$ is a function of both ${\bf r}$ and ${\bf k}$, and we focus on its ${\bf r}$-dependence at some fixed, generic value of ${\bf k}$. 
For simplicity, we work with a square skyrmion crystal with an $L\times L$ unit cell.
The spatial integral entering ${\cal Q}$ is given by 
${\cal I} = \epsilon_{abc}\int dr_x \int dr_y\, \partial_{r_x} \hat{m}_a \partial_{r_y} \hat{m}_b$, as in this regime $d$ is independent of $\bm{r}$.  The integral is over the domain 
$-L/2 \leq r_x < L/2, \, -L/2 \leq r_y < L/2$ with periodic boundary conditions (PBCs):
$m_a(r_x,L/2) = m_a(r_x,-L/2)$ and $m_a(L/2,r_y) = m_a(-L/2,r_y)$.
We first do an integration by parts in $\int dr_y$ and then in $\int dr_x$. The boundary terms vanish in both cases due to the PBCs.
This effectively interchanges the $r_x$ and $r_y$ derivatives, and we obtain 
${\cal I} = \epsilon_{abc}\int dr_x \int dr_y\, \partial_{r_y} \hat{m}_a \partial_{r_x} \hat{m}_b$. 
Interchanging the dummy indices $a$ and $b$,
we see that the integral vanishes and thus ${\cal Q} = 0$ in the strong SOC limit.

While $N_{sk}$ remains unchanged, the discontinuous change in ${\cal Q}$ going from weak to strong SOC
implies a topological phase transition in the electronic structure as a function of $(\lambda/J)$.
To understand this we need to investigate the conditions under which
the gap between the two bands ${\mathcal{E}}_{\pm}$ vanishes.
This requires all three components of ${\bf d}$ in eq.~\eqref{Rashba d} to vanish, so that 
\begin{equation}
\lambda k_x a = J m_x({\bf r}^*), \ \lambda k_y a = - J m_y({\bf r}^*), \  m_z({\bf r}^*) = 0.
\label{d=0}
\end{equation}
Here ${\bf r} = {\bf r}^*$ is a real-space contour in the skyrmion crystal u.c.~defined by the third equation in eq. \eqref{d=0}.
The topology of a skyrmion guarantees the existence of this contour on which $m_z({\bf r}^*) = 0$. 
Adding the squares of the three equations, and using $m_x^2({\bf r}^*) + m_y^2({\bf r}^*) + m_z^2({\bf r}^*) = 1$, 
we find that the gap closes when $k a = J/\lambda$.

In the weak SOC regime this leads to $ka \gg 1$, which is unphysical since the bandwidth cutoff on the dispersion
requires that $ka \lesssim \pi$. Thus we do not have to be concerned about a vanishing gap in the weak SOC regime.
Moreover, since ${\cal Q}$ is a topological invariant, by continuity it must equal $N_{sk}$ for $(\lambda/J) < (\lambda/J)_c$,.
However, as we increase the SOC, we will eventually reach a critical value of $(\lambda/J)_c = 1/(k_F a) \sim O(1)$
when the gap will collapse, the two bands ${\mathcal{E}}_{\pm}$ will become degenerate at the Fermi surface,
and $\Omega_{r_xr_y}$ will diverge.

The semiclassical approach becomes invalid when large external magnetic fields cause ``magnetic breakdown"
across narrow gaps (see, e.g., Appendix J in \cite{ashcroft1978solid}). Another condition for the validity of the semiclassical
approach is that the magnetic flux through the unit cell area $a^2$ must be $\ll \Phi_0 = h/e$, the flux quantum . 
The \textbf{r}-space Berry curvature is like a slowly varying effective magnetic field,
and its magnitude diverges as the gap $d$ vanishes, thus calling into question the entire semiclassical approach
at and near $(\lambda/J)_c$.

Finally, we turn to the large SOC regime.  Here, the gap closes on a ${\bf k}$-space contour $k a \ll 1$, which is
very far from the Fermi surfaces on both the bands, since we have asserted from the beginning that $k_F a$ is of order unity. 
(This is valid for a generic magnetic metal, but not for very low density systems.)
In the semiclassical approach, all transport is governed 
by states near the Fermi surface for $k_BT \ll E_F$, and thus the gap closure and large Berry curvatures far from $k_F$ has
no impact. Near the Fermi surface, we can use the $\Omega_{r_xr_y}$ results derived above for generic ${\bf k}$-points, 
and conclude that ${\cal Q}= 0$, and thus there is no topological Hall effect. We note that as a topological invariant ${\cal Q}$ vanishes
for all $(\lambda/J) > (\lambda/J)_c$, even though we have computed this only for strong SOC.

We must investigate the effects of the other Berry curvatures in the large SOC regime.
The mixed-space Berry curvature $\Omega_{r_xk_y}$ for the  ${\bf d}$-vector of  eq.~\eqref{Rashba d} scales like
$(J/\lambda) (a/L_s) \ll 1$ for strong SOC (in contrast to $(\lambda/J) (a/L_s) \ll 1$ for weak SOC). Here too
it is negligible compared to $\mathbb{1}$ in eq.~\eqref{Gamma matrix} in the semiclassical equations of motion.
Finally, we find that the \textbf{k}-space Berry curvature for arbitrary $\lambda/J$ is given by 
\begin{equation}
\Omega_{k_x,k_y} = -{1 \over 2d^3} \ J \lambda^2 a^2 m_z.
  \label{Omega-kxky}
\end{equation}
This expression reduces to the well-known result~\cite{xiao2010berry} for the $\bm{k}$-space Berry curvature for a 
uniform ferromagnet when we set ${\bf m} = m_z\widehat{\bm{z}}$ independent of $\bm{r}$, and generalizes that result to an arbitrary spin texture.
We see that $\Omega_{k_xk_y}$ scales like $(\lambda/J)^2 a^2$ for weak SOC and as $(J/\lambda) a^2$ for strong SOC. 
Therefore the \textbf{k}-space Berry curvature is large in the strong SOC regime and the anomalous Hall effect will
be the dominant contribution.

\section{Conclusion} 
\label{sec-conclusions}

The main results of our paper are already summarized in Section \ref{sec-intro}. We conclude the paper with
(a) order-of-magnitude estimates of the topological and anomalous contributions of the thermal Hall signal based on our results
and their comparison with experiments, and (b) a discussion of open questions that could be addressed in future work.

{\bf Estimates:} From the table in Fig.~1 we see that  
the topological contribution $\kappa^T/T\sim (k_B^2 k_F/\hbar)(\ell/L_s)^2$. 
In many chiral magnetic materials the spin texture length scale $10 \leq L_s \leq 500$nm, 
while the mean free path $1 \leq \ell \leq 100$nm~\cite{tokura2020magnetic}.  
Taking $\ell/L_s \approx 1/10$, we find $\kappa^T/T = 10^{-2}(k_B^2 k_F/\hbar)$. 
For $10 < k_F\ell < 100$, this yields $\kappa^T/T$ in the range $(10^{-6}-10^{-5})\text{W}/\text{K}^2\cdot\text{cm}$. 
Using the Wiedemann-Franz law, this is comparable to the experimental values in ref.~\cite{hirschberger2020topological}. 
Similarly, assuming $\lambda/J\approx 1/10$, we find that the anomalous thermal Hall conductivity $\kappa^A/T$ also falls within the same range. 
As shown in ref.~\cite{addison2023theory}, 
the anomalous and topological Nernst signals are both in the range $(86-860) \text{nV}/\text{K}$, 
which are also in agreement with experimental results~\cite{hirschberger2020topological}.

 {\bf Open questions:} 
First, we have focused on solving the Boltzmann equation in the semiclassical regime 
$k_F \sim a \ll \ell \ll L_s$, where we derived the simple decomposition of eq.~\eqref{sumOAT}
in the weak SOC regime. The semiclassical approach, however, is equally valid for 
$k_F \sim a  \ll L_s \ll \ell$. A unified treatment of phase-space Berry curvatures in this limit 
is left for future work, where one would need to treat the real-space Berry curvature in a manner 
analogous to the ``strong field regime" $\omega_c \tau \gg 1$ for external magnetic fields.
It is well-known that the solution of the Boltzmann equation in this regime is rather more involved than the
one presented here, but it is certainly tractable. 

Second, as noted in Section~\ref{sec-model}, we have focused here on a single band, which splits
in the presence of SOC and time-reversal breaking (non-zero magnetization) and exhibits ${\bf k}$-space Berry curvature.
The analysis of a multi-band system which can harbor ${\bf k}$-space Berry curvature even in the absence of SOC~\cite{haldane1988model}
and its interplay with real space Berry curvature is left for future investigation.

Third, we focused exclusively on the {\it intrinsic} anomalous Hall effect (AHE) that involves momentum-space Berry curvature, which is
known in many materials to be the dominant contribution~\cite{nagaosa2010anomalous, xiao2010berry}. However, there also exists
{\it extrinsic} contributions to the AHE arising from skew and side jump scattering~\cite{nagaosa2010anomalous} in the presence of SOC
scattering processes. Some progress in incorporating these effects in skyrmion materials has been reported in 
refs.~\cite{lux2020chiral, bouaziz2021transverse}. 
The interplay between the topological Hall effect and the intrinsic and extrinsic AHEs is an interesting open question.

Finally, there are several important questions that go beyond the semiclassical Boltzmann approach.
This approach becomes questionable when the skyrmion size becomes comparable to the lattice spacing $a$, 
and one must then resort to using the Kubo formula to compute conductivities. 
Some results along these lines were presented in ref.~\cite{verma2022unified},
but decomposing numerical results into anomalous and topological contributions is not straightforward.
Also note that the mixed (real/momentum-space) Berry curvature $\Omega_{r_ik_j}$ played no role in our final results because it scales like 
$a/L_s$, which is tiny in the semiclassical regime. It is an open question whether, outside this regime, 
there are observables for which the mixed-space Berry curvature plays an important role.

Another interesting set of questions relate to the topological transition as a function of $\lambda/J$. 
The band indices are constants of motion~\cite{ashcroft1978solid} in the semiclassical approach, and in the weak SOC regime
the conduction electron spins follows the local moment adiabatically~\cite{bruno2004topological, nagaosa2013topological, batista2020};
for a discussion of the opposite non-adiabatic regime, see ref.~\cite{denisov2016}.
Magnetic breakdown will cause inter-band transitions near the topological transition and an analysis of transport in this regime
using, e.g., the Kubo formalism is an open problem. 

\bigskip
\noindent
{\bf Acknowledgments}: We gratefully acknowledge support
from the NSF Materials Research Science and Engineering
Center Grant DMR-2011876.

\bibliography{Bibliography}

\begin{thebibliography}{67}%
\makeatletter
\providecommand \@ifxundefined [1]{%
 \@ifx{#1\undefined}
}%
\providecommand \@ifnum [1]{%
 \ifnum #1\expandafter \@firstoftwo
 \else \expandafter \@secondoftwo
 \fi
}%
\providecommand \@ifx [1]{%
 \ifx #1\expandafter \@firstoftwo
 \else \expandafter \@secondoftwo
 \fi
}%
\providecommand \natexlab [1]{#1}%
\providecommand \enquote  [1]{``#1''}%
\providecommand \bibnamefont  [1]{#1}%
\providecommand \bibfnamefont [1]{#1}%
\providecommand \citenamefont [1]{#1}%
\providecommand \href@noop [0]{\@secondoftwo}%
\providecommand \href [0]{\begingroup \@sanitize@url \@href}%
\providecommand \@href[1]{\@@startlink{#1}\@@href}%
\providecommand \@@href[1]{\endgroup#1\@@endlink}%
\providecommand \@sanitize@url [0]{\catcode `\\12\catcode `\$12\catcode
  `\&12\catcode `\#12\catcode `\^12\catcode `\_12\catcode `\%12\relax}%
\providecommand \@@startlink[1]{}%
\providecommand \@@endlink[0]{}%
\providecommand \url  [0]{\begingroup\@sanitize@url \@url }%
\providecommand \@url [1]{\endgroup\@href {#1}{\urlprefix }}%
\providecommand \urlprefix  [0]{URL }%
\providecommand \Eprint [0]{\href }%
\providecommand \doibase [0]{https://doi.org/}%
\providecommand \selectlanguage [0]{\@gobble}%
\providecommand \bibinfo  [0]{\@secondoftwo}%
\providecommand \bibfield  [0]{\@secondoftwo}%
\providecommand \translation [1]{[#1]}%
\providecommand \BibitemOpen [0]{}%
\providecommand \bibitemStop [0]{}%
\providecommand \bibitemNoStop [0]{.\EOS\space}%
\providecommand \EOS [0]{\spacefactor3000\relax}%
\providecommand \BibitemShut  [1]{\csname bibitem#1\endcsname}%
\let\auto@bib@innerbib\@empty
\bibitem [{\citenamefont {Thouless}\ \emph {et~al.}(1982)\citenamefont
  {Thouless}, \citenamefont {Kohmoto}, \citenamefont {Nightingale},\ and\
  \citenamefont {den Nijs}}]{thouless1982quantized}%
  \BibitemOpen
  \bibfield  {author} {\bibinfo {author} {\bibfnamefont {D.~J.}\ \bibnamefont
  {Thouless}}, \bibinfo {author} {\bibfnamefont {M.}~\bibnamefont {Kohmoto}},
  \bibinfo {author} {\bibfnamefont {M.~P.}\ \bibnamefont {Nightingale}},\ and\
  \bibinfo {author} {\bibfnamefont {M.}~\bibnamefont {den Nijs}},\ }\bibfield
  {title} {\bibinfo {title} {Quantized hall conductance in a two-dimensional
  periodic potential},\ }\href@noop {} {\bibfield  {journal} {\bibinfo
  {journal} {Physical review letters}\ }\textbf {\bibinfo {volume} {49}},\
  \bibinfo {pages} {405} (\bibinfo {year} {1982})}\BibitemShut {NoStop}%
\bibitem [{\citenamefont {Nagaosa}\ \emph {et~al.}(2010)\citenamefont
  {Nagaosa}, \citenamefont {Sinova}, \citenamefont {Onoda}, \citenamefont
  {MacDonald},\ and\ \citenamefont {Ong}}]{nagaosa2010anomalous}%
  \BibitemOpen
  \bibfield  {author} {\bibinfo {author} {\bibfnamefont {N.}~\bibnamefont
  {Nagaosa}}, \bibinfo {author} {\bibfnamefont {J.}~\bibnamefont {Sinova}},
  \bibinfo {author} {\bibfnamefont {S.}~\bibnamefont {Onoda}}, \bibinfo
  {author} {\bibfnamefont {A.~H.}\ \bibnamefont {MacDonald}},\ and\ \bibinfo
  {author} {\bibfnamefont {N.~P.}\ \bibnamefont {Ong}},\ }\bibfield  {title}
  {\bibinfo {title} {Anomalous hall effect},\ }\href@noop {} {\bibfield
  {journal} {\bibinfo  {journal} {Reviews of modern physics}\ }\textbf
  {\bibinfo {volume} {82}},\ \bibinfo {pages} {1539} (\bibinfo {year}
  {2010})}\BibitemShut {NoStop}%
\bibitem [{\citenamefont {Yao}\ \emph {et~al.}(2004)\citenamefont {Yao},
  \citenamefont {Kleinman}, \citenamefont {MacDonald}, \citenamefont {Sinova},
  \citenamefont {Jungwirth}, \citenamefont {Wang}, \citenamefont {Wang},\ and\
  \citenamefont {Niu}}]{yao2004first}%
  \BibitemOpen
  \bibfield  {author} {\bibinfo {author} {\bibfnamefont {Y.}~\bibnamefont
  {Yao}}, \bibinfo {author} {\bibfnamefont {L.}~\bibnamefont {Kleinman}},
  \bibinfo {author} {\bibfnamefont {A.}~\bibnamefont {MacDonald}}, \bibinfo
  {author} {\bibfnamefont {J.}~\bibnamefont {Sinova}}, \bibinfo {author}
  {\bibfnamefont {T.}~\bibnamefont {Jungwirth}}, \bibinfo {author}
  {\bibfnamefont {D.-s.}\ \bibnamefont {Wang}}, \bibinfo {author}
  {\bibfnamefont {.~f.~E.}\ \bibnamefont {Wang}},\ and\ \bibinfo {author}
  {\bibfnamefont {Q.}~\bibnamefont {Niu}},\ }\bibfield  {title} {\bibinfo
  {title} {First principles calculation of anomalous hall conductivity in
  ferromagnetic bcc fe},\ }\href@noop {} {\bibfield  {journal} {\bibinfo
  {journal} {Physical review letters}\ }\textbf {\bibinfo {volume} {92}},\
  \bibinfo {pages} {037204} (\bibinfo {year} {2004})}\BibitemShut {NoStop}%
\bibitem [{\citenamefont {Xiao}\ \emph {et~al.}(2010)\citenamefont {Xiao},
  \citenamefont {Chang},\ and\ \citenamefont {Niu}}]{xiao2010berry}%
  \BibitemOpen
  \bibfield  {author} {\bibinfo {author} {\bibfnamefont {D.}~\bibnamefont
  {Xiao}}, \bibinfo {author} {\bibfnamefont {M.-C.}\ \bibnamefont {Chang}},\
  and\ \bibinfo {author} {\bibfnamefont {Q.}~\bibnamefont {Niu}},\ }\bibfield
  {title} {\bibinfo {title} {{B}erry phase effects on electronic properties},\
  }\href@noop {} {\bibfield  {journal} {\bibinfo  {journal} {Reviews of modern
  physics}\ }\textbf {\bibinfo {volume} {82}},\ \bibinfo {pages} {1959}
  (\bibinfo {year} {2010})}\BibitemShut {NoStop}%
\bibitem [{\citenamefont {Haldane}(2004)}]{haldane2004berry}%
  \BibitemOpen
  \bibfield  {author} {\bibinfo {author} {\bibfnamefont {F.}~\bibnamefont
  {Haldane}},\ }\bibfield  {title} {\bibinfo {title} {Berry curvature on the
  fermi surface: Anomalous hall effect as a topological fermi-liquid
  property},\ }\href@noop {} {\bibfield  {journal} {\bibinfo  {journal}
  {Physical review letters}\ }\textbf {\bibinfo {volume} {93}},\ \bibinfo
  {pages} {206602} (\bibinfo {year} {2004})}\BibitemShut {NoStop}%
\bibitem [{\citenamefont {Onoda}\ \emph
  {et~al.}(2006{\natexlab{a}})\citenamefont {Onoda}, \citenamefont {Sugimoto},\
  and\ \citenamefont {Nagaosa}}]{onoda2006intrinsic}%
  \BibitemOpen
  \bibfield  {author} {\bibinfo {author} {\bibfnamefont {S.}~\bibnamefont
  {Onoda}}, \bibinfo {author} {\bibfnamefont {N.}~\bibnamefont {Sugimoto}},\
  and\ \bibinfo {author} {\bibfnamefont {N.}~\bibnamefont {Nagaosa}},\
  }\bibfield  {title} {\bibinfo {title} {Intrinsic versus extrinsic anomalous
  hall effect in ferromagnets},\ }\href@noop {} {\bibfield  {journal} {\bibinfo
   {journal} {Physical review letters}\ }\textbf {\bibinfo {volume} {97}},\
  \bibinfo {pages} {126602} (\bibinfo {year} {2006}{\natexlab{a}})}\BibitemShut
  {NoStop}%
\bibitem [{\citenamefont {Nagaosa}\ and\ \citenamefont
  {Tokura}(2013)}]{nagaosa2013topological}%
  \BibitemOpen
  \bibfield  {author} {\bibinfo {author} {\bibfnamefont {N.}~\bibnamefont
  {Nagaosa}}\ and\ \bibinfo {author} {\bibfnamefont {Y.}~\bibnamefont
  {Tokura}},\ }\bibfield  {title} {\bibinfo {title} {Topological properties and
  dynamics of magnetic skyrmions},\ }\href@noop {} {\bibfield  {journal}
  {\bibinfo  {journal} {Nature nanotechnology}\ }\textbf {\bibinfo {volume}
  {8}},\ \bibinfo {pages} {899} (\bibinfo {year} {2013})}\BibitemShut {NoStop}%
\bibitem [{\citenamefont {Lee}\ \emph {et~al.}(2009)\citenamefont {Lee},
  \citenamefont {Kang}, \citenamefont {Onose}, \citenamefont {Tokura},\ and\
  \citenamefont {Ong}}]{lee2009unusual}%
  \BibitemOpen
  \bibfield  {author} {\bibinfo {author} {\bibfnamefont {M.}~\bibnamefont
  {Lee}}, \bibinfo {author} {\bibfnamefont {W.}~\bibnamefont {Kang}}, \bibinfo
  {author} {\bibfnamefont {Y.}~\bibnamefont {Onose}}, \bibinfo {author}
  {\bibfnamefont {Y.}~\bibnamefont {Tokura}},\ and\ \bibinfo {author}
  {\bibfnamefont {N.~P.}\ \bibnamefont {Ong}},\ }\bibfield  {title} {\bibinfo
  {title} {Unusual {H}all effect anomaly in {M}n{S}i under pressure},\
  }\href@noop {} {\bibfield  {journal} {\bibinfo  {journal} {Physical review
  letters}\ }\textbf {\bibinfo {volume} {102}},\ \bibinfo {pages} {186601}
  (\bibinfo {year} {2009})}\BibitemShut {NoStop}%
\bibitem [{\citenamefont {Neubauer}\ \emph {et~al.}(2009)\citenamefont
  {Neubauer}, \citenamefont {Pfleiderer}, \citenamefont {Binz}, \citenamefont
  {Rosch}, \citenamefont {Ritz}, \citenamefont {Niklowitz},\ and\ \citenamefont
  {B{\"o}ni}}]{neubauer2009topological}%
  \BibitemOpen
  \bibfield  {author} {\bibinfo {author} {\bibfnamefont {A.}~\bibnamefont
  {Neubauer}}, \bibinfo {author} {\bibfnamefont {C.}~\bibnamefont
  {Pfleiderer}}, \bibinfo {author} {\bibfnamefont {B.}~\bibnamefont {Binz}},
  \bibinfo {author} {\bibfnamefont {A.}~\bibnamefont {Rosch}}, \bibinfo
  {author} {\bibfnamefont {R.}~\bibnamefont {Ritz}}, \bibinfo {author}
  {\bibfnamefont {P.}~\bibnamefont {Niklowitz}},\ and\ \bibinfo {author}
  {\bibfnamefont {P.}~\bibnamefont {B{\"o}ni}},\ }\bibfield  {title} {\bibinfo
  {title} {Topological {H}all effect in the a phase of {M}n{S}i},\ }\href@noop
  {} {\bibfield  {journal} {\bibinfo  {journal} {Physical review letters}\
  }\textbf {\bibinfo {volume} {102}},\ \bibinfo {pages} {186602} (\bibinfo
  {year} {2009})}\BibitemShut {NoStop}%
\bibitem [{\citenamefont {Kanazawa}\ \emph {et~al.}(2011)\citenamefont
  {Kanazawa}, \citenamefont {Onose}, \citenamefont {Arima}, \citenamefont
  {Okuyama}, \citenamefont {Ohoyama}, \citenamefont {Wakimoto}, \citenamefont
  {Kakurai}, \citenamefont {Ishiwata},\ and\ \citenamefont
  {Tokura}}]{kanazawa2011large}%
  \BibitemOpen
  \bibfield  {author} {\bibinfo {author} {\bibfnamefont {N.}~\bibnamefont
  {Kanazawa}}, \bibinfo {author} {\bibfnamefont {Y.}~\bibnamefont {Onose}},
  \bibinfo {author} {\bibfnamefont {T.}~\bibnamefont {Arima}}, \bibinfo
  {author} {\bibfnamefont {D.}~\bibnamefont {Okuyama}}, \bibinfo {author}
  {\bibfnamefont {K.}~\bibnamefont {Ohoyama}}, \bibinfo {author} {\bibfnamefont
  {S.}~\bibnamefont {Wakimoto}}, \bibinfo {author} {\bibfnamefont
  {K.}~\bibnamefont {Kakurai}}, \bibinfo {author} {\bibfnamefont
  {S.}~\bibnamefont {Ishiwata}},\ and\ \bibinfo {author} {\bibfnamefont
  {Y.}~\bibnamefont {Tokura}},\ }\bibfield  {title} {\bibinfo {title} {Large
  topological {H}all effect in a short-period helimagnet {M}n{G}e},\
  }\href@noop {} {\bibfield  {journal} {\bibinfo  {journal} {Physical review
  letters}\ }\textbf {\bibinfo {volume} {106}},\ \bibinfo {pages} {156603}
  (\bibinfo {year} {2011})}\BibitemShut {NoStop}%
\bibitem [{\citenamefont {Li}\ \emph {et~al.}(2013)\citenamefont {Li},
  \citenamefont {Kanazawa}, \citenamefont {Yu}, \citenamefont {Tsukazaki},
  \citenamefont {Kawasaki}, \citenamefont {Ichikawa}, \citenamefont {Jin},
  \citenamefont {Kagawa},\ and\ \citenamefont {Tokura}}]{li2013robust}%
  \BibitemOpen
  \bibfield  {author} {\bibinfo {author} {\bibfnamefont {Y.}~\bibnamefont
  {Li}}, \bibinfo {author} {\bibfnamefont {N.}~\bibnamefont {Kanazawa}},
  \bibinfo {author} {\bibfnamefont {X.}~\bibnamefont {Yu}}, \bibinfo {author}
  {\bibfnamefont {A.}~\bibnamefont {Tsukazaki}}, \bibinfo {author}
  {\bibfnamefont {M.}~\bibnamefont {Kawasaki}}, \bibinfo {author}
  {\bibfnamefont {M.}~\bibnamefont {Ichikawa}}, \bibinfo {author}
  {\bibfnamefont {X.}~\bibnamefont {Jin}}, \bibinfo {author} {\bibfnamefont
  {F.}~\bibnamefont {Kagawa}},\ and\ \bibinfo {author} {\bibfnamefont
  {Y.}~\bibnamefont {Tokura}},\ }\bibfield  {title} {\bibinfo {title} {Robust
  formation of skyrmions and topological {H}all effect anomaly in epitaxial
  thin films of {M}n{S}i},\ }\href@noop {} {\bibfield  {journal} {\bibinfo
  {journal} {Physical review letters}\ }\textbf {\bibinfo {volume} {110}},\
  \bibinfo {pages} {117202} (\bibinfo {year} {2013})}\BibitemShut {NoStop}%
\bibitem [{\citenamefont {Gallagher}\ \emph {et~al.}(2017)\citenamefont
  {Gallagher}, \citenamefont {Meng}, \citenamefont {Brangham}, \citenamefont
  {Wang}, \citenamefont {Esser}, \citenamefont {McComb},\ and\ \citenamefont
  {Yang}}]{gallagher2017robust}%
  \BibitemOpen
  \bibfield  {author} {\bibinfo {author} {\bibfnamefont {J.}~\bibnamefont
  {Gallagher}}, \bibinfo {author} {\bibfnamefont {K.}~\bibnamefont {Meng}},
  \bibinfo {author} {\bibfnamefont {J.}~\bibnamefont {Brangham}}, \bibinfo
  {author} {\bibfnamefont {H.}~\bibnamefont {Wang}}, \bibinfo {author}
  {\bibfnamefont {B.}~\bibnamefont {Esser}}, \bibinfo {author} {\bibfnamefont
  {D.}~\bibnamefont {McComb}},\ and\ \bibinfo {author} {\bibfnamefont
  {F.}~\bibnamefont {Yang}},\ }\bibfield  {title} {\bibinfo {title} {Robust
  zero-field skyrmion formation in {F}e{G}e epitaxial thin films},\ }\href@noop
  {} {\bibfield  {journal} {\bibinfo  {journal} {Physical review letters}\
  }\textbf {\bibinfo {volume} {118}},\ \bibinfo {pages} {027201} (\bibinfo
  {year} {2017})}\BibitemShut {NoStop}%
\bibitem [{\citenamefont {Ahmed}\ \emph {et~al.}(2018)\citenamefont {Ahmed},
  \citenamefont {Rowland}, \citenamefont {Esser}, \citenamefont {Dunsiger},
  \citenamefont {McComb}, \citenamefont {Randeria},\ and\ \citenamefont
  {Kawakami}}]{ahmed2018chiral}%
  \BibitemOpen
  \bibfield  {author} {\bibinfo {author} {\bibfnamefont {A.~S.}\ \bibnamefont
  {Ahmed}}, \bibinfo {author} {\bibfnamefont {J.}~\bibnamefont {Rowland}},
  \bibinfo {author} {\bibfnamefont {B.~D.}\ \bibnamefont {Esser}}, \bibinfo
  {author} {\bibfnamefont {S.~R.}\ \bibnamefont {Dunsiger}}, \bibinfo {author}
  {\bibfnamefont {D.~W.}\ \bibnamefont {McComb}}, \bibinfo {author}
  {\bibfnamefont {M.}~\bibnamefont {Randeria}},\ and\ \bibinfo {author}
  {\bibfnamefont {R.~K.}\ \bibnamefont {Kawakami}},\ }\bibfield  {title}
  {\bibinfo {title} {Chiral bobbers and skyrmions in epitaxial {F}e{G}e/{S}i
  (111) films},\ }\href@noop {} {\bibfield  {journal} {\bibinfo  {journal}
  {Physical Review Materials}\ }\textbf {\bibinfo {volume} {2}},\ \bibinfo
  {pages} {041401} (\bibinfo {year} {2018})}\BibitemShut {NoStop}%
\bibitem [{\citenamefont {Ahmed}\ \emph {et~al.}(2019)\citenamefont {Ahmed},
  \citenamefont {Lee}, \citenamefont {Bagu{\'e}s}, \citenamefont {McCullian},
  \citenamefont {Thabt}, \citenamefont {Perrine}, \citenamefont {Wu},
  \citenamefont {Rowland}, \citenamefont {Randeria}, \citenamefont {Hammel}
  \emph {et~al.}}]{ahmed2019spin}%
  \BibitemOpen
  \bibfield  {author} {\bibinfo {author} {\bibfnamefont {A.~S.}\ \bibnamefont
  {Ahmed}}, \bibinfo {author} {\bibfnamefont {A.~J.}\ \bibnamefont {Lee}},
  \bibinfo {author} {\bibfnamefont {N.}~\bibnamefont {Bagu{\'e}s}}, \bibinfo
  {author} {\bibfnamefont {B.~A.}\ \bibnamefont {McCullian}}, \bibinfo {author}
  {\bibfnamefont {A.~M.}\ \bibnamefont {Thabt}}, \bibinfo {author}
  {\bibfnamefont {A.}~\bibnamefont {Perrine}}, \bibinfo {author} {\bibfnamefont
  {P.-K.}\ \bibnamefont {Wu}}, \bibinfo {author} {\bibfnamefont {J.~R.}\
  \bibnamefont {Rowland}}, \bibinfo {author} {\bibfnamefont {M.}~\bibnamefont
  {Randeria}}, \bibinfo {author} {\bibfnamefont {P.~C.}\ \bibnamefont
  {Hammel}}, \emph {et~al.},\ }\bibfield  {title} {\bibinfo {title}
  {Spin-{H}all topological {H}all effect in highly tunable
  {P}t/ferrimagnetic-insulator bilayers},\ }\href@noop {} {\bibfield  {journal}
  {\bibinfo  {journal} {Nano letters}\ }\textbf {\bibinfo {volume} {19}},\
  \bibinfo {pages} {5683} (\bibinfo {year} {2019})}\BibitemShut {NoStop}%
\bibitem [{\citenamefont {Shao}\ \emph {et~al.}(2019)\citenamefont {Shao},
  \citenamefont {Liu}, \citenamefont {Yu}, \citenamefont {Kim}, \citenamefont
  {Che}, \citenamefont {Tang}, \citenamefont {He}, \citenamefont {Tserkovnyak},
  \citenamefont {Shi},\ and\ \citenamefont {Wang}}]{shao2019topological}%
  \BibitemOpen
  \bibfield  {author} {\bibinfo {author} {\bibfnamefont {Q.}~\bibnamefont
  {Shao}}, \bibinfo {author} {\bibfnamefont {Y.}~\bibnamefont {Liu}}, \bibinfo
  {author} {\bibfnamefont {G.}~\bibnamefont {Yu}}, \bibinfo {author}
  {\bibfnamefont {S.~K.}\ \bibnamefont {Kim}}, \bibinfo {author} {\bibfnamefont
  {X.}~\bibnamefont {Che}}, \bibinfo {author} {\bibfnamefont {C.}~\bibnamefont
  {Tang}}, \bibinfo {author} {\bibfnamefont {Q.~L.}\ \bibnamefont {He}},
  \bibinfo {author} {\bibfnamefont {Y.}~\bibnamefont {Tserkovnyak}}, \bibinfo
  {author} {\bibfnamefont {J.}~\bibnamefont {Shi}},\ and\ \bibinfo {author}
  {\bibfnamefont {K.~L.}\ \bibnamefont {Wang}},\ }\bibfield  {title} {\bibinfo
  {title} {Topological {H}all effect at above room temperature in
  heterostructures composed of a magnetic insulator and a heavy metal},\
  }\href@noop {} {\bibfield  {journal} {\bibinfo  {journal} {Nature
  Electronics}\ }\textbf {\bibinfo {volume} {2}},\ \bibinfo {pages} {182}
  (\bibinfo {year} {2019})}\BibitemShut {NoStop}%
\bibitem [{\citenamefont {Ye}\ \emph {et~al.}(1999)\citenamefont {Ye},
  \citenamefont {Kim}, \citenamefont {Millis}, \citenamefont {Shraiman},
  \citenamefont {Majumdar},\ and\ \citenamefont
  {Te{\v{s}}anovi{\'c}}}]{ye1999berry}%
  \BibitemOpen
  \bibfield  {author} {\bibinfo {author} {\bibfnamefont {J.}~\bibnamefont
  {Ye}}, \bibinfo {author} {\bibfnamefont {Y.~B.}\ \bibnamefont {Kim}},
  \bibinfo {author} {\bibfnamefont {A.}~\bibnamefont {Millis}}, \bibinfo
  {author} {\bibfnamefont {B.}~\bibnamefont {Shraiman}}, \bibinfo {author}
  {\bibfnamefont {P.}~\bibnamefont {Majumdar}},\ and\ \bibinfo {author}
  {\bibfnamefont {Z.}~\bibnamefont {Te{\v{s}}anovi{\'c}}},\ }\bibfield  {title}
  {\bibinfo {title} {{B}erry phase theory of the anomalous {H}all effect:
  application to colossal magnetoresistance manganites},\ }\href@noop {}
  {\bibfield  {journal} {\bibinfo  {journal} {Physical review letters}\
  }\textbf {\bibinfo {volume} {83}},\ \bibinfo {pages} {3737} (\bibinfo {year}
  {1999})}\BibitemShut {NoStop}%
\bibitem [{\citenamefont {Bruno}\ \emph {et~al.}(2004)\citenamefont {Bruno},
  \citenamefont {Dugaev},\ and\ \citenamefont
  {Taillefumier}}]{bruno2004topological}%
  \BibitemOpen
  \bibfield  {author} {\bibinfo {author} {\bibfnamefont {P.}~\bibnamefont
  {Bruno}}, \bibinfo {author} {\bibfnamefont {V.}~\bibnamefont {Dugaev}},\ and\
  \bibinfo {author} {\bibfnamefont {M.}~\bibnamefont {Taillefumier}},\
  }\bibfield  {title} {\bibinfo {title} {Topological {H}all effect and {B}erry
  phase in magnetic nanostructures},\ }\href@noop {} {\bibfield  {journal}
  {\bibinfo  {journal} {Physical review letters}\ }\textbf {\bibinfo {volume}
  {93}},\ \bibinfo {pages} {096806} (\bibinfo {year} {2004})}\BibitemShut
  {NoStop}%
\bibitem [{\citenamefont {Nagaosa}\ \emph {et~al.}(2012)\citenamefont
  {Nagaosa}, \citenamefont {Yu},\ and\ \citenamefont
  {Tokura}}]{nagaosa2012gauge}%
  \BibitemOpen
  \bibfield  {author} {\bibinfo {author} {\bibfnamefont {N.}~\bibnamefont
  {Nagaosa}}, \bibinfo {author} {\bibfnamefont {X.}~\bibnamefont {Yu}},\ and\
  \bibinfo {author} {\bibfnamefont {Y.}~\bibnamefont {Tokura}},\ }\bibfield
  {title} {\bibinfo {title} {Gauge fields in real and momentum spaces in
  magnets: monopoles and skyrmions},\ }\href@noop {} {\bibfield  {journal}
  {\bibinfo  {journal} {Philosophical Transactions of the Royal Society A:
  Mathematical, Physical and Engineering Sciences}\ }\textbf {\bibinfo {volume}
  {370}},\ \bibinfo {pages} {5806} (\bibinfo {year} {2012})}\BibitemShut
  {NoStop}%
\bibitem [{\citenamefont {Kim}\ \emph {et~al.}(2013)\citenamefont {Kim},
  \citenamefont {Lee}, \citenamefont {Lee},\ and\ \citenamefont
  {Stiles}}]{kim2013chirality}%
  \BibitemOpen
  \bibfield  {author} {\bibinfo {author} {\bibfnamefont {K.-W.}\ \bibnamefont
  {Kim}}, \bibinfo {author} {\bibfnamefont {H.-W.}\ \bibnamefont {Lee}},
  \bibinfo {author} {\bibfnamefont {K.-J.}\ \bibnamefont {Lee}},\ and\ \bibinfo
  {author} {\bibfnamefont {M.~D.}\ \bibnamefont {Stiles}},\ }\bibfield  {title}
  {\bibinfo {title} {Chirality from interfacial spin-orbit coupling effects in
  magnetic bilayers},\ }\href@noop {} {\bibfield  {journal} {\bibinfo
  {journal} {Physical review letters}\ }\textbf {\bibinfo {volume} {111}},\
  \bibinfo {pages} {216601} (\bibinfo {year} {2013})}\BibitemShut {NoStop}%
\bibitem [{\citenamefont {Akosa}\ \emph {et~al.}(2019)\citenamefont {Akosa},
  \citenamefont {Li}, \citenamefont {Tatara},\ and\ \citenamefont
  {Tretiakov}}]{akosa2019tuning}%
  \BibitemOpen
  \bibfield  {author} {\bibinfo {author} {\bibfnamefont {C.~A.}\ \bibnamefont
  {Akosa}}, \bibinfo {author} {\bibfnamefont {H.}~\bibnamefont {Li}}, \bibinfo
  {author} {\bibfnamefont {G.}~\bibnamefont {Tatara}},\ and\ \bibinfo {author}
  {\bibfnamefont {O.~A.}\ \bibnamefont {Tretiakov}},\ }\bibfield  {title}
  {\bibinfo {title} {Tuning the skyrmion {H}all effect via engineering of
  spin-orbit interaction},\ }\href@noop {} {\bibfield  {journal} {\bibinfo
  {journal} {Physical Review Applied}\ }\textbf {\bibinfo {volume} {12}},\
  \bibinfo {pages} {054032} (\bibinfo {year} {2019})}\BibitemShut {NoStop}%
\bibitem [{\citenamefont {Roessler}\ \emph {et~al.}(2006)\citenamefont
  {Roessler}, \citenamefont {Bogdanov},\ and\ \citenamefont
  {Pfleiderer}}]{roessler2006spontaneous}%
  \BibitemOpen
  \bibfield  {author} {\bibinfo {author} {\bibfnamefont {U.~K.}\ \bibnamefont
  {Roessler}}, \bibinfo {author} {\bibfnamefont {A.}~\bibnamefont {Bogdanov}},\
  and\ \bibinfo {author} {\bibfnamefont {C.}~\bibnamefont {Pfleiderer}},\
  }\bibfield  {title} {\bibinfo {title} {Spontaneous skyrmion ground states in
  magnetic metals},\ }\href@noop {} {\bibfield  {journal} {\bibinfo  {journal}
  {Nature}\ }\textbf {\bibinfo {volume} {442}},\ \bibinfo {pages} {797}
  (\bibinfo {year} {2006})}\BibitemShut {NoStop}%
\bibitem [{\citenamefont {Fert}\ \emph {et~al.}(2017)\citenamefont {Fert},
  \citenamefont {Reyren},\ and\ \citenamefont {Cros}}]{fert2017magnetic}%
  \BibitemOpen
  \bibfield  {author} {\bibinfo {author} {\bibfnamefont {A.}~\bibnamefont
  {Fert}}, \bibinfo {author} {\bibfnamefont {N.}~\bibnamefont {Reyren}},\ and\
  \bibinfo {author} {\bibfnamefont {V.}~\bibnamefont {Cros}},\ }\bibfield
  {title} {\bibinfo {title} {Magnetic skyrmions: advances in physics and
  potential applications},\ }\href@noop {} {\bibfield  {journal} {\bibinfo
  {journal} {Nature Reviews Materials}\ }\textbf {\bibinfo {volume} {2}},\
  \bibinfo {pages} {1} (\bibinfo {year} {2017})}\BibitemShut {NoStop}%
\bibitem [{\citenamefont {Tokura}\ and\ \citenamefont
  {Kanazawa}(2020)}]{tokura2020magnetic}%
  \BibitemOpen
  \bibfield  {author} {\bibinfo {author} {\bibfnamefont {Y.}~\bibnamefont
  {Tokura}}\ and\ \bibinfo {author} {\bibfnamefont {N.}~\bibnamefont
  {Kanazawa}},\ }\bibfield  {title} {\bibinfo {title} {Magnetic skyrmion
  materials},\ }\href@noop {} {\bibfield  {journal} {\bibinfo  {journal}
  {Chemical Reviews}\ }\textbf {\bibinfo {volume} {121}},\ \bibinfo {pages}
  {2857} (\bibinfo {year} {2020})}\BibitemShut {NoStop}%
\bibitem [{\citenamefont {Shiomi}\ \emph {et~al.}(2013)\citenamefont {Shiomi},
  \citenamefont {Kanazawa}, \citenamefont {Shibata}, \citenamefont {Onose},\
  and\ \citenamefont {Tokura}}]{shiomi2013topological}%
  \BibitemOpen
  \bibfield  {author} {\bibinfo {author} {\bibfnamefont {Y.}~\bibnamefont
  {Shiomi}}, \bibinfo {author} {\bibfnamefont {N.}~\bibnamefont {Kanazawa}},
  \bibinfo {author} {\bibfnamefont {K.}~\bibnamefont {Shibata}}, \bibinfo
  {author} {\bibfnamefont {Y.}~\bibnamefont {Onose}},\ and\ \bibinfo {author}
  {\bibfnamefont {Y.}~\bibnamefont {Tokura}},\ }\bibfield  {title} {\bibinfo
  {title} {Topological {N}ernst effect in a three-dimensional skyrmion-lattice
  phase},\ }\href@noop {} {\bibfield  {journal} {\bibinfo  {journal} {Physical
  Review B}\ }\textbf {\bibinfo {volume} {88}},\ \bibinfo {pages} {064409}
  (\bibinfo {year} {2013})}\BibitemShut {NoStop}%
\bibitem [{\citenamefont {Hirschberger}\ \emph {et~al.}(2020)\citenamefont
  {Hirschberger}, \citenamefont {Spitz}, \citenamefont {Nomoto}, \citenamefont
  {Kurumaji}, \citenamefont {Gao}, \citenamefont {Masell}, \citenamefont
  {Nakajima}, \citenamefont {Kikkawa}, \citenamefont {Yamasaki}, \citenamefont
  {Sagayama} \emph {et~al.}}]{hirschberger2020topological}%
  \BibitemOpen
  \bibfield  {author} {\bibinfo {author} {\bibfnamefont {M.}~\bibnamefont
  {Hirschberger}}, \bibinfo {author} {\bibfnamefont {L.}~\bibnamefont {Spitz}},
  \bibinfo {author} {\bibfnamefont {T.}~\bibnamefont {Nomoto}}, \bibinfo
  {author} {\bibfnamefont {T.}~\bibnamefont {Kurumaji}}, \bibinfo {author}
  {\bibfnamefont {S.}~\bibnamefont {Gao}}, \bibinfo {author} {\bibfnamefont
  {J.}~\bibnamefont {Masell}}, \bibinfo {author} {\bibfnamefont
  {T.}~\bibnamefont {Nakajima}}, \bibinfo {author} {\bibfnamefont
  {A.}~\bibnamefont {Kikkawa}}, \bibinfo {author} {\bibfnamefont
  {Y.}~\bibnamefont {Yamasaki}}, \bibinfo {author} {\bibfnamefont
  {H.}~\bibnamefont {Sagayama}}, \emph {et~al.},\ }\bibfield  {title} {\bibinfo
  {title} {Topological {N}ernst effect of the two-dimensional skyrmion
  lattice},\ }\href@noop {} {\bibfield  {journal} {\bibinfo  {journal}
  {Physical Review Letters}\ }\textbf {\bibinfo {volume} {125}},\ \bibinfo
  {pages} {076602} (\bibinfo {year} {2020})}\BibitemShut {NoStop}%
\bibitem [{\citenamefont {Kolincio}\ \emph {et~al.}(2021)\citenamefont
  {Kolincio}, \citenamefont {Hirschberger}, \citenamefont {Masell},
  \citenamefont {Gao}, \citenamefont {Kikkawa}, \citenamefont {Taguchi},
  \citenamefont {Arima}, \citenamefont {Nagaosa},\ and\ \citenamefont
  {Tokura}}]{kolincio2021large}%
  \BibitemOpen
  \bibfield  {author} {\bibinfo {author} {\bibfnamefont {K.~K.}\ \bibnamefont
  {Kolincio}}, \bibinfo {author} {\bibfnamefont {M.}~\bibnamefont
  {Hirschberger}}, \bibinfo {author} {\bibfnamefont {J.}~\bibnamefont
  {Masell}}, \bibinfo {author} {\bibfnamefont {S.}~\bibnamefont {Gao}},
  \bibinfo {author} {\bibfnamefont {A.}~\bibnamefont {Kikkawa}}, \bibinfo
  {author} {\bibfnamefont {Y.}~\bibnamefont {Taguchi}}, \bibinfo {author}
  {\bibfnamefont {T.-h.}\ \bibnamefont {Arima}}, \bibinfo {author}
  {\bibfnamefont {N.}~\bibnamefont {Nagaosa}},\ and\ \bibinfo {author}
  {\bibfnamefont {Y.}~\bibnamefont {Tokura}},\ }\bibfield  {title} {\bibinfo
  {title} {Large {H}all and {N}ernst responses from thermally induced spin
  chirality in a spin-trimer ferromagnet},\ }\href@noop {} {\bibfield
  {journal} {\bibinfo  {journal} {Proceedings of the National Academy of
  Sciences}\ }\textbf {\bibinfo {volume} {118}},\ \bibinfo {pages}
  {e2023588118} (\bibinfo {year} {2021})}\BibitemShut {NoStop}%
\bibitem [{\citenamefont {Scarioni}\ \emph {et~al.}(2021)\citenamefont
  {Scarioni}, \citenamefont {Barton}, \citenamefont {Corte-Le{\'o}n},
  \citenamefont {Sievers}, \citenamefont {Hu}, \citenamefont {Ajejas},
  \citenamefont {Legrand}, \citenamefont {Reyren}, \citenamefont {Cros},
  \citenamefont {Kazakova} \emph {et~al.}}]{scarioni2021thermoelectric}%
  \BibitemOpen
  \bibfield  {author} {\bibinfo {author} {\bibfnamefont {A.~F.}\ \bibnamefont
  {Scarioni}}, \bibinfo {author} {\bibfnamefont {C.}~\bibnamefont {Barton}},
  \bibinfo {author} {\bibfnamefont {H.}~\bibnamefont {Corte-Le{\'o}n}},
  \bibinfo {author} {\bibfnamefont {S.}~\bibnamefont {Sievers}}, \bibinfo
  {author} {\bibfnamefont {X.}~\bibnamefont {Hu}}, \bibinfo {author}
  {\bibfnamefont {F.}~\bibnamefont {Ajejas}}, \bibinfo {author} {\bibfnamefont
  {W.}~\bibnamefont {Legrand}}, \bibinfo {author} {\bibfnamefont
  {N.}~\bibnamefont {Reyren}}, \bibinfo {author} {\bibfnamefont
  {V.}~\bibnamefont {Cros}}, \bibinfo {author} {\bibfnamefont {O.}~\bibnamefont
  {Kazakova}}, \emph {et~al.},\ }\bibfield  {title} {\bibinfo {title}
  {Thermoelectric signature of individual skyrmions},\ }\href@noop {}
  {\bibfield  {journal} {\bibinfo  {journal} {Physical Review Letters}\
  }\textbf {\bibinfo {volume} {126}},\ \bibinfo {pages} {077202} (\bibinfo
  {year} {2021})}\BibitemShut {NoStop}%
\bibitem [{\citenamefont {Macy}\ \emph {et~al.}(2021)\citenamefont {Macy},
  \citenamefont {Ratkovski}, \citenamefont {Balakrishnan}, \citenamefont
  {Strungaru}, \citenamefont {Chiu}, \citenamefont {Flessa~Savvidou},
  \citenamefont {Moon}, \citenamefont {Zheng}, \citenamefont {Weiland},
  \citenamefont {McCandless} \emph {et~al.}}]{macy2021magnetic}%
  \BibitemOpen
  \bibfield  {author} {\bibinfo {author} {\bibfnamefont {J.}~\bibnamefont
  {Macy}}, \bibinfo {author} {\bibfnamefont {D.}~\bibnamefont {Ratkovski}},
  \bibinfo {author} {\bibfnamefont {P.~P.}\ \bibnamefont {Balakrishnan}},
  \bibinfo {author} {\bibfnamefont {M.}~\bibnamefont {Strungaru}}, \bibinfo
  {author} {\bibfnamefont {Y.-C.}\ \bibnamefont {Chiu}}, \bibinfo {author}
  {\bibfnamefont {A.}~\bibnamefont {Flessa~Savvidou}}, \bibinfo {author}
  {\bibfnamefont {A.}~\bibnamefont {Moon}}, \bibinfo {author} {\bibfnamefont
  {W.}~\bibnamefont {Zheng}}, \bibinfo {author} {\bibfnamefont
  {A.}~\bibnamefont {Weiland}}, \bibinfo {author} {\bibfnamefont {G.~T.}\
  \bibnamefont {McCandless}}, \emph {et~al.},\ }\bibfield  {title} {\bibinfo
  {title} {Magnetic field-induced non-trivial electronic topology in {F}e{$_{3-
  x}$} {G}e{T}e{$_2$}},\ }\href@noop {} {\bibfield  {journal} {\bibinfo
  {journal} {Applied Physics Reviews}\ }\textbf {\bibinfo {volume} {8}},\
  \bibinfo {pages} {041401} (\bibinfo {year} {2021})}\BibitemShut {NoStop}%
\bibitem [{\citenamefont {Zhang}\ \emph {et~al.}(2021)\citenamefont {Zhang},
  \citenamefont {Xu},\ and\ \citenamefont {Ke}}]{zhang2021topological}%
  \BibitemOpen
  \bibfield  {author} {\bibinfo {author} {\bibfnamefont {H.}~\bibnamefont
  {Zhang}}, \bibinfo {author} {\bibfnamefont {C.}~\bibnamefont {Xu}},\ and\
  \bibinfo {author} {\bibfnamefont {X.}~\bibnamefont {Ke}},\ }\bibfield
  {title} {\bibinfo {title} {Topological {N}ernst effect, anomalous {N}ernst
  effect, and anomalous thermal {H}all effect in the {D}irac semimetal
  {F}e{$_3$}{S}n{$_2$}},\ }\href@noop {} {\bibfield  {journal} {\bibinfo
  {journal} {Physical Review B}\ }\textbf {\bibinfo {volume} {103}},\ \bibinfo
  {pages} {L201101} (\bibinfo {year} {2021})}\BibitemShut {NoStop}%
\bibitem [{\citenamefont {Verma}\ \emph {et~al.}(2022)\citenamefont {Verma},
  \citenamefont {Addison},\ and\ \citenamefont {Randeria}}]{verma2022unified}%
  \BibitemOpen
  \bibfield  {author} {\bibinfo {author} {\bibfnamefont {N.}~\bibnamefont
  {Verma}}, \bibinfo {author} {\bibfnamefont {Z.}~\bibnamefont {Addison}},\
  and\ \bibinfo {author} {\bibfnamefont {M.}~\bibnamefont {Randeria}},\
  }\bibfield  {title} {\bibinfo {title} {Unified theory of the anomalous and
  topological {H}all effects with phase-space {B}erry curvatures},\ }\href@noop
  {} {\bibfield  {journal} {\bibinfo  {journal} {Science Advances}\ }\textbf
  {\bibinfo {volume} {8}},\ \bibinfo {pages} {eabq2765} (\bibinfo {year}
  {2022})}\BibitemShut {NoStop}%
\bibitem [{\citenamefont {Addison}\ \emph {et~al.}(2023)\citenamefont
  {Addison}, \citenamefont {Keyes},\ and\ \citenamefont
  {Randeria}}]{addison2023theory}%
  \BibitemOpen
  \bibfield  {author} {\bibinfo {author} {\bibfnamefont {Z.}~\bibnamefont
  {Addison}}, \bibinfo {author} {\bibfnamefont {L.}~\bibnamefont {Keyes}},\
  and\ \bibinfo {author} {\bibfnamefont {M.}~\bibnamefont {Randeria}},\
  }\bibfield  {title} {\bibinfo {title} {Theory of topological nernst and
  thermoelectric transport in chiral magnets},\ }\href
  {https://doi.org/10.1103/PhysRevB.108.014419} {\bibfield  {journal} {\bibinfo
   {journal} {Phys. Rev. B}\ }\textbf {\bibinfo {volume} {108}},\ \bibinfo
  {pages} {014419} (\bibinfo {year} {2023})}\BibitemShut {NoStop}%
\bibitem [{\citenamefont {Luttinger}(1964)}]{luttinger1964theory}%
  \BibitemOpen
  \bibfield  {author} {\bibinfo {author} {\bibfnamefont {J.}~\bibnamefont
  {Luttinger}},\ }\bibfield  {title} {\bibinfo {title} {Theory of thermal
  transport coefficients},\ }\href@noop {} {\bibfield  {journal} {\bibinfo
  {journal} {Physical Review}\ }\textbf {\bibinfo {volume} {135}},\ \bibinfo
  {pages} {A1505} (\bibinfo {year} {1964})}\BibitemShut {NoStop}%
\bibitem [{\citenamefont {Cooper}\ \emph {et~al.}(1997)\citenamefont {Cooper},
  \citenamefont {Halperin},\ and\ \citenamefont
  {Ruzin}}]{cooper1997thermoelectric}%
  \BibitemOpen
  \bibfield  {author} {\bibinfo {author} {\bibfnamefont {N.}~\bibnamefont
  {Cooper}}, \bibinfo {author} {\bibfnamefont {B.}~\bibnamefont {Halperin}},\
  and\ \bibinfo {author} {\bibfnamefont {I.}~\bibnamefont {Ruzin}},\ }\bibfield
   {title} {\bibinfo {title} {Thermoelectric response of an interacting
  two-dimensional electron gas in a quantizing magnetic field},\ }\href@noop {}
  {\bibfield  {journal} {\bibinfo  {journal} {Physical Review B}\ }\textbf
  {\bibinfo {volume} {55}},\ \bibinfo {pages} {2344} (\bibinfo {year}
  {1997})}\BibitemShut {NoStop}%
\bibitem [{\citenamefont {Qin}\ \emph {et~al.}(2011)\citenamefont {Qin},
  \citenamefont {Niu},\ and\ \citenamefont {Shi}}]{qin2011Energy}%
  \BibitemOpen
  \bibfield  {author} {\bibinfo {author} {\bibfnamefont {T.}~\bibnamefont
  {Qin}}, \bibinfo {author} {\bibfnamefont {Q.}~\bibnamefont {Niu}},\ and\
  \bibinfo {author} {\bibfnamefont {J.}~\bibnamefont {Shi}},\ }\bibfield
  {title} {\bibinfo {title} {Energy magnetization and the thermal hall
  effect},\ }\href {https://doi.org/10.1103/PhysRevLett.107.236601} {\bibfield
  {journal} {\bibinfo  {journal} {Phys. Rev. Lett.}\ }\textbf {\bibinfo
  {volume} {107}},\ \bibinfo {pages} {236601} (\bibinfo {year}
  {2011})}\BibitemShut {NoStop}%
\bibitem [{\citenamefont {Onoda}\ \emph
  {et~al.}(2006{\natexlab{b}})\citenamefont {Onoda}, \citenamefont {Sugimoto},\
  and\ \citenamefont {Nagaosa}}]{onoda2006theory}%
  \BibitemOpen
  \bibfield  {author} {\bibinfo {author} {\bibfnamefont {S.}~\bibnamefont
  {Onoda}}, \bibinfo {author} {\bibfnamefont {N.}~\bibnamefont {Sugimoto}},\
  and\ \bibinfo {author} {\bibfnamefont {N.}~\bibnamefont {Nagaosa}},\
  }\bibfield  {title} {\bibinfo {title} {Theory of non-equilibirum states
  driven by constant electromagnetic fields: —non-commutative quantum
  mechanics in the keldysh formalism—},\ }\href@noop {} {\bibfield  {journal}
  {\bibinfo  {journal} {Progress of theoretical physics}\ }\textbf {\bibinfo
  {volume} {116}},\ \bibinfo {pages} {61} (\bibinfo {year}
  {2006}{\natexlab{b}})}\BibitemShut {NoStop}%
\bibitem [{\citenamefont {Sugimoto}\ \emph {et~al.}(2007)\citenamefont
  {Sugimoto}, \citenamefont {Onoda},\ and\ \citenamefont
  {Nagaosa}}]{sugimoto2007gauge}%
  \BibitemOpen
  \bibfield  {author} {\bibinfo {author} {\bibfnamefont {N.}~\bibnamefont
  {Sugimoto}}, \bibinfo {author} {\bibfnamefont {S.}~\bibnamefont {Onoda}},\
  and\ \bibinfo {author} {\bibfnamefont {N.}~\bibnamefont {Nagaosa}},\
  }\bibfield  {title} {\bibinfo {title} {Gauge covariant formulation of the
  wigner representation through deformation quantization: application to
  keldysh formalism with an electromagnetic field},\ }\href@noop {} {\bibfield
  {journal} {\bibinfo  {journal} {Progress of theoretical physics}\ }\textbf
  {\bibinfo {volume} {117}},\ \bibinfo {pages} {415} (\bibinfo {year}
  {2007})}\BibitemShut {NoStop}%
\bibitem [{\citenamefont {Shitade}(2014{\natexlab{a}})}]{shitade2014heat}%
  \BibitemOpen
  \bibfield  {author} {\bibinfo {author} {\bibfnamefont {A.}~\bibnamefont
  {Shitade}},\ }\bibfield  {title} {\bibinfo {title} {Heat transport as
  torsional responses and keldysh formalism in a curved spacetime},\
  }\href@noop {} {\bibfield  {journal} {\bibinfo  {journal} {Progress of
  Theoretical and Experimental Physics}\ }\textbf {\bibinfo {volume} {2014}},\
  \bibinfo {pages} {123I01} (\bibinfo {year} {2014}{\natexlab{a}})}\BibitemShut
  {NoStop}%
\bibitem [{\citenamefont {Shitade}(2014{\natexlab{b}})}]{shitade2014theory}%
  \BibitemOpen
  \bibfield  {author} {\bibinfo {author} {\bibfnamefont {A.}~\bibnamefont
  {Shitade}},\ }\bibfield  {title} {\bibinfo {title} {Theory of charge and heat
  polarizations with the keldysh formalism},\ }\href@noop {} {\bibfield
  {journal} {\bibinfo  {journal} {Journal of the Physical Society of Japan}\
  }\textbf {\bibinfo {volume} {83}},\ \bibinfo {pages} {033708} (\bibinfo
  {year} {2014}{\natexlab{b}})}\BibitemShut {NoStop}%
\bibitem [{\citenamefont {Xiao}\ and\ \citenamefont
  {Niu}(2020)}]{xiao2020unified}%
  \BibitemOpen
  \bibfield  {author} {\bibinfo {author} {\bibfnamefont {C.}~\bibnamefont
  {Xiao}}\ and\ \bibinfo {author} {\bibfnamefont {Q.}~\bibnamefont {Niu}},\
  }\bibfield  {title} {\bibinfo {title} {Unified bulk semiclassical theory for
  intrinsic thermal transport and magnetization currents},\ }\href@noop {}
  {\bibfield  {journal} {\bibinfo  {journal} {Physical Review B}\ }\textbf
  {\bibinfo {volume} {101}},\ \bibinfo {pages} {235430} (\bibinfo {year}
  {2020})}\BibitemShut {NoStop}%
\bibitem [{\citenamefont {Fujishiro}\ \emph {et~al.}(2019)\citenamefont
  {Fujishiro}, \citenamefont {Kanazawa}, \citenamefont {Nakajima},
  \citenamefont {Yu}, \citenamefont {Ohishi}, \citenamefont {Kawamura},
  \citenamefont {Kakurai}, \citenamefont {Arima}, \citenamefont {Mitamura},
  \citenamefont {Miyake} \emph {et~al.}}]{fujishiro2019topological}%
  \BibitemOpen
  \bibfield  {author} {\bibinfo {author} {\bibfnamefont {Y.}~\bibnamefont
  {Fujishiro}}, \bibinfo {author} {\bibfnamefont {N.}~\bibnamefont {Kanazawa}},
  \bibinfo {author} {\bibfnamefont {T.}~\bibnamefont {Nakajima}}, \bibinfo
  {author} {\bibfnamefont {X.}~\bibnamefont {Yu}}, \bibinfo {author}
  {\bibfnamefont {K.}~\bibnamefont {Ohishi}}, \bibinfo {author} {\bibfnamefont
  {Y.}~\bibnamefont {Kawamura}}, \bibinfo {author} {\bibfnamefont
  {K.}~\bibnamefont {Kakurai}}, \bibinfo {author} {\bibfnamefont
  {T.}~\bibnamefont {Arima}}, \bibinfo {author} {\bibfnamefont
  {H.}~\bibnamefont {Mitamura}}, \bibinfo {author} {\bibfnamefont
  {A.}~\bibnamefont {Miyake}}, \emph {et~al.},\ }\bibfield  {title} {\bibinfo
  {title} {Topological transitions among skyrmion-and hedgehog-lattice states
  in cubic chiral magnets},\ }\href@noop {} {\bibfield  {journal} {\bibinfo
  {journal} {Nature communications}\ }\textbf {\bibinfo {volume} {10}},\
  \bibinfo {pages} {1059} (\bibinfo {year} {2019})}\BibitemShut {NoStop}%
\bibitem [{\citenamefont {You}\ \emph {et~al.}(2019)\citenamefont {You},
  \citenamefont {Gong}, \citenamefont {Li}, \citenamefont {Li}, \citenamefont
  {Zhu}, \citenamefont {Tang}, \citenamefont {Liu}, \citenamefont {Yao},
  \citenamefont {Xu}, \citenamefont {Xu} \emph {et~al.}}]{you2019angular}%
  \BibitemOpen
  \bibfield  {author} {\bibinfo {author} {\bibfnamefont {Y.}~\bibnamefont
  {You}}, \bibinfo {author} {\bibfnamefont {Y.}~\bibnamefont {Gong}}, \bibinfo
  {author} {\bibfnamefont {H.}~\bibnamefont {Li}}, \bibinfo {author}
  {\bibfnamefont {Z.}~\bibnamefont {Li}}, \bibinfo {author} {\bibfnamefont
  {M.}~\bibnamefont {Zhu}}, \bibinfo {author} {\bibfnamefont {J.}~\bibnamefont
  {Tang}}, \bibinfo {author} {\bibfnamefont {E.}~\bibnamefont {Liu}}, \bibinfo
  {author} {\bibfnamefont {Y.}~\bibnamefont {Yao}}, \bibinfo {author}
  {\bibfnamefont {G.}~\bibnamefont {Xu}}, \bibinfo {author} {\bibfnamefont
  {F.}~\bibnamefont {Xu}}, \emph {et~al.},\ }\bibfield  {title} {\bibinfo
  {title} {Angular dependence of the topological hall effect in the uniaxial
  van der waals ferromagnet fe 3 gete 2},\ }\href@noop {} {\bibfield  {journal}
  {\bibinfo  {journal} {Physical Review B}\ }\textbf {\bibinfo {volume}
  {100}},\ \bibinfo {pages} {134441} (\bibinfo {year} {2019})}\BibitemShut
  {NoStop}%
\bibitem [{\citenamefont {Wang}\ \emph {et~al.}(2019)\citenamefont {Wang},
  \citenamefont {Yan}, \citenamefont {Li}, \citenamefont {Wang}, \citenamefont
  {Song}, \citenamefont {Song}, \citenamefont {Li}, \citenamefont {Chen},
  \citenamefont {Qin}, \citenamefont {Ling} \emph {et~al.}}]{wang2019magnetic}%
  \BibitemOpen
  \bibfield  {author} {\bibinfo {author} {\bibfnamefont {Y.}~\bibnamefont
  {Wang}}, \bibinfo {author} {\bibfnamefont {J.}~\bibnamefont {Yan}}, \bibinfo
  {author} {\bibfnamefont {J.}~\bibnamefont {Li}}, \bibinfo {author}
  {\bibfnamefont {S.}~\bibnamefont {Wang}}, \bibinfo {author} {\bibfnamefont
  {M.}~\bibnamefont {Song}}, \bibinfo {author} {\bibfnamefont {J.}~\bibnamefont
  {Song}}, \bibinfo {author} {\bibfnamefont {Z.}~\bibnamefont {Li}}, \bibinfo
  {author} {\bibfnamefont {K.}~\bibnamefont {Chen}}, \bibinfo {author}
  {\bibfnamefont {Y.}~\bibnamefont {Qin}}, \bibinfo {author} {\bibfnamefont
  {L.}~\bibnamefont {Ling}}, \emph {et~al.},\ }\bibfield  {title} {\bibinfo
  {title} {Magnetic anisotropy and topological hall effect in the trigonal
  chromium tellurides cr 5 te 8},\ }\href@noop {} {\bibfield  {journal}
  {\bibinfo  {journal} {Physical Review B}\ }\textbf {\bibinfo {volume}
  {100}},\ \bibinfo {pages} {024434} (\bibinfo {year} {2019})}\BibitemShut
  {NoStop}%
\bibitem [{\citenamefont {Ge}\ \emph {et~al.}(2020)\citenamefont {Ge},
  \citenamefont {Ma}, \citenamefont {Liu}, \citenamefont {Wang}, \citenamefont
  {Li}, \citenamefont {Luo}, \citenamefont {Luo}, \citenamefont {Xing},
  \citenamefont {Yan}, \citenamefont {Mandrus} \emph
  {et~al.}}]{ge2020unconventional}%
  \BibitemOpen
  \bibfield  {author} {\bibinfo {author} {\bibfnamefont {J.}~\bibnamefont
  {Ge}}, \bibinfo {author} {\bibfnamefont {D.}~\bibnamefont {Ma}}, \bibinfo
  {author} {\bibfnamefont {Y.}~\bibnamefont {Liu}}, \bibinfo {author}
  {\bibfnamefont {H.}~\bibnamefont {Wang}}, \bibinfo {author} {\bibfnamefont
  {Y.}~\bibnamefont {Li}}, \bibinfo {author} {\bibfnamefont {J.}~\bibnamefont
  {Luo}}, \bibinfo {author} {\bibfnamefont {T.}~\bibnamefont {Luo}}, \bibinfo
  {author} {\bibfnamefont {Y.}~\bibnamefont {Xing}}, \bibinfo {author}
  {\bibfnamefont {J.}~\bibnamefont {Yan}}, \bibinfo {author} {\bibfnamefont
  {D.}~\bibnamefont {Mandrus}}, \emph {et~al.},\ }\bibfield  {title} {\bibinfo
  {title} {Unconventional hall effect induced by berry curvature},\ }\href@noop
  {} {\bibfield  {journal} {\bibinfo  {journal} {National science review}\
  }\textbf {\bibinfo {volume} {7}},\ \bibinfo {pages} {1879} (\bibinfo {year}
  {2020})}\BibitemShut {NoStop}%
\bibitem [{\citenamefont {Tan}\ \emph {et~al.}(2021)\citenamefont {Tan},
  \citenamefont {Liu},\ and\ \citenamefont {Yan}}]{tan2021unconventional}%
  \BibitemOpen
  \bibfield  {author} {\bibinfo {author} {\bibfnamefont {H.}~\bibnamefont
  {Tan}}, \bibinfo {author} {\bibfnamefont {Y.}~\bibnamefont {Liu}},\ and\
  \bibinfo {author} {\bibfnamefont {B.}~\bibnamefont {Yan}},\ }\bibfield
  {title} {\bibinfo {title} {Unconventional anomalous hall effect from
  magnetization parallel to the electric field},\ }\href@noop {} {\bibfield
  {journal} {\bibinfo  {journal} {Physical Review B}\ }\textbf {\bibinfo
  {volume} {103}},\ \bibinfo {pages} {214438} (\bibinfo {year}
  {2021})}\BibitemShut {NoStop}%
\bibitem [{\citenamefont {Zhou}\ \emph {et~al.}(2022)\citenamefont {Zhou},
  \citenamefont {Zhang}, \citenamefont {Lin}, \citenamefont {Cao},
  \citenamefont {Zhou}, \citenamefont {Jiang}, \citenamefont {Du},
  \citenamefont {Tang}, \citenamefont {Shi}, \citenamefont {Jiang} \emph
  {et~al.}}]{zhou2022heterodimensional}%
  \BibitemOpen
  \bibfield  {author} {\bibinfo {author} {\bibfnamefont {J.}~\bibnamefont
  {Zhou}}, \bibinfo {author} {\bibfnamefont {W.}~\bibnamefont {Zhang}},
  \bibinfo {author} {\bibfnamefont {Y.-C.}\ \bibnamefont {Lin}}, \bibinfo
  {author} {\bibfnamefont {J.}~\bibnamefont {Cao}}, \bibinfo {author}
  {\bibfnamefont {Y.}~\bibnamefont {Zhou}}, \bibinfo {author} {\bibfnamefont
  {W.}~\bibnamefont {Jiang}}, \bibinfo {author} {\bibfnamefont
  {H.}~\bibnamefont {Du}}, \bibinfo {author} {\bibfnamefont {B.}~\bibnamefont
  {Tang}}, \bibinfo {author} {\bibfnamefont {J.}~\bibnamefont {Shi}}, \bibinfo
  {author} {\bibfnamefont {B.}~\bibnamefont {Jiang}}, \emph {et~al.},\
  }\bibfield  {title} {\bibinfo {title} {Heterodimensional superlattice with
  in-plane anomalous hall effect},\ }\href@noop {} {\bibfield  {journal}
  {\bibinfo  {journal} {Nature}\ }\textbf {\bibinfo {volume} {609}},\ \bibinfo
  {pages} {46} (\bibinfo {year} {2022})}\BibitemShut {NoStop}%
\bibitem [{\citenamefont {Chen}\ \emph {et~al.}(2023)\citenamefont {Chen},
  \citenamefont {Zhu}, \citenamefont {Lin}, \citenamefont {Niu}, \citenamefont
  {Liu}, \citenamefont {Zhuang}, \citenamefont {Zhang}, \citenamefont {Liang},
  \citenamefont {Sun}, \citenamefont {Chen} \emph
  {et~al.}}]{chen2023observation}%
  \BibitemOpen
  \bibfield  {author} {\bibinfo {author} {\bibfnamefont {Y.}~\bibnamefont
  {Chen}}, \bibinfo {author} {\bibfnamefont {Y.}~\bibnamefont {Zhu}}, \bibinfo
  {author} {\bibfnamefont {R.}~\bibnamefont {Lin}}, \bibinfo {author}
  {\bibfnamefont {W.}~\bibnamefont {Niu}}, \bibinfo {author} {\bibfnamefont
  {R.}~\bibnamefont {Liu}}, \bibinfo {author} {\bibfnamefont {W.}~\bibnamefont
  {Zhuang}}, \bibinfo {author} {\bibfnamefont {X.}~\bibnamefont {Zhang}},
  \bibinfo {author} {\bibfnamefont {J.}~\bibnamefont {Liang}}, \bibinfo
  {author} {\bibfnamefont {W.}~\bibnamefont {Sun}}, \bibinfo {author}
  {\bibfnamefont {Z.}~\bibnamefont {Chen}}, \emph {et~al.},\ }\bibfield
  {title} {\bibinfo {title} {Observation of colossal topological hall effect in
  noncoplanar ferromagnet cr5te6 thin films},\ }\href@noop {} {\bibfield
  {journal} {\bibinfo  {journal} {Advanced Functional Materials}\ }\textbf
  {\bibinfo {volume} {33}},\ \bibinfo {pages} {2302984} (\bibinfo {year}
  {2023})}\BibitemShut {NoStop}%
\bibitem [{\citenamefont {Cao}\ \emph {et~al.}(2023)\citenamefont {Cao},
  \citenamefont {Jiang}, \citenamefont {Li}, \citenamefont {Tu}, \citenamefont
  {Zhou}, \citenamefont {Zhou},\ and\ \citenamefont {Yao}}]{cao2023plane}%
  \BibitemOpen
  \bibfield  {author} {\bibinfo {author} {\bibfnamefont {J.}~\bibnamefont
  {Cao}}, \bibinfo {author} {\bibfnamefont {W.}~\bibnamefont {Jiang}}, \bibinfo
  {author} {\bibfnamefont {X.-P.}\ \bibnamefont {Li}}, \bibinfo {author}
  {\bibfnamefont {D.}~\bibnamefont {Tu}}, \bibinfo {author} {\bibfnamefont
  {J.}~\bibnamefont {Zhou}}, \bibinfo {author} {\bibfnamefont {J.}~\bibnamefont
  {Zhou}},\ and\ \bibinfo {author} {\bibfnamefont {Y.}~\bibnamefont {Yao}},\
  }\bibfield  {title} {\bibinfo {title} {In-plane anomalous hall effect in
  pt-symmetric antiferromagnetic materials},\ }\href@noop {} {\bibfield
  {journal} {\bibinfo  {journal} {Physical Review Letters}\ }\textbf {\bibinfo
  {volume} {130}},\ \bibinfo {pages} {166702} (\bibinfo {year}
  {2023})}\BibitemShut {NoStop}%
\bibitem [{\citenamefont {Sundaram}\ and\ \citenamefont
  {Niu}(1999)}]{sundaram1999wave}%
  \BibitemOpen
  \bibfield  {author} {\bibinfo {author} {\bibfnamefont {G.}~\bibnamefont
  {Sundaram}}\ and\ \bibinfo {author} {\bibfnamefont {Q.}~\bibnamefont {Niu}},\
  }\bibfield  {title} {\bibinfo {title} {Wave-packet dynamics in slowly
  perturbed crystals: Gradient corrections and {B}erry-phase effects},\
  }\href@noop {} {\bibfield  {journal} {\bibinfo  {journal} {Physical Review
  B}\ }\textbf {\bibinfo {volume} {59}},\ \bibinfo {pages} {14915} (\bibinfo
  {year} {1999})}\BibitemShut {NoStop}%
\bibitem [{\citenamefont {Xiao}\ \emph {et~al.}(2005)\citenamefont {Xiao},
  \citenamefont {Shi},\ and\ \citenamefont {Niu}}]{xiao2005berry}%
  \BibitemOpen
  \bibfield  {author} {\bibinfo {author} {\bibfnamefont {D.}~\bibnamefont
  {Xiao}}, \bibinfo {author} {\bibfnamefont {J.}~\bibnamefont {Shi}},\ and\
  \bibinfo {author} {\bibfnamefont {Q.}~\bibnamefont {Niu}},\ }\bibfield
  {title} {\bibinfo {title} {{B}erry phase correction to electron density of
  states in solids},\ }\href@noop {} {\bibfield  {journal} {\bibinfo  {journal}
  {Physical review letters}\ }\textbf {\bibinfo {volume} {95}},\ \bibinfo
  {pages} {137204} (\bibinfo {year} {2005})}\BibitemShut {NoStop}%
\bibitem [{\citenamefont {Nakahara}(2003)}]{nakahara2003}%
  \BibitemOpen
  \bibfield  {author} {\bibinfo {author} {\bibfnamefont {M.}~\bibnamefont
  {Nakahara}},\ }\href@noop {} {\emph {\bibinfo {title} {Geometry, Topology and
  Physics}}}\ (\bibinfo  {publisher} {Taylor and Francis},\ \bibinfo {year}
  {2003})\BibitemShut {NoStop}%
\bibitem [{\citenamefont {Gromov}\ and\ \citenamefont
  {Abanov}(2015)}]{gromov2015thermal}%
  \BibitemOpen
  \bibfield  {author} {\bibinfo {author} {\bibfnamefont {A.}~\bibnamefont
  {Gromov}}\ and\ \bibinfo {author} {\bibfnamefont {A.~G.}\ \bibnamefont
  {Abanov}},\ }\bibfield  {title} {\bibinfo {title} {Thermal hall effect and
  geometry with torsion},\ }\href
  {https://doi.org/10.1103/PhysRevLett.114.016802} {\bibfield  {journal}
  {\bibinfo  {journal} {Phys. Rev. Lett.}\ }\textbf {\bibinfo {volume} {114}},\
  \bibinfo {pages} {016802} (\bibinfo {year} {2015})}\BibitemShut {NoStop}%
\bibitem [{\citenamefont {Zhang}\ \emph
  {et~al.}(2020{\natexlab{a}})\citenamefont {Zhang}, \citenamefont {Gao},\ and\
  \citenamefont {Xiao}}]{zhang2020Thermodynamics}%
  \BibitemOpen
  \bibfield  {author} {\bibinfo {author} {\bibfnamefont {Y.}~\bibnamefont
  {Zhang}}, \bibinfo {author} {\bibfnamefont {Y.}~\bibnamefont {Gao}},\ and\
  \bibinfo {author} {\bibfnamefont {D.}~\bibnamefont {Xiao}},\ }\bibfield
  {title} {\bibinfo {title} {Thermodynamics of energy magnetization},\ }\href
  {https://doi.org/10.1103/PhysRevB.102.235161} {\bibfield  {journal} {\bibinfo
   {journal} {Phys. Rev. B}\ }\textbf {\bibinfo {volume} {102}},\ \bibinfo
  {pages} {235161} (\bibinfo {year} {2020}{\natexlab{a}})}\BibitemShut
  {NoStop}%
\bibitem [{\citenamefont {Sondheimer}(1948)}]{sondheimer1948theory}%
  \BibitemOpen
  \bibfield  {author} {\bibinfo {author} {\bibfnamefont {E.}~\bibnamefont
  {Sondheimer}},\ }\bibfield  {title} {\bibinfo {title} {The theory of the
  galvanomagnetic and thermomagnetic effects in metals},\ }\href@noop {}
  {\bibfield  {journal} {\bibinfo  {journal} {Proceedings of the Royal Society
  of London. Series A. Mathematical and Physical Sciences}\ }\textbf {\bibinfo
  {volume} {193}},\ \bibinfo {pages} {484} (\bibinfo {year}
  {1948})}\BibitemShut {NoStop}%
\bibitem [{\citenamefont {Cui}\ \emph {et~al.}(2024)\citenamefont {Cui},
  \citenamefont {Li}, \citenamefont {Chen}, \citenamefont {Wu}, \citenamefont
  {Chen}, \citenamefont {Pei}, \citenamefont {Wu}, \citenamefont {Xie},
  \citenamefont {Che}, \citenamefont {Qiu} \emph
  {et~al.}}]{cui2024antisymmetric}%
  \BibitemOpen
  \bibfield  {author} {\bibinfo {author} {\bibfnamefont {Y.}~\bibnamefont
  {Cui}}, \bibinfo {author} {\bibfnamefont {Z.}~\bibnamefont {Li}}, \bibinfo
  {author} {\bibfnamefont {H.}~\bibnamefont {Chen}}, \bibinfo {author}
  {\bibfnamefont {Y.}~\bibnamefont {Wu}}, \bibinfo {author} {\bibfnamefont
  {Y.}~\bibnamefont {Chen}}, \bibinfo {author} {\bibfnamefont {K.}~\bibnamefont
  {Pei}}, \bibinfo {author} {\bibfnamefont {T.}~\bibnamefont {Wu}}, \bibinfo
  {author} {\bibfnamefont {N.}~\bibnamefont {Xie}}, \bibinfo {author}
  {\bibfnamefont {R.}~\bibnamefont {Che}}, \bibinfo {author} {\bibfnamefont
  {X.}~\bibnamefont {Qiu}}, \emph {et~al.},\ }\bibfield  {title} {\bibinfo
  {title} {Antisymmetric planar hall effect in rutile oxide films induced by
  the lorentz force},\ }\href@noop {} {\bibfield  {journal} {\bibinfo
  {journal} {Science Bulletin}\ } (\bibinfo {year} {2024})}\BibitemShut
  {NoStop}%
\bibitem [{\citenamefont {Yokouchi}\ \emph {et~al.}(2015)\citenamefont
  {Yokouchi}, \citenamefont {Kanazawa}, \citenamefont {Tsukazaki},
  \citenamefont {Kozuka}, \citenamefont {Kikkawa}, \citenamefont {Taguchi},
  \citenamefont {Kawasaki}, \citenamefont {Ichikawa}, \citenamefont {Kagawa},\
  and\ \citenamefont {Tokura}}]{yokouchi2015formation}%
  \BibitemOpen
  \bibfield  {author} {\bibinfo {author} {\bibfnamefont {T.}~\bibnamefont
  {Yokouchi}}, \bibinfo {author} {\bibfnamefont {N.}~\bibnamefont {Kanazawa}},
  \bibinfo {author} {\bibfnamefont {A.}~\bibnamefont {Tsukazaki}}, \bibinfo
  {author} {\bibfnamefont {Y.}~\bibnamefont {Kozuka}}, \bibinfo {author}
  {\bibfnamefont {A.}~\bibnamefont {Kikkawa}}, \bibinfo {author} {\bibfnamefont
  {Y.}~\bibnamefont {Taguchi}}, \bibinfo {author} {\bibfnamefont
  {M.}~\bibnamefont {Kawasaki}}, \bibinfo {author} {\bibfnamefont
  {M.}~\bibnamefont {Ichikawa}}, \bibinfo {author} {\bibfnamefont
  {F.}~\bibnamefont {Kagawa}},\ and\ \bibinfo {author} {\bibfnamefont
  {Y.}~\bibnamefont {Tokura}},\ }\bibfield  {title} {\bibinfo {title}
  {Formation of in-plane skyrmions in epitaxial mnsi thin films as revealed by
  planar hall effect},\ }\href@noop {} {\bibfield  {journal} {\bibinfo
  {journal} {Journal of the Physical Society of Japan}\ }\textbf {\bibinfo
  {volume} {84}},\ \bibinfo {pages} {104708} (\bibinfo {year}
  {2015})}\BibitemShut {NoStop}%
\bibitem [{\citenamefont {Kanazawa}\ \emph {et~al.}(2017)\citenamefont
  {Kanazawa}, \citenamefont {Seki},\ and\ \citenamefont
  {Tokura}}]{kanazawa2017noncentrosymmetric}%
  \BibitemOpen
  \bibfield  {author} {\bibinfo {author} {\bibfnamefont {N.}~\bibnamefont
  {Kanazawa}}, \bibinfo {author} {\bibfnamefont {S.}~\bibnamefont {Seki}},\
  and\ \bibinfo {author} {\bibfnamefont {Y.}~\bibnamefont {Tokura}},\
  }\bibfield  {title} {\bibinfo {title} {Noncentrosymmetric magnets hosting
  magnetic skyrmions},\ }\href@noop {} {\bibfield  {journal} {\bibinfo
  {journal} {Advanced Materials}\ }\textbf {\bibinfo {volume} {29}},\ \bibinfo
  {pages} {1603227} (\bibinfo {year} {2017})}\BibitemShut {NoStop}%
\bibitem [{\citenamefont {Wolf}\ \emph {et~al.}(2022)\citenamefont {Wolf},
  \citenamefont {Schneider}, \citenamefont {R{\"o}{\ss}ler}, \citenamefont
  {Kov{\'a}cs}, \citenamefont {Schmidt}, \citenamefont {Dunin-Borkowski},
  \citenamefont {B{\"u}chner}, \citenamefont {Rellinghaus},\ and\ \citenamefont
  {Lubk}}]{wolf2022unveiling}%
  \BibitemOpen
  \bibfield  {author} {\bibinfo {author} {\bibfnamefont {D.}~\bibnamefont
  {Wolf}}, \bibinfo {author} {\bibfnamefont {S.}~\bibnamefont {Schneider}},
  \bibinfo {author} {\bibfnamefont {U.~K.}\ \bibnamefont {R{\"o}{\ss}ler}},
  \bibinfo {author} {\bibfnamefont {A.}~\bibnamefont {Kov{\'a}cs}}, \bibinfo
  {author} {\bibfnamefont {M.}~\bibnamefont {Schmidt}}, \bibinfo {author}
  {\bibfnamefont {R.~E.}\ \bibnamefont {Dunin-Borkowski}}, \bibinfo {author}
  {\bibfnamefont {B.}~\bibnamefont {B{\"u}chner}}, \bibinfo {author}
  {\bibfnamefont {B.}~\bibnamefont {Rellinghaus}},\ and\ \bibinfo {author}
  {\bibfnamefont {A.}~\bibnamefont {Lubk}},\ }\bibfield  {title} {\bibinfo
  {title} {Unveiling the three-dimensional magnetic texture of skyrmion
  tubes},\ }\href@noop {} {\bibfield  {journal} {\bibinfo  {journal} {Nature
  nanotechnology}\ }\textbf {\bibinfo {volume} {17}},\ \bibinfo {pages} {250}
  (\bibinfo {year} {2022})}\BibitemShut {NoStop}%
\bibitem [{\citenamefont {Kurumaji}(2023)}]{kurumaji2023symmetry}%
  \BibitemOpen
  \bibfield  {author} {\bibinfo {author} {\bibfnamefont {T.}~\bibnamefont
  {Kurumaji}},\ }\bibfield  {title} {\bibinfo {title} {Symmetry-based
  requirement for the measurement of electrical and thermal hall conductivity
  under an in-plane magnetic field},\ }\href
  {https://doi.org/10.1103/PhysRevResearch.5.023138} {\bibfield  {journal}
  {\bibinfo  {journal} {Phys. Rev. Res.}\ }\textbf {\bibinfo {volume} {5}},\
  \bibinfo {pages} {023138} (\bibinfo {year} {2023})}\BibitemShut {NoStop}%
\bibitem [{\citenamefont {Keyes}(2024)}]{keyesThesis}%
  \BibitemOpen
  \bibfield  {author} {\bibinfo {author} {\bibfnamefont {L.}~\bibnamefont
  {Keyes}},\ }\emph {\bibinfo {title} {Topics in Condensed Matter Theory: Berry
  Curvature Effects in Transport and Numerical Analytic Continuation}},\
  \href@noop {} {Ph.D. thesis},\ \bibinfo  {school} {The Ohio State University}
  (\bibinfo {year} {2024})\BibitemShut {NoStop}%
\bibitem [{\citenamefont {Keyes}\ \emph {et~al.}()\citenamefont {Keyes},
  \citenamefont {Addison},\ and\ \citenamefont {Randeria}}]{keyes2025IPHE}%
  \BibitemOpen
  \bibfield  {author} {\bibinfo {author} {\bibfnamefont {L.}~\bibnamefont
  {Keyes}}, \bibinfo {author} {\bibfnamefont {Z.}~\bibnamefont {Addison}},\
  and\ \bibinfo {author} {\bibfnamefont {M.}~\bibnamefont {Randeria}},\
  }\bibfield  {title} {\bibinfo {title} {Topological and anomalous in-plane
  hall effect},\ }\href@noop {} {\bibinfo  {journal} {in preparation}\
  }\BibitemShut {NoStop}%
\bibitem [{\citenamefont {Ashcroft}\ and\ \citenamefont
  {Mermin}(1976)}]{ashcroft1978solid}%
  \BibitemOpen
\bibfield  {journal} {  }\bibfield  {author} {\bibinfo {author} {\bibfnamefont
  {N.~W.}\ \bibnamefont {Ashcroft}}\ and\ \bibinfo {author} {\bibfnamefont
  {N.~D.}\ \bibnamefont {Mermin}},\ }\href@noop {} {\emph {\bibinfo {title}
  {Solid state physics}}}\ (\bibinfo  {publisher} {Harcourt},\ \bibinfo {year}
  {1976})\BibitemShut {NoStop}%
\bibitem [{\citenamefont {Haldane}(1988)}]{haldane1988model}%
  \BibitemOpen
  \bibfield  {author} {\bibinfo {author} {\bibfnamefont {F.~D.~M.}\
  \bibnamefont {Haldane}},\ }\bibfield  {title} {\bibinfo {title} {Model for a
  quantum hall effect without landau levels: Condensed-matter realization of
  the" parity anomaly"},\ }\href@noop {} {\bibfield  {journal} {\bibinfo
  {journal} {Physical review letters}\ }\textbf {\bibinfo {volume} {61}},\
  \bibinfo {pages} {2015} (\bibinfo {year} {1988})}\BibitemShut {NoStop}%
\bibitem [{\citenamefont {Lux}\ \emph {et~al.}(2020)\citenamefont {Lux},
  \citenamefont {Freimuth}, \citenamefont {Bl{\"u}gel},\ and\ \citenamefont
  {Mokrousov}}]{lux2020chiral}%
  \BibitemOpen
  \bibfield  {author} {\bibinfo {author} {\bibfnamefont {F.~R.}\ \bibnamefont
  {Lux}}, \bibinfo {author} {\bibfnamefont {F.}~\bibnamefont {Freimuth}},
  \bibinfo {author} {\bibfnamefont {S.}~\bibnamefont {Bl{\"u}gel}},\ and\
  \bibinfo {author} {\bibfnamefont {Y.}~\bibnamefont {Mokrousov}},\ }\bibfield
  {title} {\bibinfo {title} {Chiral hall effect in noncollinear magnets from a
  cyclic cohomology approach},\ }\href@noop {} {\bibfield  {journal} {\bibinfo
  {journal} {Physical review letters}\ }\textbf {\bibinfo {volume} {124}},\
  \bibinfo {pages} {096602} (\bibinfo {year} {2020})}\BibitemShut {NoStop}%
\bibitem [{\citenamefont {Bouaziz}\ \emph {et~al.}(2021)\citenamefont
  {Bouaziz}, \citenamefont {Ishida}, \citenamefont {Lounis},\ and\
  \citenamefont {Bl{\"u}gel}}]{bouaziz2021transverse}%
  \BibitemOpen
  \bibfield  {author} {\bibinfo {author} {\bibfnamefont {J.}~\bibnamefont
  {Bouaziz}}, \bibinfo {author} {\bibfnamefont {H.}~\bibnamefont {Ishida}},
  \bibinfo {author} {\bibfnamefont {S.}~\bibnamefont {Lounis}},\ and\ \bibinfo
  {author} {\bibfnamefont {S.}~\bibnamefont {Bl{\"u}gel}},\ }\bibfield  {title}
  {\bibinfo {title} {Transverse transport in two-dimensional relativistic
  systems with nontrivial spin textures},\ }\href@noop {} {\bibfield  {journal}
  {\bibinfo  {journal} {Physical review letters}\ }\textbf {\bibinfo {volume}
  {126}},\ \bibinfo {pages} {147203} (\bibinfo {year} {2021})}\BibitemShut
  {NoStop}%
\bibitem [{\citenamefont {Zhang}\ \emph
  {et~al.}(2020{\natexlab{b}})\citenamefont {Zhang}, \citenamefont {Ishizuka},
  \citenamefont {Zhang}, \citenamefont {Hal\'asz},\ and\ \citenamefont
  {Batista}}]{batista2020}%
  \BibitemOpen
  \bibfield  {author} {\bibinfo {author} {\bibfnamefont {S.-S.}\ \bibnamefont
  {Zhang}}, \bibinfo {author} {\bibfnamefont {H.}~\bibnamefont {Ishizuka}},
  \bibinfo {author} {\bibfnamefont {H.}~\bibnamefont {Zhang}}, \bibinfo
  {author} {\bibfnamefont {G.~B.}\ \bibnamefont {Hal\'asz}},\ and\ \bibinfo
  {author} {\bibfnamefont {C.~D.}\ \bibnamefont {Batista}},\ }\bibfield
  {title} {\bibinfo {title} {Real-space berry curvature of itinerant electron
  systems with spin-orbit interaction},\ }\href
  {https://doi.org/10.1103/PhysRevB.101.024420} {\bibfield  {journal} {\bibinfo
   {journal} {Phys. Rev. B}\ }\textbf {\bibinfo {volume} {101}},\ \bibinfo
  {pages} {024420} (\bibinfo {year} {2020}{\natexlab{b}})}\BibitemShut
  {NoStop}%
\bibitem [{\citenamefont {Denisov}\ \emph {et~al.}(2016)\citenamefont
  {Denisov}, \citenamefont {Rozhansky}, \citenamefont {Averkiev},\ and\
  \citenamefont {L\"ahderanta}}]{denisov2016}%
  \BibitemOpen
  \bibfield  {author} {\bibinfo {author} {\bibfnamefont {K.~S.}\ \bibnamefont
  {Denisov}}, \bibinfo {author} {\bibfnamefont {I.~V.}\ \bibnamefont
  {Rozhansky}}, \bibinfo {author} {\bibfnamefont {N.~S.}\ \bibnamefont
  {Averkiev}},\ and\ \bibinfo {author} {\bibfnamefont {E.}~\bibnamefont
  {L\"ahderanta}},\ }\bibfield  {title} {\bibinfo {title} {Electron scattering
  on a magnetic skyrmion in the nonadiabatic approximation},\ }\href
  {https://doi.org/10.1103/PhysRevLett.117.027202} {\bibfield  {journal}
  {\bibinfo  {journal} {Phys. Rev. Lett.}\ }\textbf {\bibinfo {volume} {117}},\
  \bibinfo {pages} {027202} (\bibinfo {year} {2016})}\BibitemShut {NoStop}%
\bibitem [{\citenamefont {Addison}(2025)}]{DataRep}%
  \BibitemOpen
  \bibfield  {author} {\bibinfo {author} {\bibfnamefont {Z.}~\bibnamefont
  {Addison}},\ }\href {https://doi.org/10.7910/DVN/ENJYJP} {\bibinfo {title}
  {{Anomalous Hall Effect on Square Lattice}}} (\bibinfo {year}
  {2025})\BibitemShut {NoStop}%
\end{thebibliography}%

\clearpage 

\onecolumngrid

\appendix

\section{Semiclassical Equations of Motion}\label{app eom}

We briefly describe the semiclassical framework. A detailed description of this procedure can be found in Appendix A of Ref. \cite{addison2023theory} and in Refs. \cite{sundaram1999wave, xiao2010berry}. The standard semiclassical procedure starts by expanding the Hamiltonian \eqref{H full} in powers of $(\hat{\bm{r}}-\bm{r}_c)$ with $r_c$ being some arbitrary real space point. Because the magnetic texture varies on the scale of $L_s$, spatial derivatives scale like $\bm{\nabla}_{\bm{r}} \sim 1/L_s$. The other length scale in the system is the lattice spacing $a$ which we take to satisfy $a/L_s \ll1$. For computing transport coefficients it is sufficient to keep just the leading order correction, so that
$\widehat{H} \approx \widehat{H}_c + \Delta\widehat{H}$, where $\widehat{H}_c(\bm{r}_c) = \widehat{H}(\hat{\bm{r}}\rightarrow \bm{r}_c)$, and 
\begin{equation}
    \Delta\widehat{H}(\bm{r}_c)=-J\sum_{i,\sigma\sigma'} \hat{c}^{\dagger}_{i\sigma}\bigg((\bm{r}_i-\bm{r}_c)\cdot\bm{\nabla}_{r_c} \hat{\bm{m}}(\bm{r}_c)\cdot\bm{\sigma}^{\sigma\sigma'}\bigg)\hat{c}^{\phantom{\dagger}}_{i\sigma'}.
\end{equation}

The contribution to the Hamiltonian $\widehat{H}_{c}$ has discrete translational symmetry and thus is able to be written in the Bloch basis with states labeled by crystal momentum $\bm{q}$. The associated Bloch Hamiltonian is $\widehat{H}_{sc}(\bm{r}_c, \bm{q})=\varepsilon(\bm{q})\mathbb{1}+\bm{d}(\bm{r}_c, \bm{q})\cdot \bm{\sigma}$, where $\bm{d}(\bm{r}_c, \bm{q})$ is given by Eq. \eqref{bloch d}. The Hamiltonian $\widehat{H}_{c}$ has eigenstates $\ket{\psi_{\pm}(\bm{r}_c, \bm{q})}$, whose lattice periodic part we denote as
$\ket{u_{\pm}(\bm{r_c}, \bm{q})}$, and with energy eigenvalues $\mathcal{E}_{\pm}(\bm{r}_c, \bm{q})=\varepsilon(\bm{q})\pm |\bm{d}(\bm{r_c}, \bm{q})|$, where $\pm$ labels the two bands.  

Wavepackets $\ket{W_{\pm}(\bm{r}_c, \bm{q}_c)}=\int d\bm{q}\ \gamma_{\pm}(\bm{r}_c, \bm{q}, t)\ket{\psi_{\pm}(\bm{r}_c, \bm{q})}$ are constructed from the Bloch states $\ket{\psi_{\pm}(\bm{r}_c, \bm{q})}$ with envelope functions $\gamma_{\pm}(\bm{r}_c, \bm{q})$ such that the wavepackets are centered at $(\bm{r}_c, \bm{q}_c)$. In the semiclassical approach, 
the band index is a good quantum number and there are no interband transitions. To incorporate electromagnetic perturbations, the Hamiltonian in eq.~\eqref{H full} is modified by addition of a scalar potential $-e\phi(\hat{\bm{r}}, t)$ and incorporation of a gauge invariant momentum, $\bm{k}_c \equiv \bm{q}_c + e\bm{A}$, where $e$ is the positive electron charge and $\bm{A}$ is the vector potential. The wavepacket energies with the scalar potential are to first order 
\begin{equation}
    \bra{W_{\pm}}\widehat{H}\ket{W_{\pm}} \approx \widetilde{\mathcal{E}}_{\pm}(\bm{r}_c, \bm{k}_c) - e\phi(\bm{r}_c, t) = \mathcal{E}_{\pm}(\bm{r}_c, \bm{k}_c) + \bra{W_{\pm}}\Delta\widehat{H}\ket{W_{\pm}} - e\phi(\bm{r}_c, t)
\end{equation}
\noindent
In this case, the energetic correction $\Delta\mathcal{E} = \bra{W_{\pm}}\Delta\widehat{H}\ket{W_{\pm}}$ is the orbital magnetization energy of the wavepacket. 

Following the discussion of Ref. \cite{xiao2010berry}, the expectation value of an operator is 
\begin{equation}
    \mathcal{O} = \int_{\bm{\xi}}\mathcal{D}(\bm{\xi})f(\bm{\xi})\bra{W(\bm{\xi})}\hat{\mathcal{O}}\,\delta(\bm{r} - \hat{\bm{r}})\ket{W(\bm{\xi})}.
\end{equation}
\noindent
By expanding $\delta(\hat{\bm{r}} - \bm{r}_c)$ to first order, the expectation value of the operator consists of the operator evaluated at $\bm{r}_c$ and a dipole correction \cite{xiao2010berry}, as seen in Eq. \eqref{currents}, where the dipole correction takes the form of the orbital magnetization.

For simplicity, we drop the "c" label on the coordinates and define the phase-space vector $\bm{\xi}\equiv (r_x,r_y,r_z,k_x,k_y,k_z)$. 
Starting with the semiclassical Lagrangian
\begin{equation}
    L_\pm(\bm{\xi})=\bra{W_\pm(\bm{\xi})}i\hbar\dfrac{d}{d t}-\widehat{H}\ket{W_\pm(\bm{\xi})}.
    \label{lagrange}
\end{equation}
\noindent
we can arrive at the equations of motion of $\bm{\xi}$ 
\begin{equation}
    \sum_\beta \Gamma_{\alpha\beta}^{\pm}(\bm{\xi})\dot{\xi}_\beta=\dfrac{1}{\hbar}\nabla_{\xi_\alpha}(\widetilde{\mathcal{E}}_{\pm}(\bm{\xi})-e\phi(\bm{r},t))
\end{equation}
\noindent
using the methodology of Ref.    \cite{sundaram1999wave}.
Here
\begin{equation}
\te{\Gamma}^{\pm}(\bm{\xi}) = \begin{pmatrix}
0 & \Omega^{\pm}_{r_xr_y}-\frac{e}{\hbar}F_{xy} & \Omega^{\pm}_{r_xr_z}+\frac{e}{\hbar}F_{zx} & \Omega^{\pm}_{r_xk_x}-1 & \Omega^{\pm}_{ r_xk_y } & \Omega^{\pm}_{ r_xk_z } \\
-\Omega^{\pm}_{r_x r_y} +\frac{e}{\hbar}F_{xy} & 0 & \Omega^{\pm}_{r_yr_z} -\frac{e}{\hbar}F_{yz}& \Omega^{\pm}_{ r_yk_x} & \Omega^{\pm}_{ r_y k_y }-1 & \Omega^{\pm}_{ r_y k_z }  \\
-\Omega^{\pm}_{r_xr_z}-\frac{e}{\hbar}F_{zx} & -\Omega^{\pm}_{r_yr_z}+\frac{e}{\hbar}F_{yz} & 0 & \Omega^{\pm}_{r_zk_x} & \Omega^{\pm}_{r_zk_y} & \Omega^{\pm}_{r_zk_z}-1 \\
-\Omega^{\pm}_{r_x k_x}+1 & -\Omega^{\pm}_{ r_y k_x} & -\Omega^{\pm}_{ r_z k_x} &  0 & \Omega^{\pm}_{ k_x k_y } & \Omega^{\pm}_{ k_x k_z }  \\
- \Omega^{\pm}_{ r_x k_y } & - \Omega^{\pm}_{r_y k_y} +1& -\Omega^{\pm}_{r_zk_y} & - \Omega^{\pm}_{ k_x k_y } & 0 & \Omega^{\pm}_{ k_y k_z } \\
-\Omega^{\pm}_{r_xk_z} & -\Omega^{\pm}_{r_yk_z} & - \Omega^{\pm}_{r_zk_z}+1 & -\Omega^{\pm}_{k_xk_z} & -\Omega^{\pm}_{k_yk_z} & 0
\end{pmatrix}
\end{equation}
\noindent
The phase space Berry curvatures $\Omega^\pm_{\xi_\alpha\xi_\beta}=\partial_{\xi_\alpha}\mathcal{A}^\pm_{\xi_\beta}-\partial_{\xi_\beta}\mathcal{A}^\pm_{\xi_\alpha}$,
defined in terms of the Berry connection $\mathcal{A}^\pm_{\xi_\alpha}=i\bra{u_\pm(\bm{\xi})}\partial_{\xi_\alpha}\ket{u_\pm(\bm{\xi})}$, can be written
more conveniently as 
\begin{equation}
\Omega^\pm_{\xi_\alpha\xi_\beta}(\bm{\xi}) = \pm\frac12\hat{\bm{d}}\cdot(\partial_{\xi_{\alpha}}\hat{\bm{d}}\times\partial_{\xi_{\beta}}\hat{\bm{d}}),
    \label{curv 2}
\end{equation}
\noindent
The magnetic field is defined via the electromagnetic field tensor
\begin{equation}
    F_{ij}=\partial_{r_i}A_j-\partial_{r_j}A_i.
\end{equation}
\noindent
We have used script $\mathcal{A}$ to denote Berry connection and text $\bm{A}$ to denote electromagnetic vector potential. 
Moving forward, we will omit writing the band index unless necessary.
In order to isolate the equations of motion, $\dot{\bm{r}}_c$ and $\dot{\bm{k}}_c$, we must invert $\te{\Gamma}$. 
The inverse matrix $\te{\Gamma}^{-1}$ can be written in block form
\begin{align}
    \te{\Gamma}^{-1}(\bm{\xi})=\dfrac{1}{\mathcal{D}(\bm{\xi})}\begin{pmatrix}
\te{\mathbf{K}}(\bm{\xi}) & \te{\mathbf{S}}(\bm{\xi})\\
-\te{\mathbf{S}}^T(\bm{\xi}) & \te{\mathbf{R}}(\bm{\xi})\\
\end{pmatrix}
\label{inverse}
\end{align}
\noindent
where
\begin{align}
\mathbf{K}_{ij}(\bm{\xi}) &=\Omega_{k_ik_j}-\sum_l(\Omega_{k_ik_j}\Omega_{r_lk_l} +\Omega_{k_jk_l}\Omega_{r_lk_i} -\Omega_{k_ik_l}\Omega_{r_lk_j})|\epsilon_{ijl}| \nonumber \\
\mathbf{R}_{ij}(\bm{\xi}) &=(\Omega_{r_ir_j}-\frac{e}{\hbar}F_{ij})-\sum_l\bigg((\Omega_{r_ir_j}-\frac{e}{\hbar}F_{ij})\Omega_{r_lk_l} +(\Omega_{r_jr_l}-\frac{e}{\hbar}F_{jl})\Omega_{r_ik_l} -(\Omega_{r_ir_l}-\frac{e}{\hbar}F_{il})\Omega_{r_jk_l}\bigg)|\epsilon_{ijl}| \nonumber \\
\mathbf{S}_{ij}(\bm{\xi}) &=(1-\sum_l\Omega_{r_lk_l})\delta_{ij}+\Omega_{r_jk_i} -\frac12\delta_{ij}\sum_{nm}|\epsilon_{inm}|\bigg(\Omega_{k_nk_m}(\Omega_{r_nr_m}-\frac{e}{\hbar}F_{nm})-\Omega_{r_nk_n}\Omega_{r_mk_m}+\Omega_{r_nk_m}\Omega_{r_mk_n}\bigg)
\nonumber \\ &
+\sum_l|\epsilon_{ijl}|\bigg(\Omega_{k_ik_l}(\Omega_{r_jr_l}-\frac{e}{\hbar}F_{jl})+\Omega_{r_jk_l}\Omega_{r_lk_i}-\Omega_{r_jk_i}\Omega_{r_lk_l}\bigg) \nonumber \\
\mathcal{D}(\bm{\xi})&=\text{pf}(\Gamma(\bm{\xi}))=\bigg|1-\sum_i\Omega_{r_i k_i}+
    \frac12\sum_{i,j}\bigg(1-\delta_{ij}\bigg)\bigg(\Omega_{r_i k_i}\Omega_{r_j k_j}-(\Omega_{r_i r_j}-\frac{e}{\hbar}F_{ij})\Omega_{k_i k_j}-\Omega_{r_i k_j}\Omega_{r_j k_i}\bigg) \nonumber\\
&+\sum_{ijk,\alpha\beta\gamma}\epsilon_{ijk}\epsilon_{\alpha\beta\gamma}\bigg(\frac14\Omega_{k_{\alpha} k_{\beta}}(\Omega_{r_i r_j}-\frac{e}{\hbar}F_{ij})\Omega_{r_k k_{\gamma}}-\frac16\Omega_{r_i k_{\alpha}}\Omega_{r_j k_{\beta}}\Omega_{r_k k_{\gamma}}\bigg)\bigg|
\label{full block matrices}
\end{align}
\noindent
where $i, j, k, l \in \{x, y, z\}$. $\te{\mathbf{K}}$, $\te{\mathbf{R}}$, and $\te{\mathbf{S}}$ are each  $3 \times 3$ matrices, and thus $\te{\Gamma}^{-1}$ is $6 \times 6$.

The notation $|\epsilon_{ijk}|$ serves to enforce that $i\neq j \neq k$. Note the placement of indices explicitly in r-space and k-space in the above expressions. Terms like $\Omega_{r_ir_i}$ vanish automatically due to the antisymmetry of the Berry curvature, but a term like $\Omega_{r_i k_i}$ doesn't necessarily vanish. This is why we include the notation $|\epsilon_{ijk}|$ where appropriate.

For either $J\ll \lambda$ or $J\gg \lambda$ the leading order contribution to the Berry curvatures are small quantities. For example in the regime $\lambda\ll J$
\begin{align}
    \Omega_{k_{\alpha}k_{\beta}} &\sim a^2(\lambda/J)^2 \nonumber\\
    \Omega_{r_{\alpha}k_{\beta}} &\sim (a/L_s)(\lambda/J)\nonumber\\
    \Omega_{r_{\alpha}r_{\beta}} &\sim 1/L_s^2.
    \label{Berry scaling}
\end{align}
\noindent
where we note that $L_s$ is the largest length scale in the problem.  We therefore see that the largest part of each of these block matrices is the term with the lowest power of Berry curvatures. 
Thus, $\mathbf{K}_{ij} \approx \Omega_{k_i k_j}$, $\mathbf{R}_{ij} \approx \Omega_{r_i r_j} - \frac{e}{\hbar}F_{ij}$, and $\mathbf{S}_{ij} \approx \delta_{ij}$.

We end this Appendix by comments on why we write $\te{\Gamma}^{-1}$ in the form shown in eqs.~\eqref{gamma inverse} and \eqref{inverse} with 
the pfaffian $\mathcal{D} = \text{pf}\ \Gamma$ factored out. The usual expression for the inverse of a matrix  
$\te{\Gamma}^{-1}=(1/\det\te{\Gamma})\text{adj}\te{\Gamma}$ is in terms of the adjugate of the matrix divided by the 
determinant, rather than involving the pfaffian. We already explained in the main text that the expression we use
makes transparent the cancellation with the same pfaffian that appears in the phase space volume.

We now show why this expression is quite natural natural for the inverse of any $2n\times 2n$ real antisymmetric matrix $\te{\Gamma}$.  
While we cannot diagonalize $\te{\Gamma}$ with a real matrix, using a suitable orthogonal matrix $\te{Q}$ we can write its as
\begin{equation}
    \te{\Gamma}=\te{Q} \, \te{\Lambda}\, \te{Q}^\intercal 
\end{equation}
\noindent
where $\te{\Lambda}$ is block diagonal:
\begin{equation}
    \te{\Lambda}=\begin{pmatrix}
        0 & \lambda_1 & & & \hdots & & 0\\
        -\lambda_1 & 0 & & & & & \\
         & & 0 & \lambda_2 & & & \\
         & & -\lambda_2 & 0 & & & \\
        \vdots & & & & \ddots & & \vdots\\
         & & & & & 0 & \lambda_n \\
         0 & & & & \hdots & -\lambda_n & 0
    \end{pmatrix}
\end{equation}
\noindent
Its pfaffian is $\Lambda=\lambda_1\lambda_2...\lambda_n$, and
the inverse of $\te{\Lambda}$ is given by
\begin{equation}
    \te{\Lambda}^{-1}=\begin{pmatrix}
        0 & -1/\lambda_1 & \hdots & & & 0\\
        1/\lambda_1 & 0 & & & &  \\
        \vdots & & \ddots & & & \vdots\\
         & & & 0 & -1/\lambda_n \\
         0& & \hdots & 1/\lambda_n & 0\\
    \end{pmatrix}=\frac{1}{\lambda_1\lambda_2...\lambda_n}\begin{pmatrix}
        0 & -\lambda_2...\lambda_n & & & \hdots & & 0\\
        \lambda_2...\lambda_n & 0 &  & & \\
         \vdots & & & & \ddots & & \vdots\\
         & & & & & 0 & -\lambda_1...\lambda_{n-1} \\
         0 & & & & \hdots & \lambda_1...\lambda_{n-1} & 0
    \end{pmatrix}
\end{equation}
Because $\te{Q}$ is orthogonal, we have 
$\text{pf}\, \Gamma = \text{pf}\, \Lambda$ and
$\te{\Gamma}^{-1}=\te{Q} \, \te{\Lambda}^{-1}\, \te{Q}^\intercal$.

\section{Derivation of Phase Space Measure}\label{app measure}

Liouville's theorem states that the phase-space distribution function must remain constant along phase-space trajectories such that the phase-space volume 
\begin{equation}
\mathcal{V}=\mathcal{C}(\bm{\xi})\,\mathrm{d}r_x\wedge \mathrm{d}r_y \wedge \mathrm{d}r_z \wedge  \mathrm{d}k_x \wedge  \mathrm{d}k_y \wedge  \mathrm{d}k_z
\end{equation}
\noindent
remains invariant under time evolution; i.e., the Lie derivative with respect to the velocity field $\dot{\bm{\xi}}$ vanishes: $\mathcal{L}_{\dot{\bm{\xi}}} \mathcal{V}=0$. The velocity field is defined
\begin{equation}
    \dot{\bm{\xi}}=\frac{\mathrm{d}\xi_i}{\mathrm{d}t}\frac{\partial}{\partial \xi^i}\equiv\frac{\mathrm{d}}{\mathrm{d}t}.
\end{equation}
\noindent
We will employ the notation $\mathcal{L}_t\equiv \mathcal{L}_{\dot{\bm{\xi}}}$ to elude to the fact that the change of the volume form along the flow of $\dot{\bm{\xi}}$ measures the change of the volume form in time.

In the absence of phase-space curvatures, $\bm{k}$ and $\bm{r}$ are conjugate variables that satisfy the canonical Poisson bracket relations: $\{r_i,k_j\}=\delta_{ij}$ such that Liouville's theorem is trivially satisfied ($\mathcal{L}_t \mathcal{V}=0$) with $\mathcal{C}(\bm{\xi})=1$.  However, in the presence of phase space curvatures, $\bm{k}$ and $\bm{r}$ are no longer conjugate variables ($\{r_i,k_j\}\neq\delta_{ij})$ and thus, in order to satisfy Liouville's theorem, the phase-space measure must be altered.

Our strategy to compute $\mathcal{C}(\bm{\xi})$ in terms of phase-space curvatures is to propose the solution $\mathcal{C}(\bm{\xi})=\mathcal{D}(\bm{\xi})=\sqrt{\det(\te{\Gamma})}$ and then prove that it satisfies the constraint $\mathcal{L}_t \mathcal{V}=0$.
We define the two-form $\omega=\sum_{ij}\Gamma_{ij}\mathrm{d}\xi_i\wedge \mathrm{d}\xi_j/2$ and relate it to $\mathcal{V}$ by noting that
\begin{align}
    \frac{\omega\wedge\omega\wedge\omega}{6}& = \dfrac{1}{2^33!}\Gamma_{ij}\Gamma_{lk}\Gamma_{nm} \mathrm{d}\xi_i \wedge \mathrm{d}\xi_j \wedge \mathrm{d}\xi_l \wedge \mathrm{d}\xi_k\wedge \mathrm{d}\xi_n\wedge \mathrm{d}\xi_m \nonumber \\
    &=\sqrt{\det(\te{\Gamma})} \,\mathrm{d}r_x\wedge \mathrm{d}r_y \wedge \mathrm{d}r_z \wedge  \mathrm{d}k_x \wedge  \mathrm{d}k_y \wedge  \mathrm{d}k_z \nonumber \\
    &= \mathcal{V}.
\end{align}
\noindent
This relation leads to the Lie derivative being $\mathcal{L}_t \mathcal{V} = \omega \wedge \omega \wedge \mathcal{L}_t \omega$.
We next use the Cartan formula to rewrite the Lie derivative on $\omega$
\begin{equation}
    \mathcal{L}_t \omega= i_t (\text{d} \omega) + \text{d}  (i_t \omega)
\end{equation}
where $\text{d}$ is the exterior derivative and $i_t$ is the interior product \cite{nakahara2003}.  To prove that $\mathcal{L}_t (V)$ vanishes, we first note that
\begin{align}
    \text{d}(i_t\omega)&=\dfrac{1}{2} \sum_{ij}\text{d}(\Gamma_{ij}\dot{\xi}_i \mathrm{d}\xi_j-\Gamma_{ij}\dot{\xi}_j \mathrm{d}\xi_i) =\sum_{ij}\text{d}(\Gamma_{ij}\dot{\xi}_i \mathrm{d}\xi_j) \nonumber \\
    &=-\sum_i\dfrac{1}{\hbar}\text{d}\bigg(\partial_{\xi_i}(\varepsilon(\bm{\xi})-e\phi(\bm{r}))\bigg) \mathrm{d}\xi_i \nonumber \\
    &=-\sum_{ij}\dfrac{1}{\hbar}\partial_{\xi_j}\partial_{\xi_i}(\varepsilon(\bm{\xi})-e\phi(\bm{r})) \mathrm{d}\xi_j\wedge \mathrm{d}\xi_i 
\end{align}
which vanishes as it is a product of a symmetric and an anti-symmetric tensor in $i,j$. Secondly, we may write $\omega=\text{d}\eta$ with
\begin{equation}
    \eta=\sum_i(\mathcal{A}_{\xi_i}+P_i)d\xi_i
\end{equation}
where $\mathcal{A}_i$ are the Berry connections and $\bm{P}=(-k_x,-k_y,-k_z,r_x,r_y,r_z)/2$ such that $\text{d} \omega=\text{d}^2\eta=0$.  We have shown that $\text{d}(i_t\omega)=0$ and $\text{d}\omega=0$ so that by Eq. \eqref{cartan}, $\mathcal{L}_t\omega=0$ and hence, $\mathcal{L}_t \mathcal{V}=0$. This fixes the measure up to an overall constant that can be determined by noting that in the absence of curvatures, $\mathcal{D}(\bm{\xi}) = 1$.

\section{Solution to Boltzmann Equation}\label{app boltzmann}

There are two driving fields present in the problem, $\bm{E} - 1/e \bm{\nabla}_r \mu$ and $-\bm{\nabla}_r T$. We write the distribution function as
\begin{equation}
    f^{\pm}(\bm{\xi}) = f_{\text{leq}}^{\pm}(\bm{\xi}) + \delta f^{\pm}(\bm{\xi}),
\end{equation}
\noindent
where $\delta f^{\pm}(\bm{\xi})$ represents deviations from local equilibrium. From now on, we will omit writing the band index and $\bm{\xi}$ argument, unless necessary. Which wish to solve for $\delta f$ in the regime of linear response, so we shall drop any terms which are higher order in electric field, chemical potential gradient, or temperature gradient. 

Thus the Boltzmann equation in the relaxation time approximation can be written as
\begin{equation}
    -\frac{\delta f}{\tau} = \bm{\nabla}_k (f_{\text{leq}} + \delta f)\cdot \dot{\bm{k}} + \bm{\nabla}_r (f_{\text{leq}} + \delta f)\cdot \dot{\bm{r}}.
\end{equation}
\noindent
Analyzing spatial gradient term $\bm{\nabla}_r (f_{\text{leq}} + \delta f)$:
\begin{equation}
    \bm{\nabla}_r (f_{\text{leq}}+\delta f) = \bm{\nabla}_r \widetilde{\mathcal{E}}\partial_{\widetilde{\mathcal{E}}}(f_{\text{leq}}+\delta f) + \bm{\nabla}_r T\partial_T (f_{\text{leq}}+\delta f) + \bm{\nabla}_r \mu\partial_{\mu} (f_{\text{leq}}+\delta f).
\end{equation}
\noindent
The terms $\bm{\nabla}_r T\partial_T (\delta f)$ and $\bm{\nabla}_r \mu\partial_{\mu} (\delta f)$ are higher order and shall be dropped. Thus, to linear order in the driving fields, we have
\begin{align}
   -\frac{\delta f}{\tau} =& \bm{\nabla}_k f_{\text{eq}} \cdot \dot{\bm{k}}^{(1)} + \bm{\nabla}_k (\delta f) \cdot \dot{\bm{k}}^{(0)} + \bm{\nabla}_r \widetilde{\mathcal{E}}\partial_{\widetilde{\mathcal{E}}}f_{\text{eq}} \cdot \dot{\bm{r}}^{(1)} + \bm{\nabla}_r T \partial_T f_{\text{eq}} \cdot \dot{\bm{r}}^{(0)} + \bm{\nabla}_r \mu \partial_{\mu} f_{\text{eq}} \cdot \dot{\bm{r}}^{(0)} + \bm{\nabla}_r \widetilde{\mathcal{E}}\partial_{\widetilde{\mathcal{E}}}(\delta f)\cdot \dot{\bm{r}}^{(0)},
\end{align}
\noindent
where $\dot{\bm{r}}^{(0)}$ and $\dot{\bm{k}}^{(0)}$ indicate the equations of motion to 0th order in $\bm{E}$, and, $\dot{\bm{r}}^{(1)}$ and $\dot{\bm{k}}^{(1)}$ are to first order in $\bm{E}$, and where we have replaced $f_{\text{leq}}(\bm{\xi})$ with $f_{\text{eq}}(\bm{k})$ as we only consider small perturbation of the fields about a homogeneous background.  For example, we consider $\mu(\bm{r})\approx \mu_0+\bm{\nabla}_r \mu$, which to leading order in gradients allows us to neglect the dependence in $f_{\text{leq}}(\bm{\xi})$ of $\bm{\nabla}_r \mu$.  We collect all $\delta f$ terms:
\begin{equation}
    \bigg(1 + \tau(\dot{\bm{k}}^{(0)} \cdot \bm{\nabla}_k + \dot{\bm{r}}^{(0)} \cdot \bm{\nabla}_r\widetilde{\mathcal{E}}\partial_{\widetilde{\mathcal{E}}})\bigg)\delta f = -\tau \bigg((\dot{\bm{k}}^{(1)}\cdot \bm{\nabla}_k \widetilde{\mathcal{E}} + \dot{\bm{r}}^{(1)}\cdot \bm{\nabla}_r \widetilde{\mathcal{E}} + \dot{\bm{r}}^{(0)}\cdot\bm{\nabla}_r \mu)\partial_{\widetilde{\mathcal{E}}}+ \dot{\bm{r}}^{(0)}\cdot\bm{\nabla}_r T\partial_T \bigg) f_{\text{leq}}
\end{equation}

\noindent
Next we define the operator $1 + \mathbb{P} \equiv 1 + \tau(\dot{\bm{k}}^{(0)} \cdot \bm{\nabla}_k + \dot{\bm{r}}^{(0)} \cdot \bm{\nabla}_r\widetilde{\mathcal{E}}\partial_{\widetilde{\mathcal{E}}})$. 
To first order in driving fields we find
\begin{equation}
    \delta f = -\tau(1+\mathbb{P})^{-1} \bigg((\dot{\bm{k}}^{(1)}\cdot \bm{\nabla}_k \widetilde{\mathcal{E}} + \dot{\bm{r}}^{(1)}\cdot \bm{\nabla}_r \widetilde{\mathcal{E}} + \dot{\bm{r}}^{(0)}\cdot\bm{\nabla}_r \mu)\partial_{\widetilde{\mathcal{E}}}+ \dot{\bm{r}}^{(0)}\cdot\bm{\nabla}_r T\partial_T \bigg) f_{\text{eq}}.
\end{equation}

\textbf{Solution to Boltzmann Equation in the Small SOC Regime:} We now simplify this linear response expression by considering the regime $\lambda\ll J\ll t$ and $a\ll \ell \ll L_s$.  Recall that the energy corrections $\Delta \mathcal{E}$ are smaller than $\mathcal{E}$ by a factor of $a/L_s$, so we approximate $\widetilde{\mathcal{E}}(\bm{\xi})\approx \mathcal{E}(\bm{\xi})$. It will also be useful to keep in mind the expansion of $\mathcal{E}(\bm{\xi})$ in powers of $\lambda$:

\begin{equation}
    \mathcal{E}_{\pm}(\bm{\xi})\approx \varepsilon_{\pm}(\bm{k})\pm\frac{\lambda}{at}\hat{m}_i(\bm{r})\chi_{ji}\partial_{k_j}\varepsilon + \mathcal{O}(\lambda^2)
    \label{energy expansion}
\end{equation}
\noindent
where $\varepsilon_{\pm}(\bm{k})=\varepsilon(\bm{k}) \pm J$. This expansion shows that $\bm{\nabla}_r \mathcal{E}\sim \lambda/L_s$.  Similarly using the scaling relations in Eq. \eqref{Berry scaling} we see that $\dot{\bm{k}}^{(1)}\cdot \bm{\nabla}_k  \mathcal{E} \ll \dot{\bm{r}}^{(1)}\cdot \bm{\nabla}_r \mathcal{E}$.  The operator $\mathbb{P}$ appears in combination with the derivatives $\partial_T f_{\text{eq}}$ and $\partial_{\mathcal{E}}f_{\text{eq}}$. At temperatures much less than the Fermi temperature, these functions are peaked in a region $k_B T$ around the Fermi energy, and so we take $(1/\hbar)\bm{\nabla}_k \mathcal{E}\sim v_F$. Again using the scaling relations in Eq. \eqref{Berry scaling} and the scaling $\bm{\nabla}_r\sim (1/L_s)$ and $\bm{\nabla}_k\sim a$, and keeping in mind that $v_F\tau=\ell$, we find that the leading order contribution to the operator $\mathbb{P}$ is at most of order $\ell/L_s$.  This allows us to make the approximation  $(1+\mathbb{P})^{-1}\approx (1-\mathbb{P})$.  This analysis leads determines the leading order contribution to $\delta f$ that we may write as
\begin{equation}
    \delta f\approx -\frac{\tau}{\hbar}(1-\mathbb{P})\bm{\nabla}_k\mathcal{E} \cdot \bigg(\bm{\nabla}_r T\  \partial_T -(e \bm{E} - \bm{\nabla}_r \mu)\ \partial_{\mathcal{E}}\bigg) f_{\text{eq}}.
    \label{ZJ lin resp}
\end{equation}
\noindent
We now consider the contributions to $\delta f$ which are independent of $\mathbb{P}$ and then those which are dependent on $\mathbb{P}$.

\textbf{Contributions independent of $\mathbb{P}$:} We now consider the contribution to $\delta f$ which is independent of $\mathbb{P}$. Its contribution to the electric conductivity is
\begin{equation}
    \sigma_{ij}(\bm{r})=-\frac{e^2\tau}{\hbar^2}\sum_{\pm}\int_{\bm{k}}(\partial_{k_i}\mathcal{E}_{\pm})(\partial_{k_j}\mathcal{E}_{\pm})\partial_{\mathcal{E}}f_{\text{eq}}.
\end{equation}
\noindent
This expression is symmetric in the indices $i$ and $j$. Thus, this term can only contribute to the symmetric part of the electric conductivity. The same argument follows for the thermoelectric and thermal conductivities.  The leading order contribution in this regime is giving by Eq. \eqref{longitudinal cond}.

\textbf{Contributions dependent on $\mathbb{P}$:} Extrinsic contributions to the totally antisymmetric pieces to the conductivities must arise from the terms including $\mathbb{P}$ in $\delta f$. The operator $\mathbb{P}$ contains $\bm{\nabla}_r$ and $\bm{\nabla}_k$ which act on every term to the right of $\mathbb{P}$.  We argue that the derivatives $\bm{\nabla}_k$ and $\partial_{\mathcal{E}}$ on $(\bm{\nabla}_r T \partial_T -(e \bm{E}-\bm{\nabla}_r \mu) \partial_{\mathcal{E}}) f_{\text{eq}}$ are sub leading order and thus we only consider derivatives that act on $\bm{\nabla}_k\mathcal{E}$.  Due to this action the terms proportional to $\dot{\bm{r}}^{(0)}$ in $\mathbb{P}$ are largely suppressed in $\delta f$.

This leads to the largest terms in $\delta f$ appearing from terms in $\mathbb{P}$ that are second order in our small parameters and deriving from $\dot{\bm{k}}^{(1)}$. First, we consider terms which scale in $\mathbb{P}$ like $(\ell/L_s)(\lambda/t)$. We find
\begin{align}
    \delta f^{(\lambda^1)}&\approx -\frac{\tau^2}{\hbar^2}(\bm{\nabla}_r \mathcal{E}\cdot\bm{\nabla}_k)\bm{\nabla}_k \mathcal{E}\cdot\bigg(\bm{\nabla}_rT\partial_T-(e\bm{E}-\bm{\nabla}_r\mu)\partial_{\mathcal{E}}\bigg)f_{\text{eq}}\nonumber\\
    &\approx- \frac{\hbar\tau^2\lambda}{a t} (\partial_{r_l}\hat{m_i})\chi_{ji}v_j M_{ln}^{-1}\bigg(\partial_{r_n}T\partial_T -(e E_n-\partial_{r_n}\mu)\partial_{\mathcal{E}}\bigg)f_{\text{eq}}
\end{align}
\noindent
where the second line used the expansion of the energy \eqref{energy expansion} to first order in $\lambda$. Adopting the notation $\zeta\in\{\sigma,\alpha,\kappa\}$, the contribution to the conductivities which arises from $\delta f^{(\lambda^1)}$ is 

\begin{equation}
    \zeta_{mn}=\frac{\hbar\tau^2\lambda}{at}\sum_{\pm}\int_{\bm{k}}s\, \chi_{ji}v_j M_{ln}^{-1}v_m(\partial_{\varepsilon} X_{\pm}^{\zeta})\int_{\bm{r}}\partial_{r_l}\hat{m}_i
    \label{div m}
\end{equation}
\noindent
where $X^{\zeta}$ is defined in eq. \eqref{dist fns}. The total derivative in the real space integral would vanish for periodic textures and at most is a boundary contribution to the conductivities. As a reminder these terms scale in $\mathbb{P}$ as $(\ell/L_s)(\lambda/t)$.
We next turn to the terms in $\mathbb{P}$ which scale like $(\ell/L_s) (a/L_s)$.

Letting $\lambda=0$, spatial derivatives on the energy vanish and $\mathcal{E}_{\pm}(\bm{\xi})=\varepsilon_{\pm}(\bm{k})$. We therefore find that $\mathbb{P}\approx \tau\dot{\bm{k}}^{(0)}_{\lambda=0}\cdot\bm{\nabla}_k$, where $(\dot{\bm{k}}^{(0)}_{\lambda=0})_i=(\Omega_{r_i r_j}-(e/\hbar)F_{ij})\partial_{k_j} \varepsilon/\hbar$. The largest term in this limit in $\delta f$ arising from $\mathbb{P}$ is 

\begin{equation}
      \delta f^{(\lambda^0)}=\hbar\tau^2 \bigg(\Omega_{r_i r_j} - \frac{e}{\hbar}F_{ij}\bigg) v_j M^{-1}_{im}  \bigg(\partial_{r_m} T \partial_T -(e E_m - \partial_{r_m}\mu) \partial_{\varepsilon}\bigg) f_{\text{eq}}.
\label{df soln}
\end{equation}
\noindent
Recall that $\partial_T f_{\text{eq}}=-(\varepsilon-\mu_0)/T_0\,\partial_{\varepsilon}f_{\text{eq}}$. Thus, the final expression for the $\tau^2$ term is

\begin{equation}
    \delta f^{(\lambda^0)}=-\hbar\tau^2 \bigg(\Omega_{r_i r_j} - \frac{e}{\hbar}F_{ij}\bigg) v_j M^{-1}_{im}  \bigg(\frac{(\varepsilon - \mu_0)}{T_0}\partial_{r_m} T + (e E_m - \partial_{r_m}\mu) \bigg)\partial_{\varepsilon} f_{\text{eq}}.
\end{equation}

\section{Relating Equilibrium Current to Bound Current}\label{app jeq} 

We start with the expression in Eq. \eqref{currents}. In global equilibrium, the distribution function is the Fermi-Dirac distribution, and $\dot{\bm{r}}^{(0)}$ is given by Eq. \eqref{EOM} without the electric field.
We have
\begin{align}
    \bm{j}_{\text{eq}}^e(\bm{r})&=-e\sum_{\pm}\int_{\bm{k}}\bigg[\mathcal{D}f_{\text{eq}}(\bm{\xi}) \dot{\bm{r}}^{(0)}+\bm{\nabla}_r \times \bigg( \mathcal{D}f_{\text{eq}}(\bm{\xi})\bm{\mathfrak{m}}(\bm{\xi})\bigg)\bigg] \nonumber\\
    \bm{j}_{\text{eq}}^Q(\bm{r})&=\sum_{\pm}\int_{\bm{k}}\bigg[\mathcal{D}f_{\text{eq}}(\bm{\xi}) \dot{\bm{r}}^{(0)}(\mathcal{E}-\mu_0)+\bm{\nabla}_r \times \bigg(\mathcal{D}f_{\text{eq}}(\bm{\xi})(\mathcal{E}-\mu_0)\bm{\mathfrak{m}}(\bm{\xi})\bigg)\bigg]
\end{align}

Our goal is to manipulate the equilibrium currents so that they take a manifestly solenoidal form: $\bm{j}\sim \bm{\nabla}\times \bm{\mathcal{F}}$ for some vector field $\bm{\mathcal{F}}$. Since the second terms in both currents are already written in this manner, we only need to focus on rewriting the first terms. Let us first focus on the electric current. Recall that $\partial g_{\text{eq}}/\partial \mathcal{E} = f_{\text{eq}}$ and $(\partial/\partial \mathcal{E}) (h_{\text{eq}}-\mu_0 g_{\text{eq}}) = (\mathcal{E} - \mu_0)f_{\text{eq}}$.  We have
\begin{align}
   \bigg(-e\int_{\bm{k}}\mathcal{D}(\partial_{\mathcal{E}} g_{\text{eq}})\dot{\bm{r}}^{(0)}\bigg)_i&=-\frac{e}{\hbar}\int_{\bm{k}}\bigg( \mathbf{K}_{ij}\partial_{r_j} \mathcal{E}\partial_{\mathcal{E}}g_{\text{eq}} + \mathbf{S}_{ij}\partial_{k_j} \mathcal{E}\partial_{\mathcal{E}}g_{\text{eq}}\bigg) \nonumber\\
    &=-\frac{e}{\hbar}\int_{\bm{k}}\bigg( \mathbf{K}_{ij}\partial_{r_j}g_{\text{eq}} + \mathbf{S}_{ij}\partial_{k_j}g_{\text{eq}}\bigg)
\end{align}
\noindent
We use the chain rule to rewrite these terms:
\begin{align}
    \mathbf{K}_{ij}\partial_{r_j}g_{\text{eq}} &= \partial_{r_j}(\mathbf{K}_{ij} g_{\text{eq}}) - (\partial_{r_j}\mathbf{K}_{ij}) g_{\text{eq}} \nonumber\\
    \mathbf{S}_{ij}\partial_{k_j}g_{\text{eq}} &= \partial_{k_j}(\mathbf{S}_{ij} g_{\text{eq}}) - (\partial_{k_j}\mathbf{S}_{ij}) g_{\text{eq}}
\end{align}
\noindent
The term $\partial_{k_j}(\mathbf{S}_{ij} g_{\text{eq}})$ can be neglected in the integral because the Billouin zone has no boundary.  To leading order in the curvatures $\mathbf{K}_{ij}\approx \Omega_{k_ik_j}$, which affords us the relation $\partial_{r_j}\mathbf{K}_{ij} = -\partial_{k_j}\mathbf{S}_{ij}$.  Using these results we have
\begin{equation}
    -e\int_{\bm{k}}\mathcal{D}(\partial_{\mathcal{E}}g_{\text{eq}})\dot{r}_i^{(0)} \approx -\frac{e}{\hbar}\partial_{r_j}\int_{\bm{k}} (\Omega_{k_i k_j}g_{\text{eq}})
\end{equation}
Writing the antisymmetric k-space Berry curvature in its vector form we may express the electric charge current as
\begin{equation}
    \bm{j}_{\text{eq}}^e(\bm{r})=-\frac{e}{\hbar}\bm{\nabla}_{r}\times\sum_{\pm}\int_{\bm{k}} \bigg(g_{\text{eq}}\bm{\Omega}_k +\hbar\mathcal{D}f_{\text{eq}}\bm{\mathfrak{m}}\bigg)
\end{equation}
\noindent
The analysis proceeds in the same way for the thermal currents, with $(h_{\text{eq}}-\mu_0 g_{\text{eq}})$ in this case taking the place of $g_{\text{eq}}$.

\section{Anomalous Transport Coefficients}\label{app anom}
In this appendix, we show how the anomalous Hall transport coefficients can be rewritten to assume the form shown in Eq. \eqref{transport coeffs}. We show the calculation for the anomalous Hall conductivity as an example. Starting with the expression for the anomalous contribution to $\bm{\sigma}_H$ from Eq. \eqref{full Hall}:
\begin{equation}
    \sigma_i^A = -\frac{e^2}{\hbar}\sum_{\pm}\int_{\bm{\xi}}f_{\text{eq}}\Omega_{k_i}
\end{equation}
We substitute the expression for the k-space Berry curvature, to lowest order in $(\lambda/J)$:
\begin{align}
    \Omega_{k_i} = \frac12 \epsilon_{ijk}\Omega_{k_j k_k} &= \frac14 \epsilon_{ijk}\epsilon_{lmn} \hat{d}_l (\partial_{k_j}\hat{d}_m)(\partial_{k_k}\hat{d}_n)\nonumber\\
    &= -\frac{\hbar^4}{4} \bigg(\frac{\lambda}{Jat}\bigg)^2\epsilon_{ijk} \epsilon_{lmn} \hat{m}_l(\bm{r}) \chi_{m \mu}\chi_{n \nu}M_{j \mu}^{-1}M_{k \nu}^{-1}
\end{align}
At this order of the calculation, the only spatial dependence is coming from the magnetic texture in the berry curvature. Noting that the elements of $\te{\chi}$ are constants, we find
\begin{equation}
    \sigma_i^A = \frac{e^2\hbar^3}{4}  \bigg(\frac{\lambda}{Jat}\bigg)^2 \epsilon_{ijk}\epsilon_{lmn} \chi_{m \mu}\chi_{n \nu}\sum_{\pm}\int_{\bm{r}} \hat{m}_l(\bm{r}) \int_{\bm{k}}  f_{\text{eq}} M_{j \mu}^{-1}M_{k\nu}^{-1}
\end{equation}
\noindent
We focus on manipulating the k-space integral:
\begin{align}
f_{\text{eq}}M_{j\mu}^{-1}M_{k\nu}^{-1}=& \frac{1}{\hbar^2}\bigg[\partial_{k_j}(f_{\text{eq}} v_{\mu}\partial_{k_k}v_\nu)- (\partial_{k_j}f_{\text{eq}}) v_{\mu}\partial_{k_k}v_\nu - f_{\text{eq}} v_{\mu}(\partial_{k_j}\partial_{k_k}v_\nu)\bigg]
\end{align}
The first term vanishes upon integrating over k-space.  The term containing $\partial_{k_j}\partial_{k_k}v_\mu$ is symmetric in the $j, k$ indices. Combined with the antisymmetric $\epsilon_{ijk}$, this term must also vanish. Thus, the k-space integral becomes
\begin{equation}
    \int_{\bm{k}}  f_{\text{eq}} M_{j \mu}^{-1}M_{k\nu}^{-1}=-\frac{1}{\hbar}\int_{\bm{k}} (\partial_{k_j}f_{\text{eq}}) v_{\mu} M_{k\nu}^{-1}
\end{equation}
\noindent
We note that $\partial_{k_j}f_{\text{eq}} = \hbar v_j \partial_{\varepsilon}f_{\text{eq}}$ such that we may express the anamolous contribution to the electric conductivity as
\begin{equation}
    \sigma_i^A = -\frac{e^2\hbar^3}{4}\bigg(\frac{\lambda}{Jat}\bigg)^2 \epsilon_{ijk}\epsilon_{lmn}\chi_{m \mu}\chi_{n \nu}\sum_{\pm}\int_{\bm{r}}\hat{m}_l(\bm{r}) \int_{\bm{k}} (\partial_{\varepsilon}f_{\text{eq}})v_j M_{k \nu}^{-1} v_{\mu}
\end{equation}
\noindent
Next, we use $\epsilon_{lmn}\chi_{m\mu}\chi_{n\nu}=-\epsilon_{lmn}\chi_{n\mu}\chi_{m\nu}$,
which allows us to write
$\epsilon_{lmn}\chi_{m\mu}\chi_{n\nu}=\frac12\epsilon_{lmn}\epsilon_{\gamma\alpha\beta}\epsilon_{\gamma\mu\nu}\chi_{m\alpha}\chi_{n\beta}.$
Inserting this into our expression for the conductivity,
\begin{align}
    \sigma_i^A &= -\frac{e^2\hbar^3}{8}\bigg(\frac{\lambda}{Jat}\bigg)^2 \epsilon_{ijk}\epsilon_{lmn}\epsilon_{\gamma\alpha\beta}\epsilon_{\gamma\mu\nu}\chi_{m \alpha}\chi_{n \beta}\sum_{\pm}\int_{\bm{r}}\hat{m}_l(\bm{r}) \int_{\bm{k}} (\partial_{\varepsilon}f_{\text{eq}})v_j M_{k \nu}^{-1} v_{\mu}
    = F_{il}\Lambda_l
\end{align}

\noindent
where $\bm{\Lambda}$ is defined in Eq. \eqref{lambda}. The procedure for expressing $\alpha_H^A$ and $\kappa_H^A$ in this form follows from a similar analysis.

\section{Kelvin, Wiedemann-Franz, and Mott Relations}\label{app Wiedemann}

The full expressions (including both longitudinal and transverse parts) for the transport coefficients are
\begin{align}
    \sigma_{lm}&=-e^2\sum_{\pm}\int_{\bm{\xi}} \bigg[\tau\bigg(v_m-\hbar\tau\sum_{ij}(s\Omega_{r_ir_j} - \frac{e}{\hbar}F_{ij})v_j M_{im}^{-1}\bigg)\bigg( \partial_{\varepsilon}f_{\text{eq}}\bigg)v_l + \epsilon_{lnm}s\frac{\Omega_{k_n}}{\hbar}\partial_{\mu_0}g_{\text{eq}}\bigg] \nonumber\\
    \alpha_{lm} &= e\sum_{\pm}\int_{\bm{\xi}}\bigg[\tau\bigg(v_m-\hbar\tau\sum_{ij}(s\Omega_{r_ir_j} - \frac{e}{\hbar}F_{ij})v_j M_{im}^{-1}\bigg)\bigg(\frac{(\varepsilon-\mu_0)}{T_0} \partial_{\varepsilon}f_{\text{eq}}\bigg)v_l + \epsilon_{lnm}s\frac{\Omega_{k_n}}{\hbar}\partial_{T_0}g_{\text{eq}}\bigg]\nonumber\\
    \beta_{lm} &= e\sum_{\pm}\int_{\bm{\xi}}\bigg[\tau\bigg(v_m - \hbar \tau\sum_{ij}(s\Omega_{r_ir_j}-\frac{e}{\hbar}F_{ij})v_j M_{im}^{-1}\bigg)\bigg((\varepsilon - \mu_0)\partial_{\varepsilon}f_{\text{eq}}\bigg)v_l +  \epsilon_{lnm}s\frac{\Omega_{k_n}}{\hbar}T_0 \partial_{T_0}g_{\text{eq}}\bigg] \nonumber\\
    \kappa_{lm} &= - \sum_{\pm}\int_{\bm{\xi}} \bigg[\tau\bigg(v_m-\hbar\tau\sum_{ij}(s\Omega_{r_ir_j} - \frac{e}{\hbar}F_{ij})v_j M_{im}^{-1}\bigg)\bigg(\frac{(\varepsilon - \mu_0)^2}{T_0} \partial_{\varepsilon}f_{\text{eq}}\bigg)v_l + \epsilon_{lnm}s\frac{\Omega_{k_n}}{\hbar}\partial_{T_0}(h_{\text{eq}}-\mu_0 g_{\text{eq}})\bigg]
    \label{full transport} 
\end{align}

\noindent
where for simplicity we have suppressed the dependence of various quantities on the band index $s=\pm$ and phase space variables $\bm{\xi}$. The Kelvin relation ($\te{\bm{\beta}} = T_0 \te{\bm{\alpha}}$) is clearly seen from the expressions above. 

\medskip
\textbf{Wiedemann-Franz.} The Wiedemann-Franz relationship states that $\te{\bm{\kappa}} \approx \frac{\pi^2}{3}\frac{k_B^2 T}{e^2}\te{\bm{\sigma}}$. We start by comparing the intrinsic contributions

\begin{equation}
    \kappa_{lm}^{(\text{int})}= -\frac{1}{\hbar}\sum_{\pm}\int_{\bm{\xi}}\partial_{T_0}(h_{\text{eq}}-\mu_0 g_{\text{eq}})\epsilon_{lnm}s\Omega_{k_n}; \quad\quad \sigma_{lm}^{(\text{int})}= -\frac{e^2}{\hbar}\sum_{\pm}\int_{\bm{\xi}}\partial_{\mu_0}g_{\text{eq}}\epsilon_{lnm}s\Omega_{k_n}.
\end{equation}
\noindent
First we rewrite  $\kappa_{lm}^{(\text{int})}$ as

\begin{equation}
    \kappa_{lm}^{(\text{int})}= -\frac{1}{\hbar}\sum_{\pm}\int d\eta\,\int_{\bm{\xi}}\partial_{T_0}\bigg(h_{\text{eq}}(\eta)-\mu_0 g_{\text{eq}}(\eta)\bigg)\partial_{\eta}\Theta(\eta-\varepsilon)\epsilon_{lnm}s\Omega_{k_n},
\end{equation}

\noindent
where $\Theta(x)$ is the Heaviside step function, and integrate by parts to write $\kappa_{lm}^{(\text{int})}$ in terms of a function $G_{lm}(\eta)$ as

\begin{equation}
    \kappa_{lm}^{(\text{int})}=\sum_{\pm}\int d\eta\,G_{lm}(\eta)\partial_{T_0}f_\text{eq}(\eta)(\eta-\mu_0)
\end{equation}

\noindent
with

\begin{equation}
    G_{lm}(\eta)=\frac{1}{\hbar}\int_{\bm{\xi}}\Theta(\eta-\varepsilon)\epsilon_{lnm}s\Omega_{k_n}.
\end{equation}

\noindent
Similarly, we may write $\sigma_{lm}^{(\text{int})}$ as

\begin{equation}
\sigma_{lm}^{(\text{int})}= -e^2\sum_{\pm}\int d\eta\, G_{lm}(\eta) \partial_{\eta}f_{\text{eq}}(\eta).
\end{equation}

\noindent

Next, we note that derivatives with respect to $T_0$ and $\eta$ of $f_{\text{eq}}(\eta)$ satisfy the relation $\partial_{T_0}f(\eta)=-(\eta-\mu_0)\partial_{\eta}f(\eta)/T_0$. At low temperatures $k_B T_0 \ll E_F$, the function $\partial_{\eta}f_{\text{eq}}(\eta)$ is sharply peaked in the small region of width $k_B T_0$ about $\mu_0$. We therefore perform a low temperature expansion around $\mu_0$. The intrinsic contributions can be expanded as follows:

\begin{align}
    \kappa_{lm}^{(\text{int})} &= -\sum_{\pm}\int d\eta\ \bigg[G_{lm}(\mu_0) + (\eta - \mu_0)\partial_{\mu_0}G_{lm}\bigg] \frac{(\eta - \mu_0)^2}{T_0} \partial_{\eta}f_{\text{eq}}(\eta)+...\nonumber\\
    \sigma_{lm}^{(\text{int})} &= -e^2\sum_{\pm}\int d\eta\ \bigg[G_{lm}(\mu_0) + (\eta - \mu_0)\partial_{\mu_0}G_{lm}\bigg]\partial_{\eta}f_{\text{eq}}(\eta)+...
\end{align}
\noindent
The term involving $(\eta - \mu_0)^3$ in $\kappa_{lm}^{(\text{int})}$ and the term involving $(\eta - \mu_0)$ in $\sigma_{lm}^{(int)}$ both vanish upon integration. The expressions are thus approximately

\begin{align}
    \kappa_{lm}^{(\text{int})} &\approx -\sum_{\pm} G_{lm}(\mu_0)\int d\eta\ \frac{(\eta - \mu_0)^2}{T_0}  \partial_{\eta}f_{\text{eq}}(\eta) = -k_B^2 T_0 \frac{\pi^2}{3}\sum_{\pm} G_{lm}(\mu_0)\nonumber\\
    \sigma_{lm}^{(\text{int})} &\approx -e^2\sum_{\pm} G_{lm}(\mu_0)\int d\eta\  \partial_{\eta}f_{\text{eq}}(\eta) = -e^2 \sum_{\pm} G_{lm}(\mu_0).
\end{align}
\noindent
Note the relationship between the expressions for $\kappa_{lm}^{(int)}$ and $\sigma_{lm}^{(int)}$.

Next, we analyze the extrinsic contributions:

\begin{align}
    \sigma_{lm}^{\text{(ext)}}&=-e^2\sum_{\pm}\int_{\bm{\xi}} \bigg[\tau\bigg(v_m-\hbar\tau\sum_{ij}(s\Omega_{r_ir_j} - \frac{e}{\hbar}F_{ij})v_j M_{im}^{-1}\bigg)\bigg( \partial_{\varepsilon}f_{\text{eq}}\bigg)v_l\bigg] \nonumber\\
    \kappa_{lm}^\text{(ext)} &= - \sum_{\pm}\int_{\bm{\xi}} \bigg[\tau\bigg(v_m-\hbar\tau\sum_{ij}(s\Omega_{r_ir_j} - \frac{e}{\hbar}F_{ij})v_j M_{im}^{-1}\bigg)\bigg(\frac{(\varepsilon - \mu_0)^2}{T_0} \partial_{\varepsilon}f_{\text{eq}}\bigg)v_l\bigg].
    \label{full transport} 
\end{align}

\noindent
We rewrite these contributions in terms of a function $G'_{lm}(\eta)$ as

\begin{equation}
    \kappa_{lm}^{(\text{ext})}=\sum_{\pm}\int d\eta\, G'_{lm}(\eta)\frac{(\eta-\mu_0)^2}{T_0}\partial_{\eta}f_{\text{eq}}(\eta); \quad\quad \sigma_{lm}^{(\text{ext})}=e^2\sum_{\pm}\int d\eta\, G'_{lm}(\eta)\partial_{\eta}f_{\text{eq}}(\eta)
\end{equation}
\noindent
with 

\begin{equation}
    G'_{lm}(\eta)=-\int_{\bm{\xi}} \bigg[\tau\bigg(v_m-\hbar\tau\sum_{ij}(s\Omega_{r_ir_j} - \frac{e}{\hbar}F_{ij})v_j M_{im}^{-1}\bigg)\delta(\eta-\varepsilon)
\end{equation}

\noindent
where $\delta(x)$ is the Dirac delta function. Again, we perform a low temperature expansion. The extrinsic contributions are approximated as

\begin{align}
\kappa_{lm}^{(\text{ext})}&\approx\sum_{\pm}G_{lm}(\mu_0)\int d\eta\, \frac{(\eta-\mu_0)^2}{T_0}\partial_{\eta}f_{\text{eq}}(\eta)=\dfrac{\pi^2}{3}k_B^2 T_0\sum_\pm G_{lm}(\mu_0) \nonumber\\
    \sigma_{lm}^{(\text{ext})}&\approx e^2\sum_{\pm}G_{lm}(\mu_0)\int d\eta\, \partial_{\eta}f_{\text{eq}}(\eta)=e^2\sum_{\pm}G_{lm}(\mu_0).
\end{align}
\noindent
Overall, we therefore find that the Wiedemann-Franz relation $\kappa_{lm}/\sigma_{lm}=(\pi^2/3)(k_B^2 T_0/e^2)$ is satisfied for both the intrinsic and extrinsic contributions.

\medskip
\textbf{The Mott Relation.} The Mott relation states $\te{\bm{\alpha}} \approx -\frac{\pi^2}{3}\frac{k_B^2 T}{e}\frac{\partial}{\partial \mu_0}\te{\bm{\sigma}}$. We shall follow a procedure similar to the one shown for the Wiedemann-Franz relation. We start by comparing the intrinsic contributions:

\begin{equation}
    \alpha_{lm}^{(\text{int})}= \frac{e}{\hbar}\sum_{\pm}\int_{\bm{\xi}}\epsilon_{lnm}s\Omega_{k_n}\partial_{T_0}g_{\text{eq}}; \quad\quad \sigma_{lm}^{(\text{int})}= \frac{e^2}{\hbar}\sum_{\pm}\int_{\bm{\xi}}\epsilon_{lnm}s\Omega_{k_n}f_{\text{eq}}.
\end{equation}
\noindent
\noindent
Following the same procedure as was used in the Wiedemann-Franz section, we rewrite $\alpha_{lm}^{(\text{int})}$ and $\sigma_{lm}^{(\text{int})}$ as

\begin{equation}
    \alpha_{lm}^{(\text{int})}=-\dfrac{e}{\hbar}\sum_{\pm}\int d\eta \, \mathcal{G}_{lm}(\eta)\partial_{T_0} f_{\text{eq}}(\eta); \quad\quad \sigma_{lm}^{(\text{int})}=-\dfrac{e^2}{\hbar}\sum_{\pm}\int d\eta \, \mathcal{G}_{lm}(\eta)\partial_{\eta} f_{\text{eq}}(\eta)
\end{equation}

\noindent
with 

\begin{equation}
\mathcal{G}_{lm}(\eta)=\int_{\bm{\xi}}\epsilon_{lnm}s\Omega_{k_n}\Theta(\eta-\varepsilon).
\end{equation}

\noindent
Again using the fact
that $\partial_{T_0}f_{\text{eq}}(\eta)=-(\eta-\mu_0)\partial_{\eta}f_{\text{eq}}(\eta)/T_0$, and employing a low temperature expansion about $\mu_0$, we find

\begin{equation}
    \alpha_{lm}^{(\text{int})} \approx \frac{e}{\hbar}k_B^2 T_0 \frac{\pi^2}{3}\sum_{\pm}\partial_{\mu_0}\mathcal{G}_{lm}(\mu_0); \quad\quad
    \sigma_{lm}^{(\text{int})} \approx -\frac{e^2}{\hbar}\sum_{\pm}\mathcal{G}_{lm}(\mu_0).
\end{equation}

A similar form can be determined for the extrinsic contributions:

\begin{equation}
    \alpha_{lm}^{(\text{ext})} = \frac{e}{\hbar}\sum_{\pm}\int d\eta\, \mathcal{G}'_{lm}(\eta)\frac{(\eta-\mu_0)}{T_0}\partial_{\eta}f_{\text{eq}}(\eta); \quad\quad
    \sigma_{lm}^{(\text{ext})} = -\frac{e^2}{\hbar}\sum_{\pm}\int d\eta\, \mathcal{G}'_{lm}(\eta)\partial_{\eta}f_{\text{eq}}(\eta)
\end{equation}
\noindent
with

\begin{equation}
    \mathcal{G}'_{lm}(\eta)=e\sum_{\pm}\int_{\bm{\xi}}\tau\bigg(v_m-\hbar\tau\sum_{ij}(s\Omega_{r_ir_j} - \frac{e}{\hbar}F_{ij})v_j M_{im}^{-1}\bigg)\delta(\eta-\varepsilon).
\end{equation}
\noindent
After the low temperature expansion, we have

\begin{equation}
    \alpha_{lm}^{(\text{ext})} \approx \frac{e}{\hbar}\sum_{\pm}\frac{\partial \mathcal{G}'_{lm}}{\partial \mu_0}\int d\eta\, \frac{(\eta-\mu_0)^2}{T_0}\partial_{\eta}f_{\text{eq}}(\eta); \quad\quad \sigma_{lm}^{(\text{ext})}\approx -\frac{e^2}{\hbar}\sum_{\pm}\mathcal{G}'_{lm}(\mu_0)\int d\eta\, \partial_{\eta}f_{\text{eq}}(\eta)
\end{equation}
\noindent
which integrates to 
\begin{equation}
    \alpha_{lm}^{(\text{ext})} \approx \frac{e}{\hbar}k_B^2 T_0\frac{\pi^2}{3}\sum_{\pm}\frac{\partial \mathcal{G}'_{lm}(\mu_0)}{\partial \mu_0}; \quad\quad \sigma_{lm}^{(\text{ext})} \approx -\frac{e^2}{\hbar}\sum_{\pm}\mathcal{G}'_{lm}(\mu_0)
\end{equation}

\noindent
We see that $\alpha_{lm} \approx -\frac{\pi^2}{3}\frac{k_B^2 T}{e}\frac{\partial}{\partial \mu_0}\sigma_{lm}$ and the Mott relation is indeed satisfied for both the intrinsic and extrinsic contributions.

\section{Scaling of Transport Coefficients}\label{app scaling}

Here we take the regime of interest to be defined by the inequalities $a \ll \ell \ll L_s$ and $\lambda \ll J < t \sim E_F$. 
 The resistivity scaling shown in Table \ref{scaling table} derives from the approximations $\rho_{xx}\approx(1/\sigma_{xx})$ and $\rho_{xy}\approx \sigma_{xy}/\sigma_{xx}^2$. The ordinary conductivities are in agreement with the usual 
Drude-Boltzmann results, meaning that $\sigma_{xx}\sim(e^2/\hbar)(k_F^2 \ell)$ and $\sigma_{xy}^O=\omega_c\tau\sigma_{xx}$, which leads to the results for the longitudinal and ordinary Hall resistivities. The scaling of the anomalous Hall conductivity is most easily seen from Eq. \eqref{an hall}, which leads to the scaling for the anomalous Hall resistivity. The scaling for the topological Hall conductivity is found directly from Eq. \eqref{transport coeffs}. Using the fact that $n^{\text{sk}}\sim(1/L_s^2)$, this analysis proceeds as

\begin{equation}
    \sigma^T\sim e^2\tau^2\hbar \frac{1}{L_s^2}\int_{\bm{k}}\frac{\partial f_{\text{eq}}}{\partial \varepsilon}v^2 M^{-1}\sim \frac{e^2}{\hbar}\frac{1}{L_s^2} (\tau v_F)^2 k_F^2 a \sim\bigg(\frac{e^2}{\hbar}k_F\bigg)\bigg(\frac{\ell}{L_s}\bigg)^2, 
\end{equation}
\noindent
and the topological Hall resistivity scaling follows.  To determine the Seebeck and Nernst scaling, we employ the Mott relation:

\begin{equation}
    S\sim\frac{k_B^2 T_0}{e}\frac{(\partial/\partial \mu)\sigma_{xx}}{\sigma_{xx}}; \quad\quad\quad N\sim\frac{k_B^2 T_0}{e}\frac{\partial}{\partial \mu}\bigg(\frac{\sigma_{xy}}{\sigma_{xx}}\bigg)
_0\end{equation}
\noindent
The derivative of the conductivities with respect to the chemical potential is governed by the distribution function $f_{\text{eq}}$. The higher the temperature $T_0$ compared to the Fermi temperature $T_F$, the more electronic states are available to participate in transport. The correct factor is thus $T_0/T_F$. This combined with the conductivity scaling leads to the entries for $S$ and $N$ found in Table \eqref{scaling table}. 
 Finally, the scaling of the thermal conductivities is determined from the Wiedemann-Franz law (see Appendix \ref{app Wiedemann}).

\section{Dependence of Transport Coefficients on the Chemical Potential}\label{hallC}

\begin{figure}
    \centering    \includegraphics[width=0.65\textwidth]{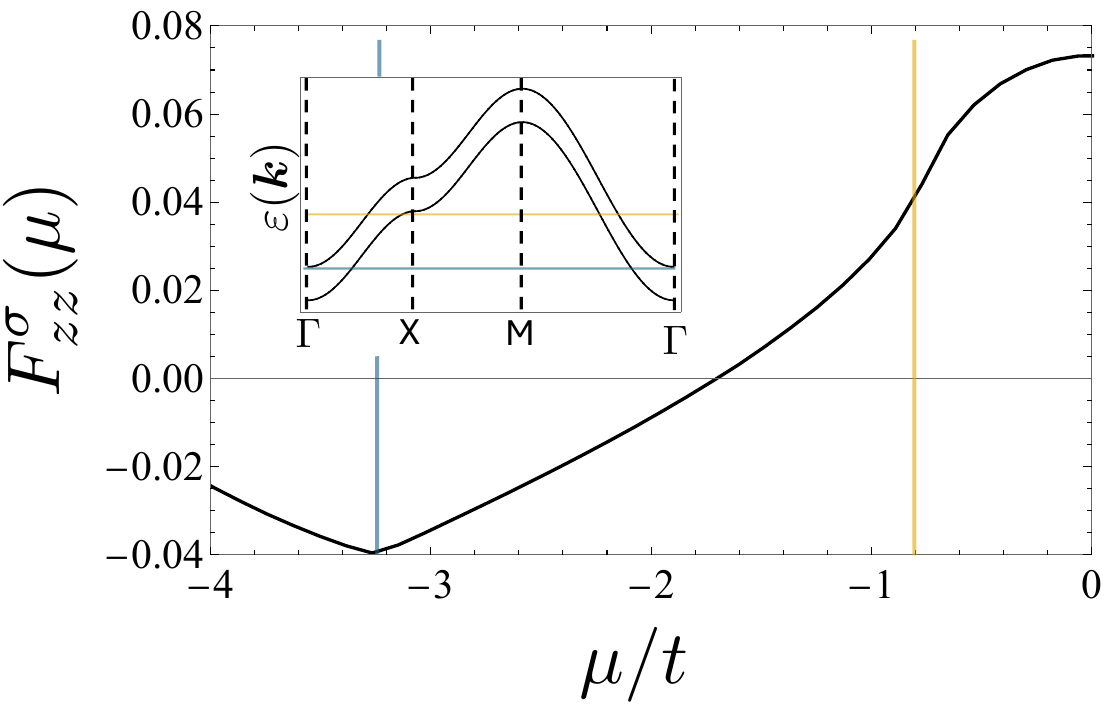}
    \caption{{\bf Chemical Potential Dependence of Hall Conductivity.}  The Hall conductivity for a planar 2D system is proportional to $F_{zz}^\sigma$.  Here we plot $F_{zz}^\sigma$ in units of $e^2t^2a^2/\hbar^4$ for a system of nearest neighbor interactions, t, on the 2D square lattice with lattice constant $a$.  Here we take $t=3J/4$.  Vertical lines denote critical values where $\mu$ crosses a band edge (blue) or Van Hove singularity (gold) (see inset for the band structure of the model along the high symmetry lines of the Brillouin zone).  Data available at \cite{DataRep}.}
    \label{fig:HallC}
\end{figure}

Having shown the validity of the Kevlin, Mott, and Wiedemann-Franz relations (see Section \ref{sec-transport}), determining the chemical potential (or density) dependence of the transport coefficients can all be determined from the dependence of the Hall conductivity with respect $\mu$.  While the Kelvin relation is true at all temperatures the Mott and Wiedemann-Franz relations are valid in the regime $k_BT_0\ll E_F$.  Here we give an illustrative example of the chemical potential dependence of the Hall conductivity using a tight-binding model of nearest neighbor hopping $t$ on the 2D square lattice.  The anomalous and topological electric Hall coefficients in the plane of the material are proportional to $F^\sigma_{zz}$ (see eq. \eqref{transport coeffs}).  The contributions that enter $F_{zz}^\sigma$ couple to the electronic dispersion in the absence of the systems spin-orbit coupling: $\varepsilon_{\pm}(\bm{k})=-2t(\cos(k_x a)+\cos(k_ya))\pm J$.  Figure \ref{fig:HallC} show the dependence of $F^\sigma_{zz}$ with respect to the chemical potential.  This model contains a particle hole like symmetry that restricts $F^\sigma_{zz}$ to be an odd function of $\mu$ and therefore for simplicity we only show the dependence for $\mu<0$.

Due to the validity of the Wiedemann-Franz relations the thermal conductivity $\te{\kappa}$ in the low temperature limit is expected to follow a similar dependence (see \eqref{wf}).  On the other hand, the thermoelectric conductivity is obtained via a Mott relation by differentiation with respect to $\mu$ ($\te{\alpha}\sim\partial_\mu\te{\sigma}$) (see \cite{addison2023theory} for details).

\end{document}